\documentclass[prb,twocolumn,showpacs,superscriptaddress,nofootinbib,floatfix]{revtex4}

\usepackage{epsfig}
\usepackage{amsmath,amssymb}
\usepackage{graphicx}
\usepackage{footnote}
\usepackage{grffile}
\usepackage{color}
\usepackage{dsfont}
\usepackage{mathtools}
\usepackage{verbatim}

\newcommand{\tw}{\textwidth}
\newcommand{\cw}{\columnwidth}

\newcommand{\mc}[1]{\mathcal{#1}}

\newcommand{\IM}{\operatorname{Im}}

\newcommand{\ceq}[1]{Eq.~(\ref{eq:#1})}
\newcommand{\cfg}[1]{Fig.~\ref{fig:#1}}

\newcommand{\ud}{{ \uparrow\downarrow }}		
\newcommand{\udc}{{ \overline{\uparrow\downarrow} }}	
\newcommand{\uu}{{ \uparrow\uparrow }}			

\newcommand{\xph}{{ \overline{ph} }}			

\newcommand{\K}{\mc{K}_1}				
\newcommand{\KUD}{\mc{K}_{1,\ud}}			
\newcommand{\KPP}{\mc{K}_{1, pp}}			
\newcommand{\KPPUD}{\mc{K}_{1, pp, \ud}}		
\newcommand{\KPHUD}{\mc{K}_{1, ph, \ud}}		
\newcommand{\KXPHUD}{\mc{K}_{1, \xph, \ud}}		
\newcommand{\KK}{\mc{K}_2}				
\newcommand{\KKUD}{\mc{K}_{2,\ud}}			
\newcommand{\KKB}{\overline{\mc{K}}_2}			
\newcommand{\KKPP}{\mc{K}_{2, pp}}			
\newcommand{\KKBPP}{\overline{\mc{K}}_{2, pp}}		
\newcommand{\KKPH}{\mc{K}_{2, ph}}			
\newcommand{\KKPPUD}{\mc{K}_{2, pp, \ud}}		
\newcommand{\KKPHUD}{\mc{K}_{2, ph, \ud}}		
\newcommand{\KKXPHUD}{\mc{K}_{2, \xph, \ud}}		
\newcommand{\R}{\mc{R}} 				
\newcommand{\RUD}{\mc{R}_{\ud}}				
\newcommand{\RPP}{\mc{R}_{pp}} 				

\DeclareMathOperator*{\sumint}{%
   \mathchoice%
  {\ooalign{$\displaystyle\sum$\cr\hidewidth$\displaystyle\int$\hidewidth\cr}}
  {\ooalign{\raisebox{.14\height}{\scalebox{.7}{$\textstyle\sum$}}\cr\hidewidth$\textstyle\int$\hidewidth\cr}}
  {\ooalign{\raisebox{.2\height}{\scalebox{.6}{$\scriptstyle\sum$}}\cr$\scriptstyle\int$\cr}}
  {\ooalign{\raisebox{.2\height}{\scalebox{.6}{$\scriptstyle\sum$}}\cr$\scriptstyle\int$\cr}}
}

\begin{document}

\title{High-frequency asymptotics of the vertex function: \mbox{diagrammatic parametrization and algorithmic implementation}}

\newcommand{\TUVienna}{\affiliation{Institute for Solid State Physics, Vienna University of Technology, 1040 Vienna, Austria}}
\newcommand{\UniTueb}{\affiliation{Institut f\"ur Theoretische Physik and Center for Quantum Science, Universit\"at T\"ubingen, Auf der Morgenstelle 14, 72076 T\"ubingen, Germany}}
\newcommand{\CCQ}{\affiliation{Center for Computational Quantum Physics, Flatiron institute, Simons Foundation, 162 5th Ave., New York, 10010 NY, USA}}
\newcommand{\MPIStutt}{\affiliation{Max-Planck-Institute for Solid State Research, 70569 Stuttgart, Germany}}
\newcommand{\RussianQC}{\affiliation{Russian Quantum Center, 143025 Skolkovo, Russia}}
\newcommand{\UniHamburg}{\affiliation{I. Institute of Theoretical Physics, University of Hamburg, 20355 Hamburg, Germany}}

\author{Nils Wentzell} 		\TUVienna \UniTueb \CCQ

\author{Gang Li} 		\TUVienna

\author{Agnese Tagliavini} 	\TUVienna \UniTueb

\author{Ciro Taranto} 		 \MPIStutt

\author{Georg Rohringer} 	\TUVienna \RussianQC \UniHamburg

\author{Karsten Held} 		\TUVienna

\author{Alessandro Toschi}  	\TUVienna

\author{Sabine Andergassen} 	\UniTueb

\begin{abstract}
Vertex functions are a crucial ingredient of several forefront many-body algorithms in condensed matter physics. However, the full treatment of their frequency and momentum dependence severely restricts numerical calculations.
A significant advancement requires an efficient treatment of the high-frequency asymptotic behavior of the vertex functions. In this work, we first provide a detailed diagrammatic analysis of the high-frequency structures and their physical interpretation. Based on these insights, we propose a parametrization scheme, which captures the whole high-frequency domain for arbitrary values of the Coulomb interaction and electronic density, and we discuss the details of its algorithmic implementation in many-body solvers based on parquet-equations as well as functional renormalization group schemes. Finally, we assess its validity by comparing our results for a single impurity Anderson model with exact diagonalization calculations.  The proposed parametrization is pivotal for the algorithmic development of all quantum many-body methods based on vertex functions arising from both local and non-local static microscopic interactions as well as effective dynamic interactions which uniformly approach a static value for large frequencies. In this way, our present technique can substantially improve vertex-based diagrammatic approaches including spatial correlations beyond dynamical mean-field theory.


\end{abstract}

\maketitle

\section{Introduction} 					
\label{sec:introduction}

One of the most challenging aspects in contemporary condensed matter research is the theoretical treatment of correlation effects in the non-perturbative regime. While the state-of-the-art theoretical tools allow for an accurate treatment of quantum many-body correlations in specific cases, their reliability is not guaranteed in general and is often limited to particular parameter regimes. 
In the last decade, several promising quantum field theoretical schemes have been proposed, but their actual implementation calls for a significant improvement of the current algorithmic procedures. 
In particular, most of the novel non-perturbative schemes are based on a Feynman diagrammatic expansion around a correlated starting point. 
This means to replace the bare electronic interaction with a dynamical effective one and the bare propagators by dressed ones,
which includes non-perturbatively, through the two-particle vertex function and the self-energy, a significant part of the correlations from the very beginning. 

From a more general perspective, the two-particle vertex has been the object of several focused studies\cite{Bickers2004,Tahvildar-Zadeh1997,Kunes2011,Rohringer2012,Kinza2013,Rohringer2014,Hummel2014,Li2016} in the last two decades. Apart from its technical importance for the above-mentioned development of new many-body approaches, it contains crucial physical information on its own. In fact, the main frequency and momentum structures of the vertex functions can be related to physical observables such as susceptibilities\cite{Rohringer2012}. Vertex corrections for the spin susceptibility\cite{Toschi2012} and the optical conductivity\cite{Kauch2020}, on the other hand, often play a crucial role in correlated electron systems. Moreover, vertex functions have also been used for analyzing certain features in spectral functions by means of the ``fluctuations diagnostics'' technique\cite{Gunnarsson2015} and recently unexpected divergences have been discovered in the low-frequency regime of the irreducible vertex functions\cite{Schaefer2013,PhysRevB.90.045143,Schaefer2016,Gunnarsson2016,PhysRevB.93.195105,Thunstroem2018,Chalupa2018, Springer2019, Chalupa2020b} in strongly disordered or correlated electron systems.

The importance of the two-particle vertices is unfortunately contrasted with the tremendous difficulties to treat these correlation functions numerically, which becomes even more challenging when turning to
multi-orbital systems in the future\cite{Kaufmann2017,PhysRevB.95.115107,Otsuki2019}. Hence, the development of efficient ways to include them in the current algorithms is mandatory. Two-particle vertex functions depend in general on three independent frequencies and momenta, and additionally on spin and orbital variables. Even in the $SU(2)$ symmetric case, the efficient computation of two-particle vertices becomes very challenging for a reasonably large system at low temperatures. Besides the storage, the inclusion of the asymptotic behavior during the computation represents a major issue and requires to exploit a detailed understanding of its underlying structure\cite{Rohringer2012,Kinza2013,Rohringer2014,Hummel2014} in order to reduce the numerical effort\cite{Li2016}.

To this aim, we present a detailed diagrammatic analysis of the frequency and momentum structures of the vertex functions, focusing on the algorithmic aspects relevant for the development of improved parametrization schemes. 
After defining the general guidelines of the algorithmic implementation, we present applications for many-body solvers based on functional renormalization group\cite{Metzner2012} (fRG) schemes as well as parquet equations\cite{Dominicis1964,Dominicis1964a,Bickers2004,Yang2009,Li2016}.
In particular, the validity of the proposed parametrization algorithm could be quantitatively assessed by comparing our results for the single impurity Anderson model (SIAM) obtained by means of fRG and the parquet approximation\cite{Janis1999,Yang2009,Tam2013} (PA) with exact diagonalization (ED) calculations.
We emphasize that the identification of the relevant asymptotic structures of the vertex functions in frequency and momentum space and the resulting reduced parametrizations are extremely valuable for an efficient implementation of
several other many-body approaches beyond fRG and PA, such as the dynamical vertex approximation\cite{Toschi2007,Katanin2009a,Rohringer2011,Schaefer2015,Valli2015,Li2016,DelRe2019} (D$\Gamma$A), the one-particle irreducible (1PI) approach\cite{Rohringer2013}, DMF$^{2}$RG,\cite{Taranto2014,Vilardi2019} dual fermion\cite{Antipov2014,Rubtsov2008,Rubtsov2009,Hafermann2009a,Wentzell2015,Li2015} (DF), TRILEX\cite{Ayral2015,Ayral2016a} and QUADRILEX\cite{Ayral2016}.

The paper is organized as follows: We introduce the formalism and notation at the two-particle level in Sec.~\ref{sec:summary}, and the parametrization of the asymptotics
in Sec.~\ref{sec:param}. The generic implementation is presented in Sec.~\ref{sec:implem}, with some technical details specified in the appendices. 
After a short discussion of analytical results obtained in the atomic limit, we describe the specific implementations for the fRG and the parquet solvers. 
In Sec.~\ref{sec:compare_SIAM} we then provide a discussion of the obtained results together with a comparison to exact results of the SIAM. A conclusion and outlook is eventually provided in Sec.~\ref{sec:conclusion}.

\section{Diagrammatic formalism at the two-particle level}
\label{sec:summary}

In this section we define the various two-particle vertex functions which will be employed in this work, and recall the basic idea for investigating their frequency (and momentum) dependence. In particular, in Sec.~\ref{subsec:formalism} we introduce all vertex functions (reducible, irreducible in one channel, and fully irreducible) that constitute the parquet equations. Then, in Sec.~\ref{subsec:asympt}, we will concisely review the technique for describing the main frequency (and momentum) structures of all these vertex functions by means of lowest-order perturbation theory, as it was developed in Refs.~\onlinecite{Rohringer2012,Hummel2014}.

\subsection{Definitions and two-particle formalism}
\label{subsec:formalism}

In the following, we will restrict ourselves, for the sake of clarity, to one-band systems with a local Coulomb interaction. Let us, however, note that the methods presented later can be straightforwardly extended to more general systems including, e.g., more orbitals or non-local interactions. An application to models with retarded (frequency- or time-dependent) interactions, on the other hand, would require significant modifications.

We consider the Hamiltonian:
\begin{equation}
 \label{eq:defhamilt}
 \hat{\mathcal{H}}=\sum_{ij , \sigma}t_{ij}(c^{\dagger}_{i\sigma}c_{j\sigma}+c^{\dagger}_{j\sigma}c_{i\sigma})+\sum_i U_in_{i\uparrow}n_{i\downarrow},
\end{equation}
where $c^{(\dagger)}_{i\sigma}$ annihilates (creates) an electron with spin $\sigma$ at the lattice site $i$ and $n_{i\sigma}=c^{\dagger}_{i\sigma}c_{i\sigma}$. The hopping amplitude for an electron between the lattice sites $i$ and $j$ is denoted by $t_{ij}$ (for $i=j$ this corresponds to setting the energy-level for an electron at site $i$), while $U_i$ is a (site-dependent) local interaction between electrons of opposite spin.

From the Hamiltonian in \ceq{defhamilt} one retains the standard Hubbard model by choosing the parameters $t_{ij} = -t$ if $i$ and $j$ are nearest neighbors and $t_{ij}=0$ otherwise, and $U_i= U$ (site independent). The restriction $t_{ij}=V_j \delta_{i0}$, $t_{ii}=\varepsilon_i/2$ and $U_i=U\delta_{i0}$, on the other hand, corresponds to the SIAM, where the lattice site $i=0$ is the impurity. For these two situations let us introduce the notation used throughout this paper by considering the one-particle Green's function which, in the $SU(2)$ symmetric, time- and space-translational invariant case, is defined as
\begin{equation}
\label{equ:def1pgf}
 G(k)=-\sumint dx \frac{1}{Z} \langle {\mc T}[c_{i\sigma}(\tau)c^{\dagger}_{0\sigma}(0)]\rangle e^{ixk}.
\end{equation}
Here, $x$ denotes a generalized (imaginary) time/space index which, for the Hubbard model, corresponds to $x\!=\!(\tau,\mathbf{R})$ where $\tau$ is the imaginary time and $\mathbf{R}$ a lattice vector. For the SIAM we consider only local correlation functions for the impurity site (i.e., $i\!=\!0$) and, hence, $x\!=\!\tau$. $\langle {\mc T}\ldots \rangle\!=\!\text{Tr}({\mc T}e^{-\beta\hat{\mathcal{H}}}\ldots)$ indicates the thermal average, where $\beta\!=\!1/T$ is the inverse temperature and ${\mc T}$ the time-ordering operator. Further $Z\!=\!\text{Tr}(e^{-\beta\hat{\mathcal{H}}})$ denotes the partition function. The generalized fermionic frequency/momentum index $k$ is, for the Hubbard model, given by $k\!=\!(\nu,\mathbf{k})$ where $\nu$ is a fermionic Matsubara frequency and $\mathbf{k}$ denotes a momentum vector in the first Brillouin zone.
For the SIAM we have $k\!=\!\nu$. 
In the case of the Hubbard model, the integration/summation in Eq.~(\ref{equ:def1pgf}) is a generalized four-vector integration/summation over imaginary times/lattice sites $\sumint dx\equiv\sum_{\mathbf{R}}\int_0^{\beta}d\tau$. Correspondingly, we will consider in the following a generalized summation/integration over Matsubara frequencies/momenta $\sumint dk\equiv\frac{1}{\beta}\sum_{\nu}\frac{1}{V_{\text{BZ}}}\int_{\text{BZ}}d\mathbf{k}$ where BZ denotes the first Brillouin zone with the volume $V_{\text{BZ}}$. Finally, $xk$ is a short-hand notation for $xk\!\equiv\!\nu\tau\! - \!\mathbf{k}\!\cdot\!\mathbf{R}$ in the Hubbard model while for the SIAM $xk\!\equiv\!\nu\tau$.

\begin{figure*}
   \centering
   \begin{minipage}{0.185\textwidth}
      \hspace{0.1cm}
      \includegraphics[width=0.8\textwidth]{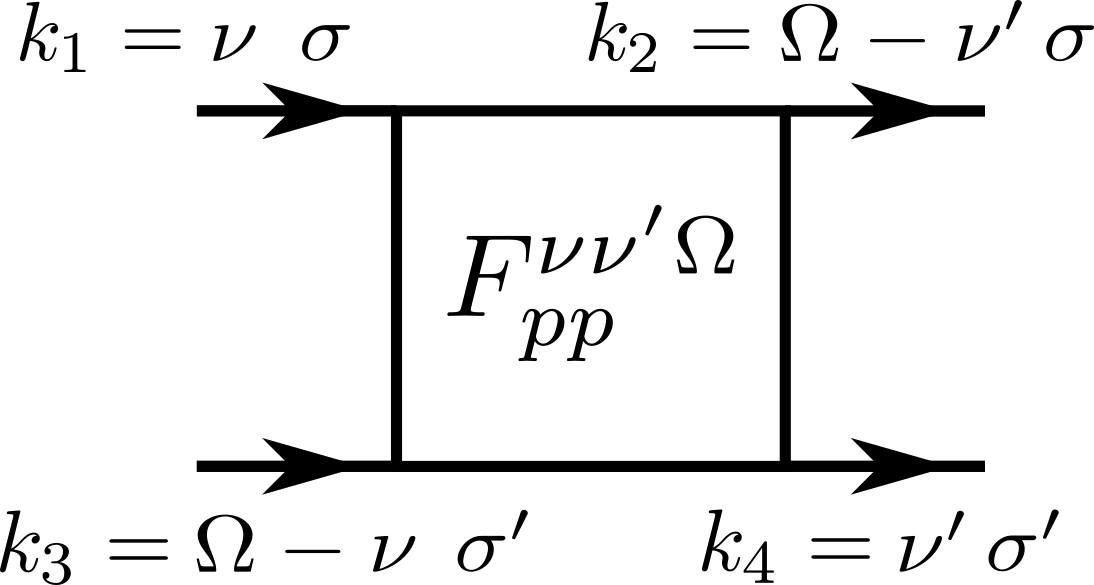}
   \end{minipage}
   \begin{minipage}{0.06\columnwidth}
      \textit{e.g.}
   \end{minipage}
   \begin{minipage}{0.17\textwidth}
      \hspace{-0.1cm}
      \includegraphics[width=0.55\textwidth]{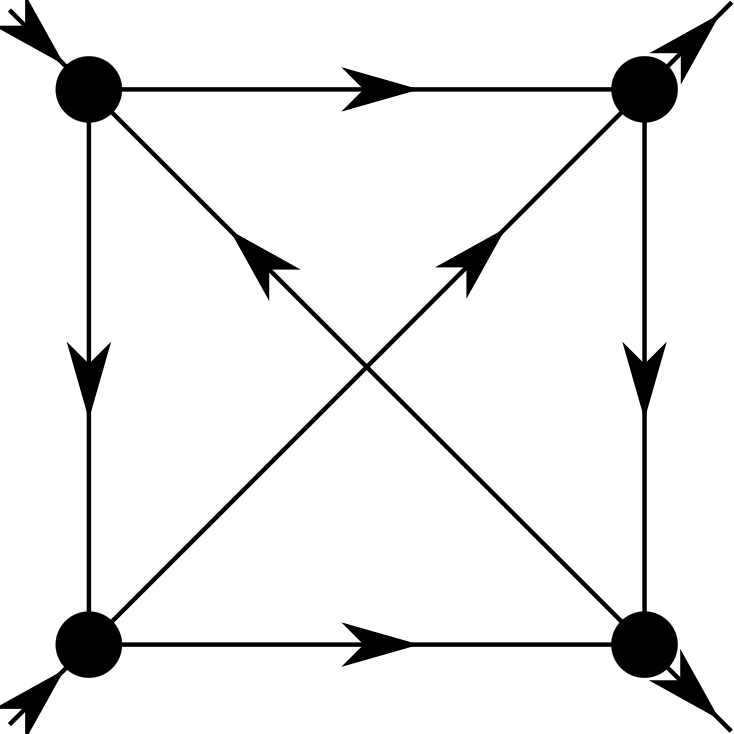}
   \end{minipage}
   \begin{minipage}{0.02\textwidth}
      \hspace{1.0\textwidth}
   \end{minipage}
   \begin{minipage}{0.17\textwidth}
      \hspace{-0.1cm}
      \includegraphics[width=0.85\textwidth]{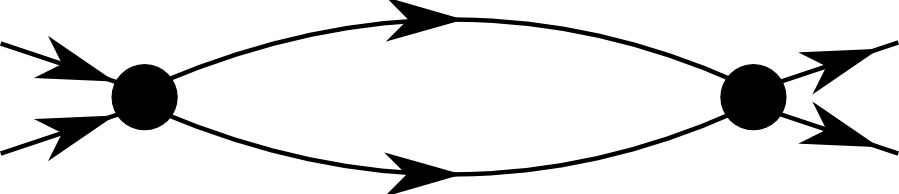}
   \end{minipage}
   \begin{minipage}{0.02\textwidth}
      \hspace{1.0\textwidth}
   \end{minipage}
   \begin{minipage}{0.17\textwidth}
      \hspace{-0.1cm}
      \includegraphics[width=0.29\textwidth]{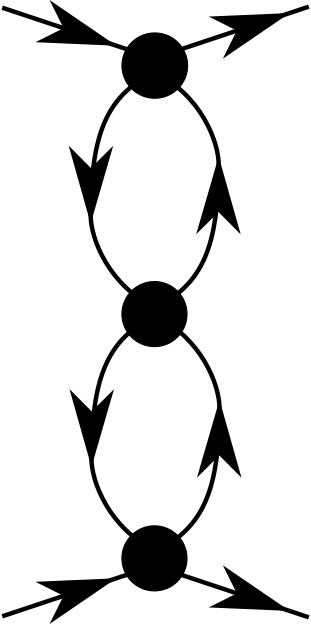}
   \end{minipage}
   \begin{minipage}{0.02\textwidth}
      \hspace{1.0\textwidth}
   \end{minipage}
   \begin{minipage}{0.17\textwidth}
      \hspace{-0.15cm}
      \includegraphics[width=0.59\textwidth]{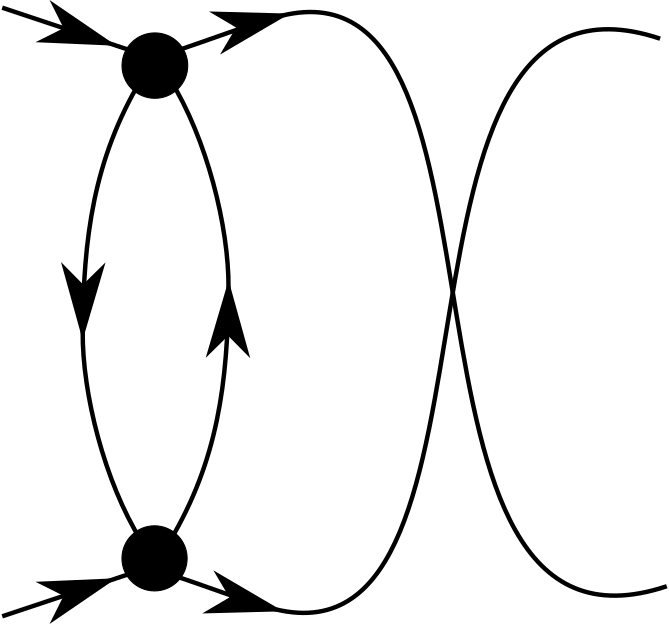}
   \end{minipage}
   \\[2.5ex] 
   \centering
   \begin{minipage}{0.185\textwidth}
      \includegraphics[width=1.0\textwidth]{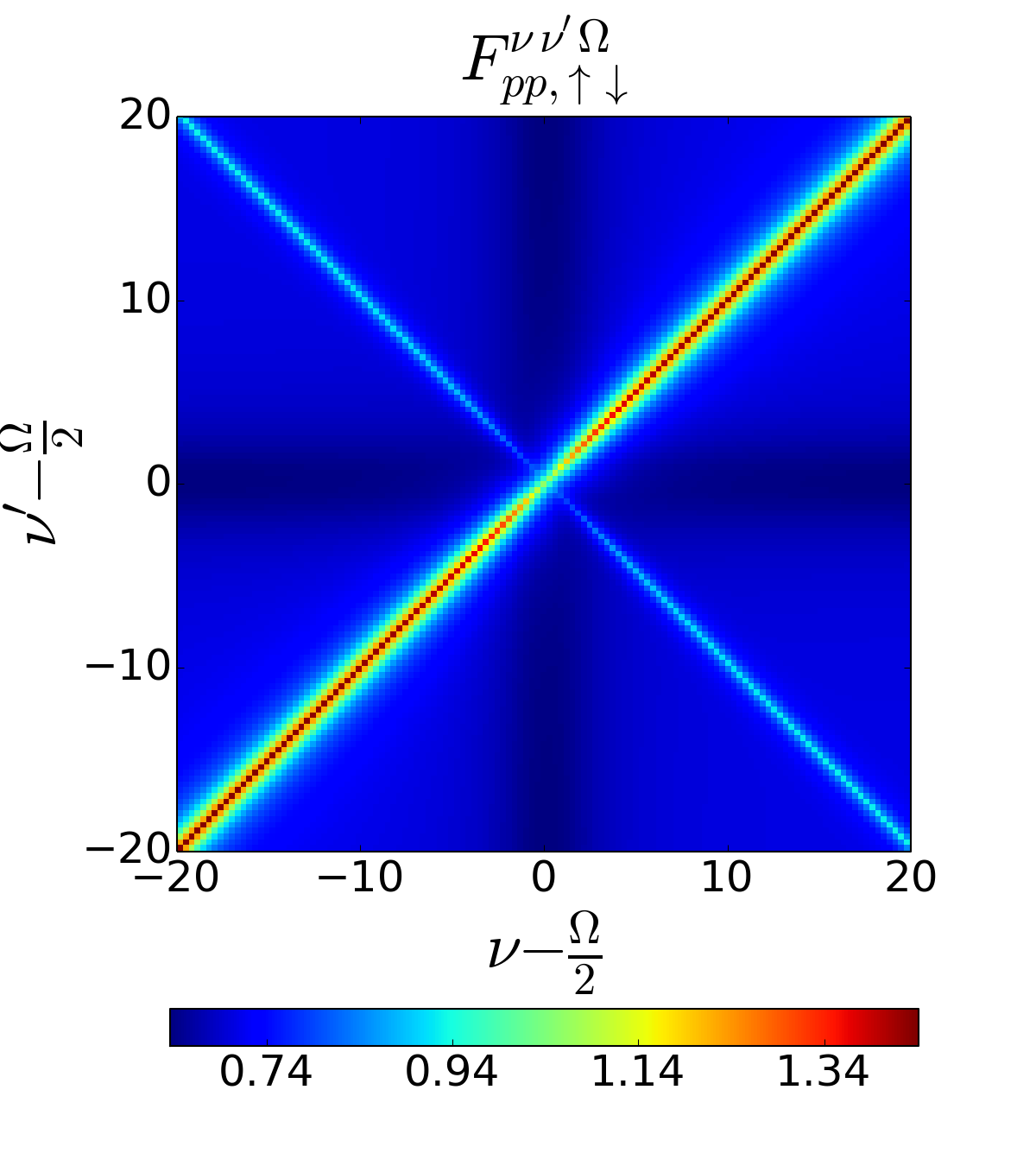}
   \end{minipage}
   \begin{minipage}{0.06\columnwidth}
      \vspace{-0.45cm} 
      $=$
   \end{minipage}
   \begin{minipage}{0.17\textwidth}
      \hspace{-0.25cm}
      \includegraphics[width=1.0\textwidth]{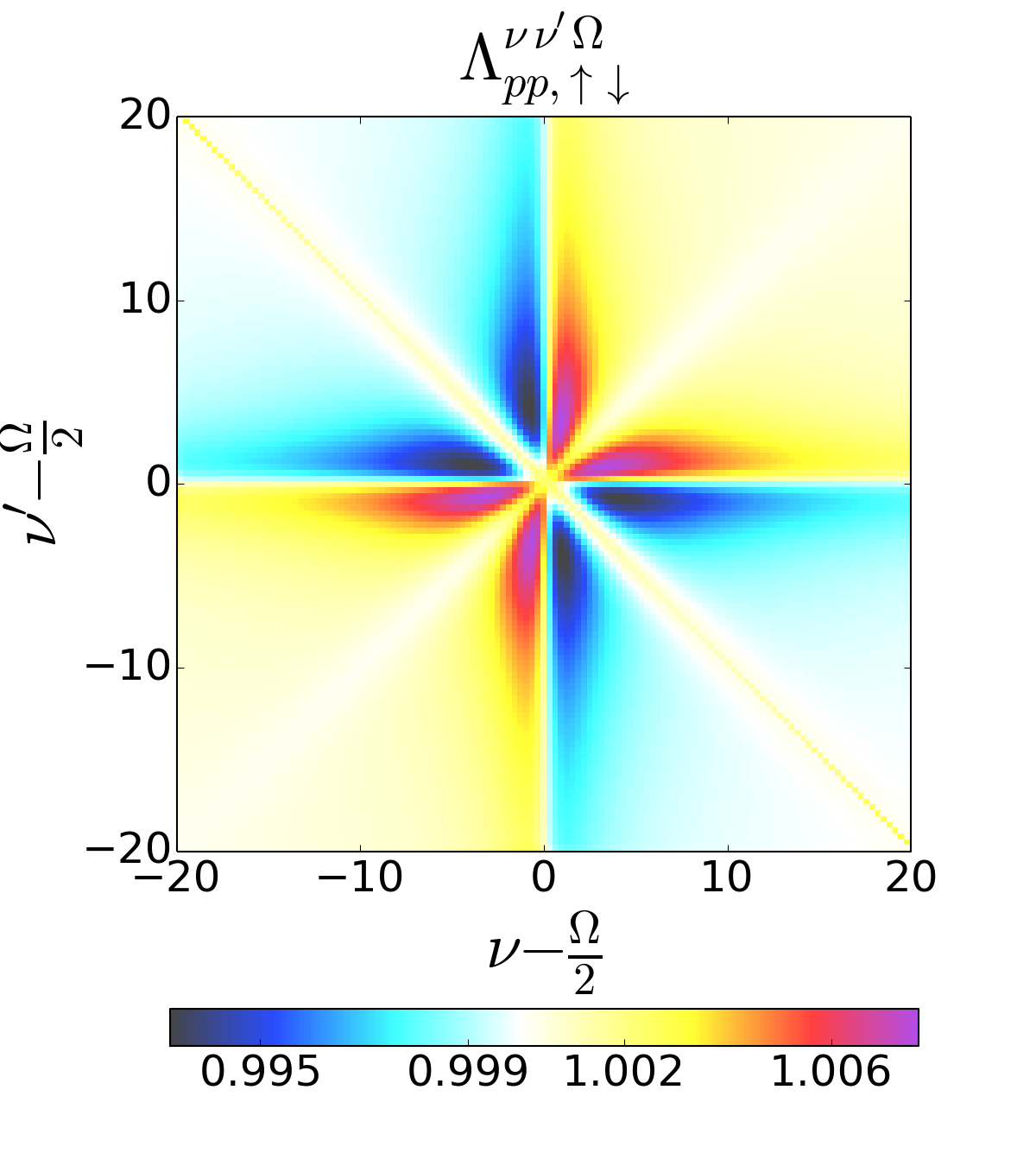}
   \end{minipage}
   \begin{minipage}{0.02\textwidth}
      \vspace{-0.45cm} 
      $+$
   \end{minipage}
   \begin{minipage}{0.17\textwidth}
      \hspace{-0.25cm}
      \includegraphics[width=1.0\textwidth]{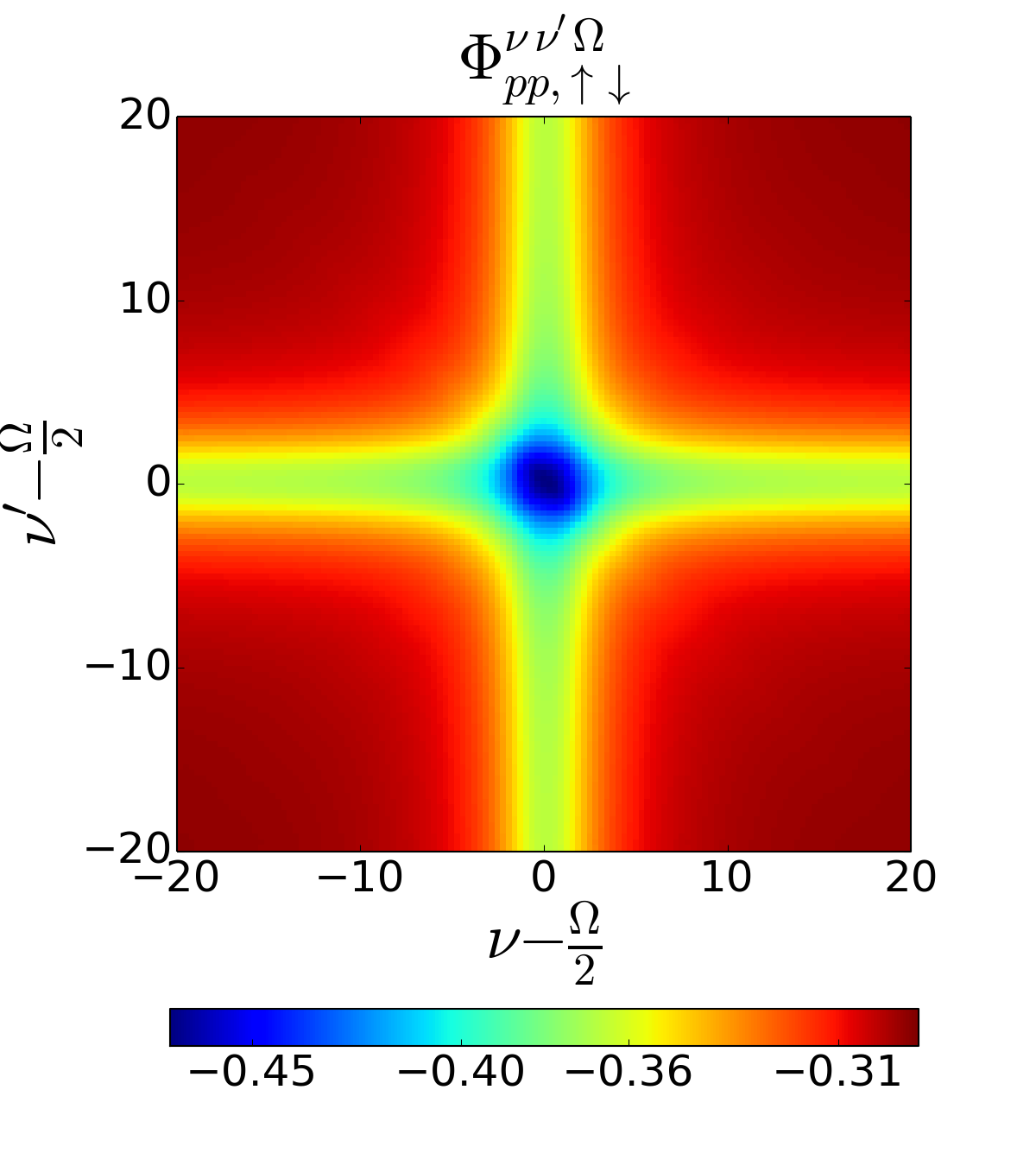}
   \end{minipage}
   \begin{minipage}{0.02\textwidth}
      \vspace{-0.45cm} 
      $+$
   \end{minipage}
   \begin{minipage}{0.17\textwidth}
      \hspace{-0.25cm}
      \includegraphics[width=1.0\textwidth]{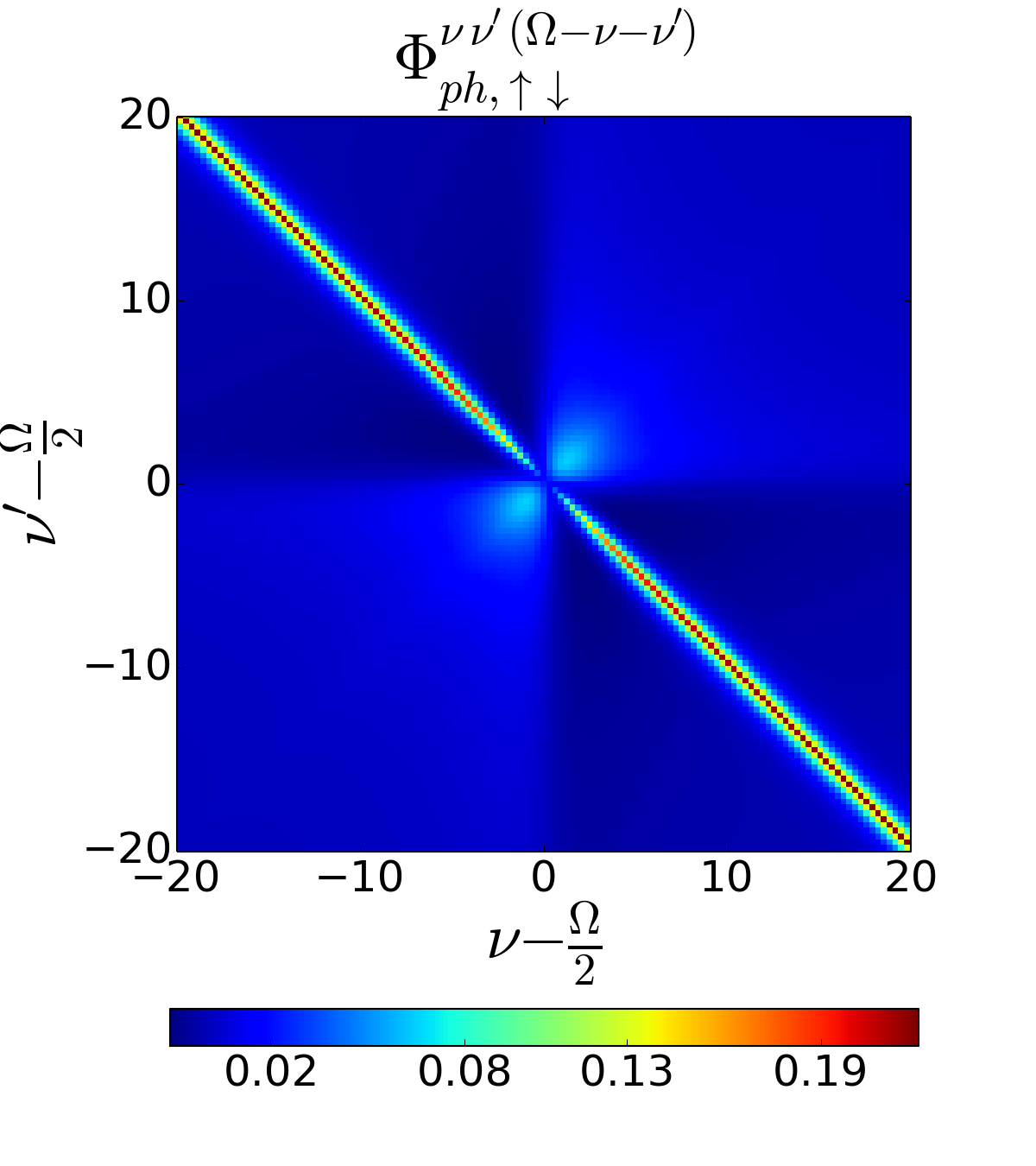}
   \end{minipage}
   \begin{minipage}{0.02\textwidth}
      \vspace{-0.45cm} 
      $+$
   \end{minipage}
   \begin{minipage}{0.17\textwidth}
      \hspace{-0.25cm}
      \includegraphics[width=1.0\textwidth]{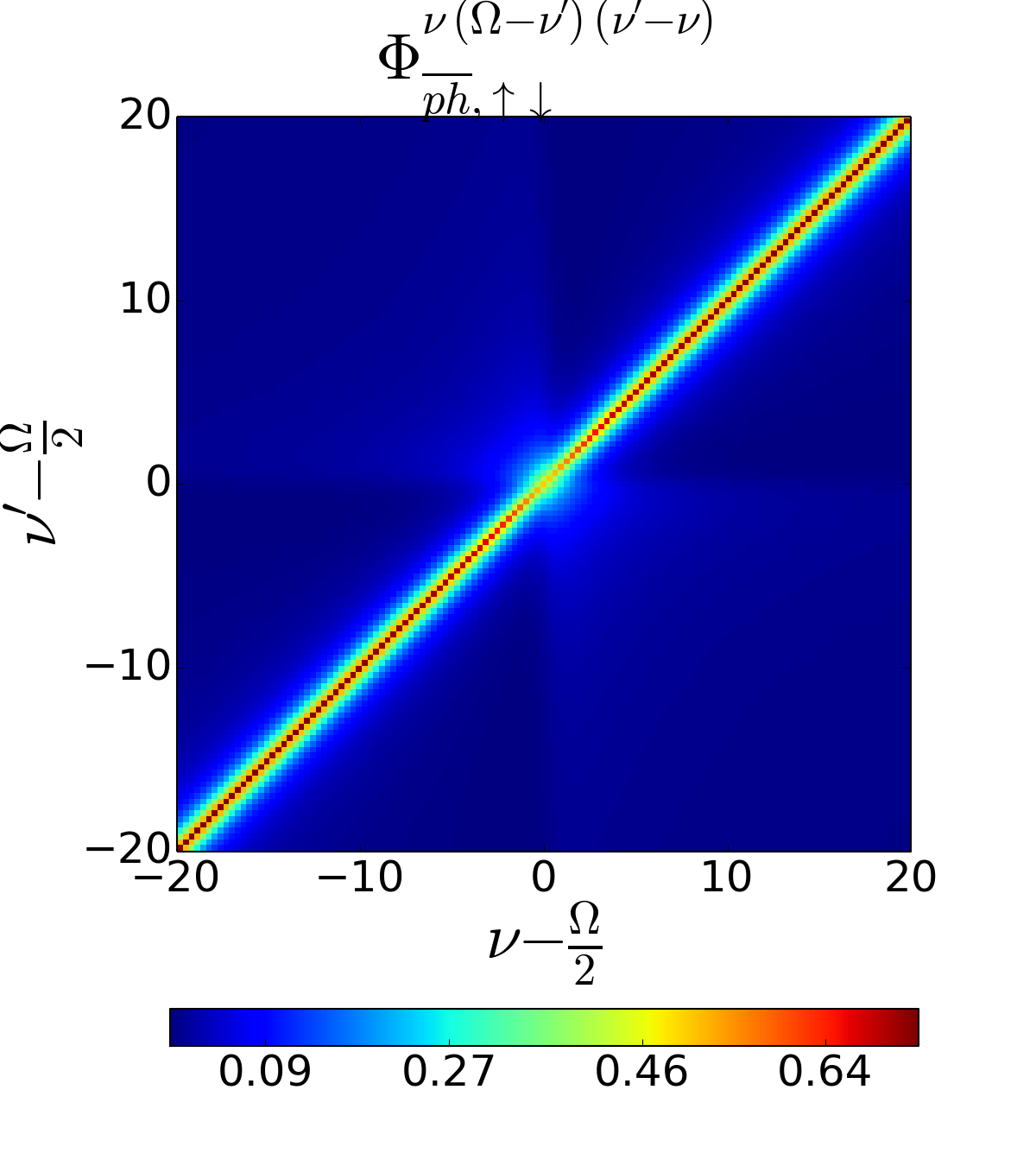}
   \end{minipage}
   \caption{The upper row shows the full vertex $F_\ud$ in $pp$-notation, and the lowest order diagrams (excluding the bare interaction) for each of the four contributions to the parquet Eq.~(\ref{eq:parquet}).
      The bottom row shows the numerical SIAM results for these vertices ($U=1$, $\beta=20$, $D=1$) and the corresponding decomposition through the parquet equation (\ref{eq:parquet}) using $pp$-notation
      for vanishing transfer frequency $\Omega=0$. 
      In the diagrammatic representations used throughout this paper, all external legs are to be considered as the remainder of the amputated propagators.}
   \label{fig:parquet}
\end{figure*}

\begin{figure}
   \centering
   \includegraphics[width=0.29\columnwidth]{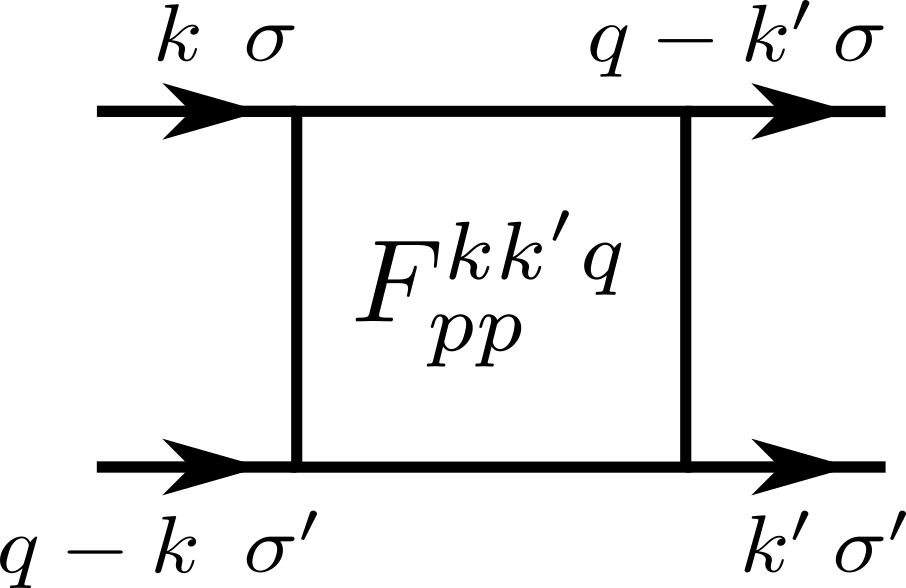} 
   \hspace{0.04\columnwidth}
   \includegraphics[width=0.29\columnwidth]{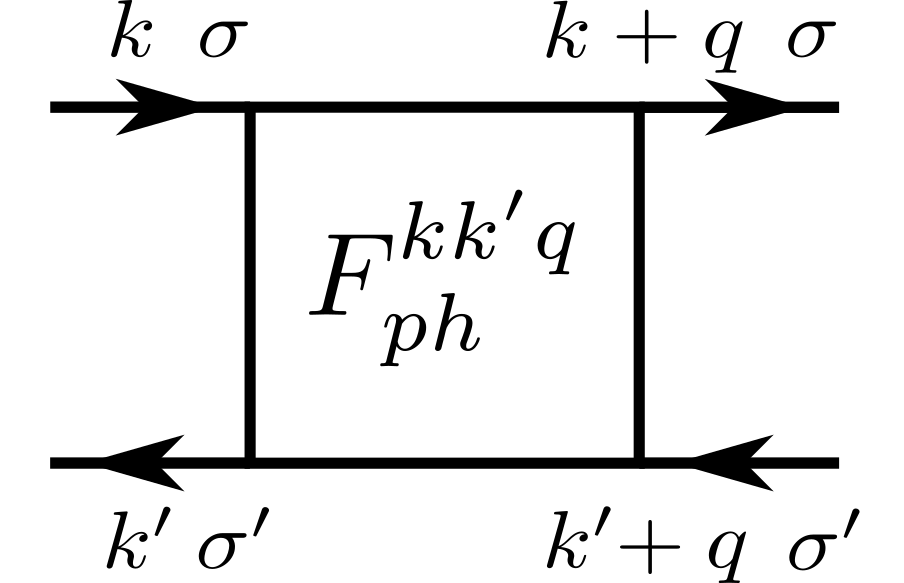} 
   \hspace{0.04\columnwidth}
   \includegraphics[width=0.29\columnwidth]{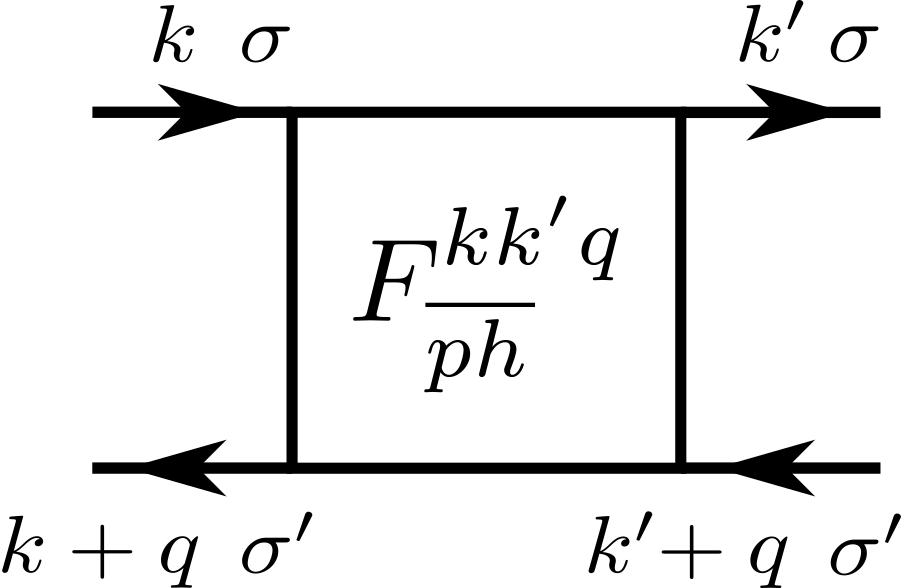}
   \caption{Notations of the vertex functions in the three different scattering channels.}
   \label{fig:F_notations}
\end{figure}

Following the notation introduced above for the one-particle Green's function, the two-particle Green's function in frequency/momentum space reads\cite{Rohringer2012}
 
\begin{align}
\label{eq:G2_iw_pf}
 G_{2,\sigma\sigma'}^{k_1 k_2 k_3}=&\sumint dx_1dx_2dx_3\;e^{ix_1k_1}e^{-ix_2k_2}e^{ix_3k_3}\nonumber\\ \times&\frac{1}{Z}\langle {\mc T}[c^{\dagger}_{\sigma}(x_1)c_{\sigma}(x_2)c^{\dagger}_{\sigma'}(x_3)c_{\sigma'}(0)]\rangle.
\end{align}
From the two-particle Green's function one readily obtains the (full) vertex $F$ by removing all unconnected (``bubble-like'') contributions and amputating the outer legs\cite{Rohringer2012}
\begin{equation}
\label{equ:defF}
F_{\sigma\sigma'}^{k_1 k_2 k_3} = - \frac{G_{2,\sigma\sigma'}^{k_1 k_2 k_3} - G(k_1)G(k_3) \left( \delta_{k_1,k_2} - \delta_{k_2,k_3}\delta_{\sigma,\sigma'} \right)}{ G(k_1)G(k_3)G(k_2)G(k_1+k_3-k_2) }. 
\end{equation}

These definitions are given in a notation that depends solely on fermionic frequencies, momenta and spins.
In this paper we will, however, mainly consider two-particle quantities in the particle-particle ($pp$), particle-hole ($ph$) and transverse particle-hole ($\xph$) notation, which can be obtained from the purely fermionic one as (compare \cfg{F_notations}) 
\begin{equation}
   \begin{split}
  F_{pp, \sigma \sigma'}^{k k' q} &= F_{\sigma \sigma'}^{k( q-k')( q-k )}, \\
  F_{ph, \sigma \sigma'}^{k k' q} &= F_{\sigma \sigma'}^{k( k+q)( k'+q)}, \\
  F_{\xph, \sigma \sigma'}^{k k' q} &= F_{\sigma \sigma'}^{kk'(k'+q)}. 
  \label{eq:notations}
  \end{split}
\end{equation}
which corresponds to adopt a mixed representation in terms of fermionic $(\nu,\mathbf{k})$ and bosonic $(\Omega,\mathbf{q})$ variables. 
Note that the aforementioned notations can be defined accordingly for all other two-particle quantities. 
The full vertex $F$ for the $\uparrow\downarrow$ spin-combination is depicted diagrammatically (in $pp$-notation) in the upper leftmost panel of Fig.~\ref{fig:parquet}, while numerical results for the SIAM detailed in Sec.~\ref{sec:compare_SIAM} are shown in the bottom-left panel.

\begin{figure}[b]
   \centering
   \begin{minipage}{0.55\columnwidth}
      \begin{minipage}{0.2\textwidth}
	 $\Phi^{kk'q}_{pp} =$
      \end{minipage}
      \hspace{0.03\textwidth}
      \begin{minipage}{0.7\textwidth}
	 \includegraphics[width=1.0\textwidth]{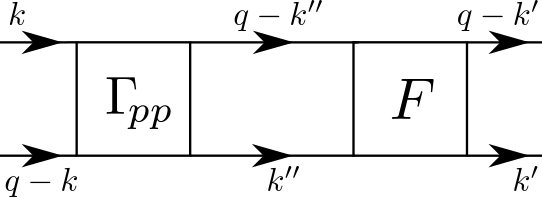}
      \end{minipage}
      \vspace{0.5cm} \\
      \begin{minipage}{0.2\textwidth}
	 $\Phi^{kk'q}_{\xph} =$
      \end{minipage}
      \hspace{0.03\textwidth}
      \begin{minipage}{0.7\textwidth}
	 \includegraphics[width=1.0\textwidth]{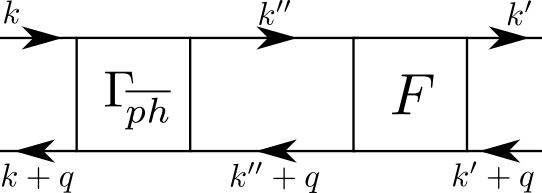}
      \end{minipage}
   \end{minipage}
   \hspace{0.04\columnwidth}
   \begin{minipage}{0.35\columnwidth}
      \begin{minipage}{0.3\textwidth}
	 $\Phi^{kk'q}_{ph} =\,$
      \end{minipage}
      \hspace{0.01\textwidth}
      \begin{minipage}{0.6\textwidth}
	 \includegraphics[width=1.0\textwidth]{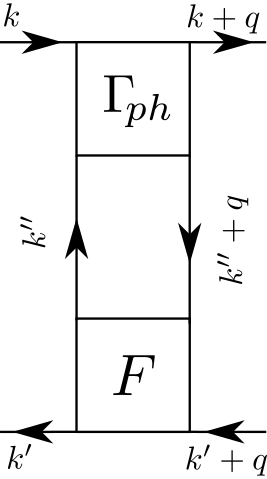}
      \end{minipage}
   \end{minipage}
   \caption{Compact diagrammatic representation of the BSE in all scattering channels. Each $k''$ is integrated over. }
   \label{fig:BetheSalpeter}
\end{figure}

The set of all Feynman diagrams for $F$ can be decomposed into four different classes regarding their two-particle (ir)reducibility\cite{Bickers2004,Dominicis1964,Dominicis1964a}: They are either fully two-particle irreducible (2PI) ($\Lambda_{\text{2PI}}$, see second diagram in the upper panel of Fig.~\ref{fig:parquet} for a lowest-order example) or reducible ($\Phi_r$) in one of the three channels $r\in \{pp,ph,\xph\}$ (see the third, fourth and fifth diagrams in the upper panel of Fig.~\ref{fig:parquet}, respectively). As each diagram belongs to precisely one of these four classes $F$ can be decomposed in the following way:
\begin{equation} 
   F=\Lambda_{\rm 2PI}+\Phi_{pp}+\Phi_{ph}+\Phi_{\xph},
   \label{eq:parquet}
\end{equation} 
which is known as parquet equation. It has to be supplemented by three so-called Bethe-Salpeter equations (BSE) which is presented here in a symbolical notation omitting the frequency-, momentum-, and spin-dependencies
\begin{equation}
 \label{eq:bethe_salpeter}
 F=\Gamma_{r}+\Phi_{r}=\Gamma_r+\underset{\Phi_r}{\underbrace{\sumint \Gamma_r(GG)_rF}},
\end{equation}
where $(GG)_r$ denotes a product of two one-particle Green's functions which frequency/momentum arguments have to be chosen according to the channel $r\in \{pp,ph,\xph\}$ (for the details see Appendix~\ref{app:equations} and Refs.~\onlinecite{Rohringer2012,Rohringer2014,Li2016}). $\Gamma_r$ is the set of diagrams which are irreducible in channel $r$. Let us point out an important difference between $F_r,\Lambda_{\text{2PI},r}$ on the one and $\Phi_r,\Gamma_r$ on the other side regarding the index $r$. The former have no channel dependence and the index $r\in \{pp,ph,\xph\}$ just denotes the representation in which these vertex functions are given. For the latter, on the contrary, the index $r$ refers actually to three different functions (i.e. built on sums of different diagrams). Hence, in principle, the latter could also be expressed in each of the three notations. This would require, apart from the index for the channel, another one to specify the notation\cite{Ayral2016}. For simplicity, however, we assume $\Gamma_r$ and $\Phi_r$ always to be written in their {\sl natural} notation, which is the $pp$-one for $r=pp$ and analogously for the $ph$-cases. This is convenient also because in the natural notation the BSE~(\ref{eq:bethe_salpeter}) can be written as simple matrix multiplications in frequency/momentum space\cite{Bickers2004,Rohringer2012,Rohringer2014} with a transferred bosonic frequency/momentum $q$ (see Fig.~\ref{fig:BetheSalpeter}). For the parquet equation(s)~(\ref{eq:parquet}) on the other hand, all vertices have to be transformed to the same representation which requires the corresponding shifts of arguments in $\Phi_r$ and $\Gamma_r$.  

\subsection{Asymptotics of $F$ from perturbation theory}
\label{subsec:asympt}

While an extensive discussion of the general properties of the two-particle vertex function can be found in Refs.~\onlinecite{Rohringer2012,Rohringer2014}, because of the relevance for the algorithm and results of this paper we will concisely recall here how the high-frequency behavior of the various vertex functions can be qualitatively understood already by means of lowest order perturbation theory. In the lower panels of \cfg{parquet}, the dependence of the local vertex functions $F_{\ud}$, $\Lambda_{{\rm 2PI},\ud}$ and $\Phi_{r,\ud}$ (of a SIAM) on the fermionic frequency arguments $\nu$ and $\nu'$ is shown, where the $pp$-notation for vanishing transfer frequency $\Omega=0$ is used. While we restrict ourselves here to the $\ud$ spin combination, we stress that analogous features are found also in $F_{pp,\uu}^{\nu\nu'(\Omega=0)}$ and even at finite transfer frequencies $\Omega\neq0$. The full vertex $F_{pp,\ud}^{\nu\nu'(\Omega=0)}$ (leftmost panel) has three main features: (i) A constant background {\sl different} from the (constant) bare Hubbard interaction $U$; (ii) Two diagonal structures which we will refer to as main (for $\nu=\nu'$) and secondary (for $\nu=-\nu'$) diagonal; (iii) A ``plus''-like structure, i.e., an enhanced scattering rate along the lines $\nu=\pm \pi/\beta$ and $\nu'=\pm\pi/\beta$. These features do {\sl not} decay in the limit of large fermionic frequencies and give rise to a highly non-trivial asymptotic behavior of the vertex functions. This can be understood by considering the frequency dependence of the lowest order diagrams for the building blocks of $F$, i.e., $\Lambda_{\text{2PI}}$ and $\Phi_r$ [see Eq.~(\ref{eq:parquet})].  

The fully irreducible vertex $\Lambda_{{\rm 2PI},pp,\ud}^{\nu\nu'(\Omega=0)}$ decays uniformly in all directions of the two-dimensional (Matsubara) frequency space (second plot in lower panel of Fig.~\ref{fig:parquet}). Hence, $\Lambda_{{\rm 2PI},pp,\ud}^{\nu\nu'(\Omega=0)}$ does not contribute to the asymptotic structures of the two-particle scattering amplitude $F_{pp,\ud}^{\nu\nu'(\Omega=0)}$ (except for the trivial constant background given by the interaction $U$). This is visible already from the lowest order fully irreducible diagrammatic contribution, the so-called ``envelope'' diagram (second upper panel of Fig.~\ref{fig:parquet}), which depends explicitly on $\nu$ and $\nu'$ (and $\Omega$) and, hence, decays for a large value of any of its frequency arguments\footnote{Every frequency dependence of a diagram originates from the frequency dependence of its internal Green's functions. The latter decay in the high-frequency regime as $1/i\nu$.}.

\begin{figure}
   \includegraphics[width=0.35\cw]{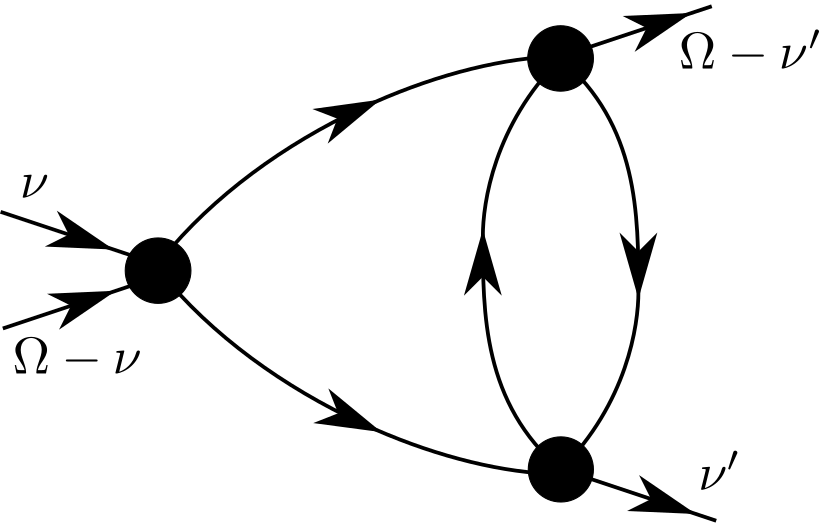}
   \hspace*{0.15\cw}   
   \includegraphics[width=0.35\cw]{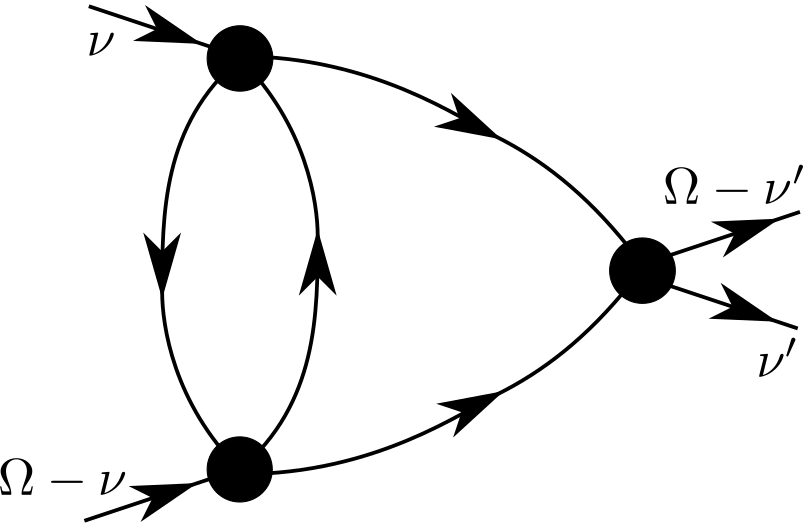}
   \caption{The so-called ``eye''-diagrams in the $pp$ channel.}
   \label{fig:eyediagram}
\end{figure}

The vertex reducible in the particle-particle scattering channel, i.e., $\Phi_{pp,\ud}^{\nu\nu'(\Omega=0)}$ (third lower panel in \cfg{parquet}) exhibits a constant background and a well-defined ``plus''-shaped structure in the $\nu,\nu'$ space. The former originates from bubble diagrams such as
$ \frac{U^2}{\beta}\sum_{\nu_1}G(\nu_1)G(\Omega-\nu_1)$
(third upper panel of Fig.~\ref{fig:parquet}), which does not depend explicitly on $\nu$ and $\nu'$. The latter is generated, in lowest order, from the so-called ``eye''-diagrams (see \cfg{eyediagram}), which either on their left or on their right hand side collapse into a bare vertex $U$. For this reason (as detailed in Sec.~\ref{sec:param}), they cannot explicitly depend on both $\nu$ and $\nu'$, thus remaining constant upon increasing the corresponding (unnecessary) frequency\cite{Rohringer2012}.

The vertex function reducible in the particle-hole (longitudinal) channel, i.e., $\Phi_{ph,\ud}^{\nu\nu'(\Omega-\nu-\nu')}$ (fourth lower panel of \cfg{parquet}) exhibits a secondary diagonal structure along $\nu'=-\nu$. This originates, in lowest (here third) order, from the diagrammatic contribution 
 $\frac{U^3}{\beta^2}\left[\sum_{\nu_1}G(\nu_1)G(\nu_1-\Omega+\nu+\nu')\right]^2$ (fourth upper panel of Fig.~\ref{fig:parquet}),
which depends only on the $ph$ transfer frequency $\Omega-\nu-\nu'$. For fixed $pp$ transfer $\Omega$, its value remains constant along a line $\Omega-\nu-\nu'=\text{const}$ and generates the secondary diagonal structure.

Finally, the vertex reducible in the transverse particle-hole channel (rightmost lower panel of \cfg{parquet}), $\Phi_{\xph,\ud}^{\nu(\Omega-\nu')(\nu'-\nu)}$ accounts for the main diagonal in the full scattering amplitude $F_{pp,\ud}^{\nu\nu'(\Omega=0)}$. Its lowest order (bubble) contribution (rightmost upper panel of \cfg{parquet}) reads
 
 $\frac{U^2}{\beta}\sum_{\nu_1}G(\nu_1)G(\nu_1+\nu'-\nu)$,
and depends only on the $\xph$ transfer frequency $\nu'-\nu$, hence its value remains constant along a line $\nu'-\nu=\text{const}$.

The above analysis demonstrates how the high-frequency asymptotic features of the vertex functions in the weak coupling regime are determined at the second and third order in $U$ by two-particle reducible bubble- and ``eye''-like diagrams. A generalization of these conclusions to the non-perturbative regime will be discussed in the following Section. 

\section{Parametrization of the asymptotics}		
\label{sec:param}

In the following, we will generalize the intuitive discussion of the previous section about the main asymptotic structures of the various vertex functions to the non-perturbative situation. To this end, we first recall\cite{Rohringer2012} that the reduced complexity of specific diagrams regarding their frequency and momentum dependence is {\sl not} a peculiarity of low(est) order perturbation theory but rather a general consequence of the frequency and momentum independence of the bare Coulomb (Hubbard) interaction $U$. In fact, if any two external lines of the vertex, e.g., the incoming momenta and frequencies $k_1$ and $k_3$, are attached to the same bare vertex $U$ (which is possible only for two-particle reducible diagrams), energy and momentum conservation requires $k_1+k_3=k'+k''$ where $k'$ and $k''$ denote internal frequencies/momenta which are summed. Obviously, in this situation the entire diagram does depend only on the linear combination $k_1+k_3$ rather than $k_1$ and $k_3$ separately. Such behavior has already been observed for lowest order perturbative (bubble and ``eye'') diagrams in the previous section, and does not change, as a matter of course, upon dressing these diagrams by means of vertex corrections. These observations\cite{Rohringer2012} hence suggest to introduce the following subdivision of the reducible vertex function $\Phi_{pp}^{kk'q}$ (and correspondingly for the other two channels) into three distinct classes, that are depicted diagrammatically in \cfg{Asymptotic_Diagrams}:

\begin{itemize}
   \item Class 1: The incoming \textit{and} outgoing frequencies/momenta are each attached to a single bare vertex. These diagrams correspond to (dressed) bubble diagrams (see first line of \cfg{Asymptotic_Diagrams}), and can hence be parametrized by a {\sl single} (bosonic) transfer frequency and momentum $q=k_1+k_3$. The sum of all diagrams of this class will be denoted by $\KPP^{q}$. 
   \item Class 2: Either the incoming {\sl or} the outgoing frequencies/momenta are attached to the same bare vertex. These diagrams correspond to (dressed) eye diagrams (see, e.g., \cfg{eyediagram} and first two diagrams in the second line of \cfg{Asymptotic_Diagrams}). These diagrams depend on the bosonic transfer frequency/momentum $q=k_1+k_3$ and one fermionic frequency $k=k_1$ or $k'=k_4$, respectively. The sum of such types of diagrams will be denoted as $\KKPP^{kq}$ and $\KKBPP^{k'q}$. 
	\item Class 3: Every external frequency/momentum is attached to a different bare vertex. These diagrams depend independently on {\sl all three} external arguments. Their sum will in the following be referred to as the ``rest'' function, denoted by $\RPP^{kk'q}$. It is illustrated diagrammatically by the last diagram in the second row of \cfg{Asymptotic_Diagrams}.
\end{itemize}
\begin{figure*}
   \centering
   \begin{minipage}{0.1\textwidth}
      $\KPP =$
   \end{minipage}
   \begin{minipage}{0.22\textwidth}
      \includegraphics[width=1.0\textwidth]{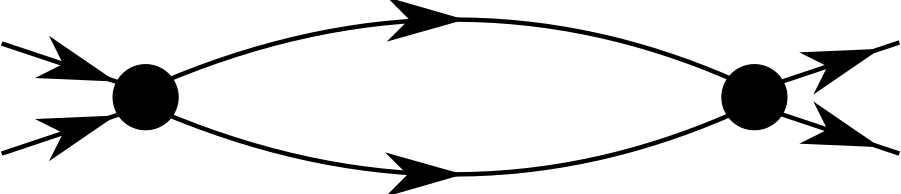}
   \end{minipage}
   \begin{minipage}{0.05\textwidth}
      $+$
   \end{minipage}
   \begin{minipage}{0.22\textwidth}
      \includegraphics[width=1.0\textwidth]{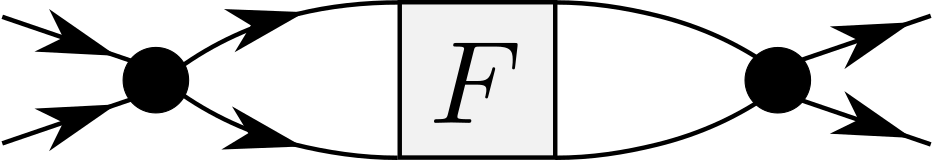}
   \end{minipage}
   \\[4ex]
   \begin{minipage}{0.05\textwidth}
      $\KKPP =$
   \end{minipage}
   \begin{minipage}{0.2\textwidth}
      \includegraphics[width=0.9\textwidth]{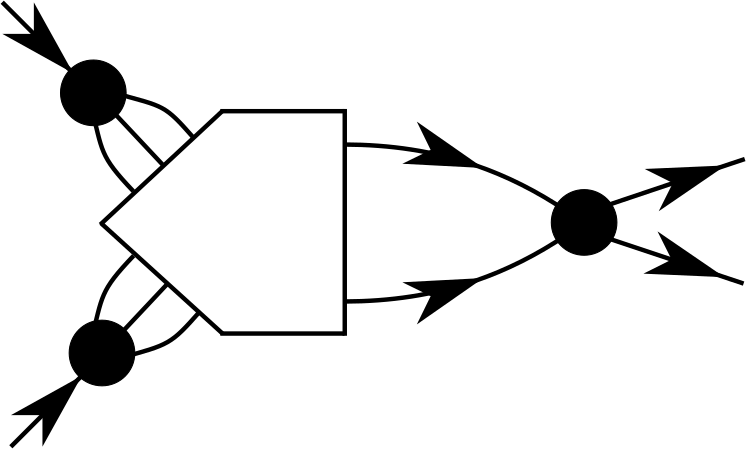}
   \end{minipage}
   \begin{minipage}{0.05\textwidth}
      \hspace{1.0\textwidth}
   \end{minipage}
   \begin{minipage}{0.05\textwidth}
      $\KKBPP =$
   \end{minipage}
   \begin{minipage}{0.2\textwidth}
      \includegraphics[width=0.9\textwidth]{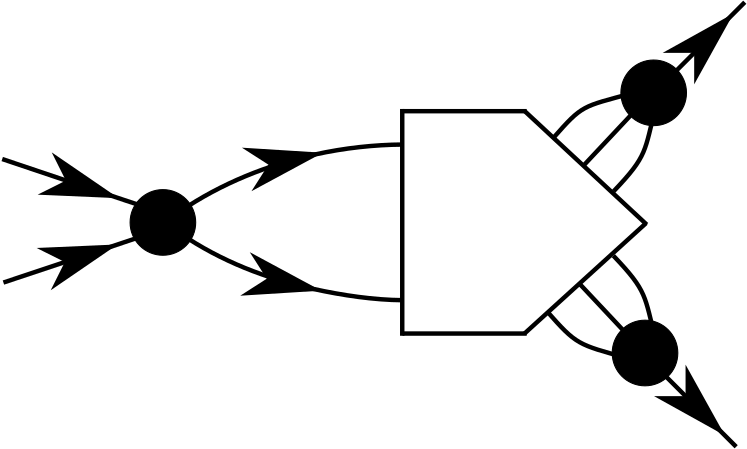}
   \end{minipage}
   \begin{minipage}{0.05\textwidth}
      \hspace{1.0\textwidth}
   \end{minipage}
   \begin{minipage}{0.04\textwidth}
      $\RPP =$
   \end{minipage}
   \begin{minipage}{0.2\textwidth}
      \includegraphics[width=0.9\textwidth]{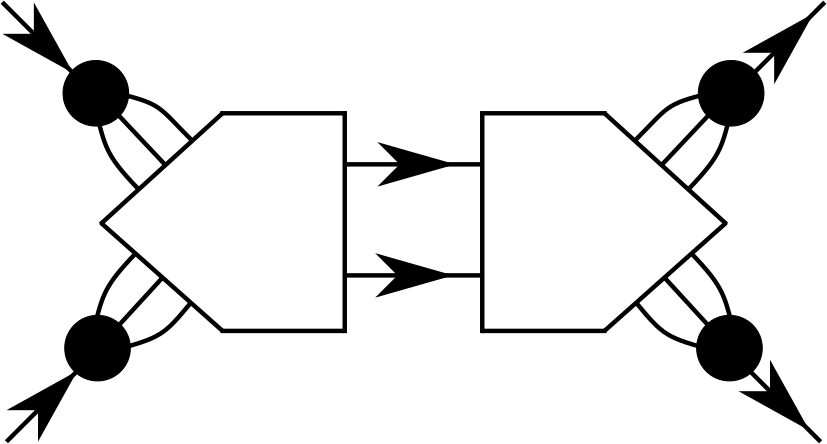}
   \end{minipage}
   \caption{Diagrammatic representation of the asymptotic functions for the particle-particle channel. For a more rigorous definition see Appendix \ref{app:asympt_ED}. }
   \label{fig:Asymptotic_Diagrams}
\end{figure*}
Based on this classification, we can thus introduce an (a priori exact) decomposition of each reducible $\Phi$-function into these four terms\footnote{Note that for $\K$, $\KK$ and $\KKB$ respectively, the index denotes the reduced number of external arguments required to describe them. These shall in the following be referred to as 'necessary' arguments for the corresponding term.}. In the particle-particle channel it reads\footnote{Let us remark that the concrete form of the argument(s) for $\KPP$/$\KKPP$ depend(s) on the chosen frequency/momentum convention. The dependence on one/two {\sl single} argument(s) becomes apparent only in its natural notation, while for other conventions, $\KPP$/$\KKPP$ will depend on one/two linear combination(s) of all frequencies/momenta. Nevertheless, these functions will be constant along two-dimensional planes/one-dimensional lines in the space of three frequencies/momenta (in the natural notation these planes/lines are parallel to the coordinate axes).}
\begin{equation}
   \Phi_{pp}^{ k k' q } = \KPP^q + \KKPP^{ k q } + \KKBPP^{ k'q } + \RPP^{ k k' q }.
\label{eq:phi_decomp}
\end{equation}
In the same way we can decompose also the other scattering channels $ph$ and $\xph$. It is important to note that the structures arising due to $\K$, $\KK$ and $\KKB$ extend to infinitely large frequencies and, hence, generate a highly non-trivial high-frequency asymptotic behavior of the corresponding vertex function.  

On the contrary, the diagrammatic content of $\R$ implies a decay in all frequency directions, since each external fermionic frequency will enter directly one of the inner diagrammatic propagator lines by means of the frequency conservation at its attached bare vertex. These decay properties are verified numerically in Sec.~\ref{sec:compare_SIAM}, and motivate our proposed approximation for treating the vertex asymptotics. Our strategy will be the following: We will consider the full frequency dependence of $\Phi_{pp}^{kk'q}$ only within a small (``inner'') box in the three-dimensional frequency space which is spanned by the intervals $[N_\text{min}^\text{bose},N_\text{max}^\text{bose}]$ for the bosonic ($\Omega$) and $[N_\text{min}^\text{fermi},N_\text{max}^\text{fermi}]$ for the fermionic ($\nu$ and $\nu'$) Matsubara frequencies, where typically $N_\text{min}^\text{bose}\!=\!-N_\text{max}^\text{bose}$ and the fermionic interval is centered around $\frac{\Omega}{2}$ (see Ref.~\onlinecite{Tagliavini2018}). If the value of one or more of these three frequency arguments is located outside of the intervals, we neglect only the contributions to $\Phi_{pp}^{kk'q}$ which decay with increasing values of this (these) frequency argument(s) and keep the terms which remain finite in the respective large-frequency limit. This approximation can be summarized as
\begin{equation}
\Phi^{kk'q}_{pp, {\rm asympt.}} \approx \KPP^q + \KKPP^{ kq } + \KKBPP^{ k'q }.
\label{eq:asympt}
\end{equation}
where $\KPP^q$, $\KKPP^{ kq }$ and $\KKBPP^{ k'q }$ are truncated to $0$ outside of their respective frequency grids. Hence, outside the ``inner'' frequency box, the reducible vertex $\Phi^{kk'q}_{pp}$ is described by functions of {\sl at most two} arguments in its asymptotic regime, which drastically lowers the cost for its numerical treatment. Let us stress that our approach represents a clear improvement over a simple cutoff in frequencies where $\Phi_{pp}^{kk'q}\!=\!0$ outside of the inner frequency box. By restoring the correct high-frequency asymptotic behavior of this vertex function, we are able to substantially mitigate the problems arising from boundary effects due to finite-size frequency grids in vertex based numerical algorithms.


Let us now discuss the physical content of the asymptotic functions $\K$ and $\KK$. 
The former is directly related to the susceptibility in the corresponding scattering channel\cite{Kunes2011,Rohringer2012,Rohringer2013,Hummel2014} as
\begin{equation}
\K^q = -U^2 \chi^q.
\label{eq:relation_chi}
\end{equation}
We recall that the susceptibilities in the different channels are defined as
\begin{subequations}
\label{eq:chi_iw}
\begin{align}
   \begin{split}
\chi_{pp, \sigma \sigma'}^{q} =  \phantom{+} &(1-\delta_{\sigma\sigma'})\sumint dk \, dk' \, G_{2,c,pp, \sigma \sigma'}^{k k' q} \\
+&(1-\delta_{\sigma\sigma'}) \sumint dk \, G(q-k) G(k), 
\label{eq:chi_iw_pp} 
   \end{split}
\\[1ex]
\chi_{ph, \sigma \sigma'}^{q} = \phantom{-} & \sumint dk \, dk' \, G_{2,c,ph, \sigma \sigma'}^{k k' q} - \delta_{\sigma,\sigma'}\sumint dk \, G(k) G(k+q),
\label{eq:chi_iw_ph} 
\\[1ex]
\chi_{\xph, \sigma \sigma'}^{q} = \phantom{+} & \sumint dk \, dk' \, G_{2,c,\xph, \sigma \sigma'}^{k k' q} + \sumint dk \, G(k) G(k+q),\label{eq:chi_iw_xph}
\end{align}
\end{subequations}
where $G_{2,c,r}$ denotes the connected part of the two-particle Green function. 

$\KK$ on the other hand encodes information about how the electrons couple to different bosonic degrees of freedom. For instance, for the generalized density in Fourier space $n_q = \sumint dk \hspace{0.1cm} \sum_\sigma c^\dagger_{\sigma}(k)c_{\sigma}(k+q)$, we find the relation
\begin{equation}
   \begin{split}
      &U \times \langle {\mc T} n_q c_\sigma(k+q) c_{\sigma}^{\dagger}(k) \rangle_c = \\
      G(k) &G(k+q) \sum_{ \sigma' } \biggl(\mathcal{K}_{1,ph,\sigma\sigma'}^q + \mathcal{K}_{2,ph,\sigma\sigma'}^{kq}\biggr).
   \end{split}
\end{equation}
Here, $\langle \ldots \rangle_c$ considers only connected contractions, and the imaginary time-ordering acts inside the Fourier-integrals. 
The above equation identifies the sum of $\K$ and $\KK$ with the expectation value $\langle {\mc T} n_q c_\sigma(k+q) c_{\sigma}^{\dagger}(k) \rangle_c$, which is directly related to the electron-boson coupling (three-point or Hedin) vertex as used in the ladder version of D$\Gamma$A\cite{Toschi2007,Rohringer2016} and in the recently introduced TRILEX\cite{Ayral2015,Ayral2016a} approach.

Let us now turn our attention to the momentum dependence of the vertex and, in particular, of its asymptotic functions $\K$ and $\KK$. For the case of a purely local Hubbard interaction $U$ which is considered here, the momentum dependence of $\K$ and $\KK$ follows exactly the corresponding frequency dependence (which has actually allowed us to use a combined four-vector notation for frequencies and momenta). Hence, in this situation the vertices $\Phi_r$ exhibit their full dependence on the three momenta $\mathbf{k}$, $\mathbf{k'}$ and $\mathbf{q}$ only in the domain of small frequencies while at larger frequencies the  momentum structure is reduced alongside the corresponding frequency dependence. Since the fully irreducible vertex $\Lambda_{\rm 2PI}$ decays to the bare interaction $U$ in all frequency directions, strong momentum dependent parts of the full vertex $F$, e.g.~the contributions responsible for a $d$-wave scattering amplitude, have to be localized in the frequency domain.

The situation is different for a non-local (but instantaneous) interaction which (for a translational invariant system) can depend on three momenta, i.e., $U\equiv U_{\mathbf{kk'q}}$. In fact, while the reduced dependence of $\K$ and $\KK$ on the frequencies still holds, the momentum dependence is more complicated as it obviously contains contributions from the bare interaction $U_{\mathbf{kk'q}}$. The rest function $\R$, which exhibits a strong momentum dependence even in the case of local interactions, will however still strongly decay for large frequency arguments. Hence, Eq.~(\ref{eq:asympt}) and the methods for the improved treatment of the high-frequency asymptotic regime of the vertex, which are discussed below, remain applicable. Of course, the asymptotic functions become more involved in this case as their explicit full momentum dependence has to be considered (which also prevents the use of the compact four-vector notation as the dependence on frequencies and momenta is now different).

Let us finally also consider the case of a retarded, i.e., time- or frequency-dependent, interaction $U\equiv U^{\nu\nu'\omega}$. While formally a diagrammatic decomposition of $\Phi_r$ into its contributions $\K$, $\KK$ and $\R$ is still possible, a reduced frequency dependence of these objects can no longer be expected. In fact, they explicitly depend on all (three) frequency arguments due to the frequency dependence of the bare interaction. Even more importantly, it is in general unclear whether the rest function $\R$ decays for large frequency arguments which prevents the application of Eq.~(\ref{eq:asympt}). The asymptotic methods outlined in the next sections are nevertheless applicable if the effective interaction $U^{\nu\nu'\omega}$ decays in all directions of the frequency space. In this respect, we want to stress that our high frequency treatment is also applicable to methods such as D$\Gamma$A and QUADRILEX, which use the local fully irreducible vertex of DMFT as input to the parquet equations, since the latter approaches uniformly the static bare Hubbard interaction $U$ in the high frequency regime.


Finally, a comment is in order regarding the choice of basis for both the one- and two-particle Green's functions: In this work, we have restricted ourselves to the common frequency and momentum representation which is particularly suited when adopting the parquet equations. In this case, the BSE ~(\ref{eq:bethe_salpeter}) are just matrix multiplications and the parquet Eq.~(\ref{eq:parquet}) corresponds to a simple algebraic equation with frequency shifts in the various vertex functions [see Eq.~(\ref{eq:parquet_pp})]. While the parquet equations might become more complicated in other basis representations, it is nevertheless an interesting question to what extent our parametrization scheme could be combined with the use of alternate basis functions in both momenta\cite{Husemann2009,Lichtenstein2017} and frequencies\cite{Boehnke2011, shinaoka2017, shinaoka2018, Otsuki2020} to further improve the numerical performance. The investigation of these combinations is an interesting future research direction.

\section{Implementation}				
\label{sec:implem}

In this section, we describe how the ideas presented in the previous section can be practically exploited in analytical and numerical calculations based on two-particle vertex functions. After a general presentation of the main concepts, we will explicitly discuss the application of our scheme for analytic calculations based on the atomic limit vertex, and for numerical implementations of the fRG in its second order truncation and the parquet approximation. 

The observation 
that any diagram vanishes if one of its necessary frequency arguments is taken to infinity allows us to select the different diagrammatic contributions by taking the corresponding limits in the frequency domain, i.e.
\begin{subequations}
\label{eq:limits_phi}
\begin{align}
\lim_{|\nu| \to \infty} \lim_{|\nu'| \to \infty} &\Phi_{r, \sigma \sigma'}^{k k' q} = \mathcal{K}_{1, r, \sigma \sigma'}^{q},  \label{eq:limit_phi_1} \\
\lim_{|\nu'| \to \infty}&\Phi_{r, \sigma \sigma'}^{k k' q} =  \mathcal{K}_{1, r, \sigma \sigma'}^{q}  + \mathcal{K}_{2, r, \sigma \sigma'}^{k q},  \label{eq:limit_phi_2} \\
\lim_{|\nu| \to \infty}&\Phi_{r,\sigma \sigma'}^{k k' q}  = \mathcal{K}_{1, r, \sigma \sigma'}^{q}  + \overline{\mathcal{K}}_{2, r, \sigma \sigma'}^{k' q}, \label{eq:limit_phi_3}
\end{align}
\end{subequations}
where $r \in \{ pp, ph, \xph\}$. 
We recall again that for a local interaction $U$, by taking limits in the frequency domain, we find alongside the reduced frequency dependence also a reduced momentum dependence.
The remaining diagrammatic class 3 introduced in Sec.~\ref{sec:param}, or rest function $\R$, which requires the full dependence on all arguments, can then be acquired by inverting \ceq{phi_decomp}
\begin{equation}
   \R_{r, \sigma \sigma'}^{k k' q} = \Phi_{r, \sigma \sigma'}^{k k' q} - \mathcal{K}_{1, r, \sigma \sigma'}^{q} - \mathcal{K}_{2,r, \sigma \sigma'}^{k q} - \overline{\mathcal{K}}_{2,r, \sigma \sigma'}^{k' q}.
   \label{eq:R}
\end{equation}
One advantage of performing this limiting procedure based on the reducible vertex, is that \ceq{limit_phi_1} holds equally if $|\nu|$ and $|\nu'|$ are taken to infinity at the same time, i.e.
\begin{equation}
\lim_{\substack{|\nu\phantom{'}| \to \infty \\ |\nu'| \to \infty}} \Phi_{r, \sigma \sigma'}^{k k' q} = \mathcal{K}_{1, r, \sigma \sigma'}^{q}.
\label{eq:double_limit}
\end{equation}
This property allows for a simplified scanning procedure to numerically extract asymptotic functions, which, depending on the frequency ranges and parameters, provide a good approximation.

\begin{figure}[b]
\centering
\includegraphics[width=0.8\linewidth]{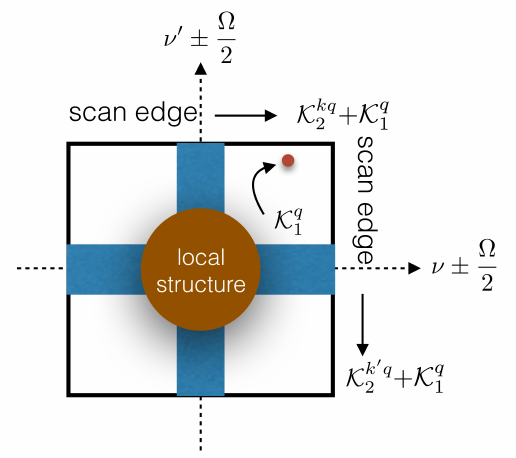}
\caption{Sketch of a reducible vertex function in frequency space as a function of $k$ and $k'$ for fixed $q$. It consists mainly of two extensive stripes and a more dynamical localized structure centered at a position determined by the transfer frequency. The two stripes are described by ${\cal K}_{2,r}^{kq}$ and ${\cal K}_{2,r}^{k'q}$, the local structure is contained in the rest function ${\cal R}$. The nearly constant background is described by ${\cal K}_{1,r}^{q}$. }
\label{fig:kernel12}
\end{figure}
The procedure is straightforward and applicable in all channels (see also \cfg{kernel12} and Ref.~\onlinecite{Li2016}):
\begin{itemize}
   \item I: For large $|\nu|$ and $|\nu'|$ vary the transfer four-vector $q$ to acquire $\K^q$.
   \item II: For large $|\nu'|$, vary $k$ and the transfer four-vector $q$ and subtract $\K^q$ in order to obtain $\KK^{kq}$.
   \item III: Repeat II by replacing $\nu' \to \nu$  and $k \to k'$ to determine $\KKB^{k'q}$.
   
\end{itemize}

The above described procedure proposed to determine $\K$ and $\KK$ has some limitations. Firstly, one can easily see that if the scanning is not performed at sufficiently large $|\nu|$ ($|\nu'|$), the rest function might not be fully decayed, giving rise to an error in the $\K$ and $\KK$ extraction. We found this error to be particularly pronounced in the strong coupling regime ($U=4$ for the comparisons in Sec.~\ref{sec:compare_SIAM}) where the rest function becomes comparable with the asymptotic functions in the domain of small frequencies. 
Secondly, the scanning procedure requires the knowledge of the reducible vertex functions $\Phi_r$, which are not directly available in some algorithms, as e.g.~for the exact diagonalization.
This raises the question whether a similar set of limits can be formulated also for $F$. 
In fact, as will be clarified in the following, the limits presented in \ceq{limits_phi} still hold, i.e.
\begin{subequations}
\label{eq:limits_F}
\begin{align}
\lim_{|\nu| \to \infty} \lim_{|\nu'| \to \infty} F_{r, \sigma \sigma'}^{k k' q} - (1- \delta_{\sigma,\sigma'}) U = \mathcal{K}_{1, r, \sigma \sigma'}^{q}, \label{eq:limit_F_1}\\
\lim_{|\nu'| \to \infty}F_{r, \sigma \sigma'}^{k k' q} - (1- \delta_{\sigma,\sigma'}) U = \mathcal{K}_{1, r, \sigma \sigma'}^{q}  + \mathcal{K}_{2, r, \sigma \sigma'}^{k q}, \label{eq:limit_F_2}\\
\lim_{|\nu| \to \infty}F_{r,\sigma \sigma'}^{k k' q} -  (1- \delta_{\sigma,\sigma'}) U = \mathcal{K}_{1, r, \sigma \sigma'}^{q}  + \overline{\mathcal{K}}_{2, r, \sigma \sigma'}^{k' q},\label{eq:limit_F_3}
\end{align}
\end{subequations}
where again $F_r$ denotes the representation of $F$ in one of the three mixed notations. 
However, the numerical equivalent of the limiting procedure, i.e.~the scanning procedure previously described for the $\Phi$-functions, is not feasible in the case of $F$, which is directly related to the fact that \ceq{double_limit} does not hold equally for $F$. 
In order to numerically extract the asymptotics from $F$ directly we thus suggest an alternative approach detailed in Appendix \ref{app:asympt_ED}. We implemented this diagrammatic extraction to determine the exact asymptotic functions, as presented in Sec.~\ref{sec:compare_SIAM}, from ED calculations. Let us also remark, that the asymptotic functions $\K$ and $\KK$ can be also directly obtained from the impurity solver as discussed in Ref.~\onlinecite{Kaufmann2017} for a quantum Monte-Carlo based impurity solver and in Ref.~\onlinecite{Tagliavini2018} for ED.

The limiting procedure \ceq{limits_F} is particularly suited in the case that analytical expressions for $F$ are available, as demonstrated for the atomic limit case in Sec.~\ref{subsec:atomic_limit}. Let us in the following argue why the generalization of \ceq{limits_phi} holds. It relies on the property that any reducible diagram vanishes if the corresponding transfer frequency, being a necessary argument, is sufficiently large, i.e.
\begin{equation}
\lim_{|\Omega| \to \infty} \Phi_{r, \sigma \sigma'}^{k k' q} = 0.
\end{equation}
We have to further consider, that in order to take the limits in \ceq{limits_F}, we should formulate \ceq{parquet} in the corresponding mixed notation. E.g.~for the particle-particle channel we have to translate $\Phi_{ph}$ and $\Phi_{\xph}$ to the $pp$-notation as follows
\begin{equation}
\label{eq:parquet_pp}
\begin{split}
F_{pp, \sigma \sigma'}^{k k' q} =& \, \Phi_{pp, \sigma \sigma'}^{k k' q} + \Phi_{ph, \sigma \sigma'}^{k k' (q-k'-k)} + \Phi_{\xph, \sigma \sigma'}^{k (q-k') (k'-k)}\\
&+\Lambda_{{\rm 2PI},pp, \sigma \sigma'}^{k k' q}.
\end{split}
\end{equation}
It now becomes clear that for fixed $\Omega$ and $\nu'$, the bosonic frequencies of the $ph$ and $\xph$ channel, that is $\Omega-\nu'-\nu$ and $\nu'-\nu$, will lead to a vanishing of the respective scattering channels for $|\nu| \to \infty$. This behavior can also be observed in \cfg{parquet}, and holds equally for the other scattering channels. Since $\Lambda_{\rm 2PI}$ decays in all frequency directions to the bare interaction, we conclude that $\lim_{|\nu| \to \infty}F_{r,\sigma \sigma'}^{k k' q} - (1- \delta_{\sigma,\sigma'}) U = \lim_{|\nu| \to \infty}\Phi_{r,\sigma \sigma'}^{k k' q}$, while the same argument can be made for the other limits in \ceq{limits_F}.

  \subsection{The atomic limit}				
  \label{subsec:atomic_limit}

As a first showcase of these ideas we discuss the vertex decomposition for a system 
that can be treated analytically, i.e., the atomic limit, whose Hamiltonian reads
\begin{equation}
\hat{\mc{H}}=U\left[\hat{n}_\uparrow \hat{n}_\downarrow-\frac{1}{2}(\hat{n}_\uparrow+\hat{n}_\downarrow)\right].
\end{equation}  
Here, $\hat{n}_\sigma=\hat{c}^\dag_\sigma \hat{c}_\sigma$ is the number operator for fermions of spin $\sigma$, and we have imposed the half-filling (particle-hole symmetry) condition $\mu=U/2$.  
The Hilbert space is spanned by the four eigenstates $|0\rangle$, $|\uparrow\rangle$, $|\downarrow \rangle$ and $|\ud\rangle$, allowing for a direct calculation of the 
two-particle Green's functions by means of the Lehmann representation.
The resulting two-particle vertex function\cite{Hafermann2008,Rohringer2012,Kinza2013} is, for our purposes, split into four terms\footnote{We consider here only the $\ud$ component, as it allows to calculate all the remaining ones, given that $SU(2)$ symmetry holds.} (note $\mc{F}_r \neq F_r$)
\begin{equation}
\label{eq:atlimit} 
F_{\uparrow\downarrow} = \mc{F}_{\rm odd} + \mc{F}_{pp} + \mc{F}_{ph} + \mc{F}_{\xph},
\end{equation}
which are defined in the following.
The first term contains only odd orders in the interaction, and takes the most compact form using a function dependent on four fermionic frequencies
\begin{equation}
\mc{F}^{\nu_1\nu_2\nu_3\nu_4}_{\rm odd} = U - \frac{U^3}{8} \frac{\sum_i \nu_i^2}{\prod_i \nu_i} - \frac{3U^5}{16}\prod_i \frac{1}{\nu_i}, 
\label{eq:Fodd}
\end{equation} 
while frequency conservation is implicitly assumed. 
The link to the purely fermionic notation introduced previously is then given as $\mc{F}^{\nu_1\nu_2\nu_3}_{\rm odd} =  \mc{F}^{\nu_1\nu_2\nu_3(\nu_4=\nu_1-\nu_2+\nu_3)}_{\rm odd}$. 
The functions $\mc{F}_{r}$ with $r\in\{pp,ph,\xph\}$, however, are more conveniently expressed in their respective mixed notation (see Sec.~\ref{subsec:formalism})
\begin{subequations}
\label{eq:fatall}
\begin{align} 
\mc{F}_{pp}^{\nu\nu'\Omega}&=-\beta\, \delta_{\Omega,0}\frac{U^2}{2} \mc{D}^{\nu\nu'} f\left(\frac{U}{2}\right), 
\label{eq:fatpp} \\                      
\mc{F}_{ph}^{\nu\nu'\Omega}&=-\beta\, \delta_{\Omega,0} \frac{U^2}{4} \mc{D}^{\nu\nu'} \left[ f\left(\frac{U}{2}\right) - f\left(-\frac{U}{2}\right) \right],
\label{eq:fatph} \\
\mc{F}_{\xph}^{\nu\nu'\Omega}&=\phantom{-}\beta\, \delta_{\Omega,0}\frac{U^2}{2} \mc{D}^{\nu\nu'} f\left(-\frac{U}{2}\right),
\label{eq:fatxph}
\end{align}
\end{subequations}
with $\mc{D}^{\nu\nu'}=\frac{1}{\nu^2 \nu'^2} \left(\nu^2+\frac{U^2}{4} \right) \left( \nu'^2+\frac{U^2}{4} \right)$ and the Fermi function $f(\epsilon)=\frac{1}{1+e^{\beta\epsilon}}$.
Note that, at this stage, the decomposition for the full vertex $F$ is motivated solely by algebraic reasons, while the connection to the physical scattering channels will be established in the following.

Let us now use the limits in Eqs.~(\ref{eq:limits_F}) to identify the contributions arising from the different diagrammatic classes.
This task can be performed by considering each term in \ceq{atlimit} separately. Let us illustrate this procedure for the $pp$-channel,
beginning with the first term, $\mc{F}_{\rm odd}$. Here, we have to translate from the purely fermionic notation to the mixed $pp$-notation:
\begin{equation}
\begin{split}
\mc{F}_{\rm odd,pp}^{\nu\nu'\Omega}=& \,\mc{F}_{\rm odd}^{\nu,\Omega-\nu',\Omega-\nu,\nu'} = U - \frac{3U^5}{16}\frac{1}{\nu(\Omega-\nu')(\Omega-\nu)\nu'} \\
&- \frac{U^3}{8} \frac{\nu^2+(\Omega-\nu')^2+(\Omega-\nu)^2+\nu'^2}{\nu(\Omega-\nu')(\Omega-\nu)\nu'}.  
\end{split}
\end{equation}
The large frequency limits then result in
\begin{subequations}
\begin{align}
\lim_{|\nu|\to \infty} \lim_{|\nu'|\to\infty} \mc{F}_{\rm odd,pp}^{\nu\nu'\Omega}& = U, \\ 
\lim_{|\nu'|\to\infty} \mc{F}_{\rm odd,pp}^{\nu \nu' \Omega}& = U - \frac{U^3}{4}\frac{1}{\nu }\frac{1}{\nu -\Omega}, \\ 
\lim_{|\nu|\to\infty}  \mc{F}_{\rm odd,pp}^{\nu \nu' \Omega}& = U - \frac{U^3}{4}\frac{1}{\nu'}\frac{1}{\nu'-\Omega}. 
\end{align}
\end{subequations}
As for the limits of the second term, $\mc{F}_{pp}$, we have
\begin{subequations}
\begin{align}
\lim_{|\nu|\to \infty} \lim_{|\nu'|\to\infty} \mc{F}_{pp}^{\nu\nu'\Omega}& = -\beta \, \delta_{\Omega,0}\frac{U^2}{2} f\left(\frac{U}{2}\right), \\ 
\lim_{|\nu'|\to\infty} \mc{F}_{pp}^{\nu \nu' \Omega}& = -\beta \, \delta_{\Omega,0}\frac{U^2}{2}\left[ 1 + \frac{U^2}{4} \frac{1}{\nu^2}  \right] f\left(\frac{U}{2}\right), \\
\lim_{|\nu|\to\infty}  \mc{F}_{pp}^{\nu \nu' \Omega}& = -\beta \, \delta_{\Omega,0}\frac{U^2}{2}\left[ 1 + \frac{U^2}{4} \frac{1}{\nu'^2} \right] f\left(\frac{U}{2}\right).
\end{align}
\end{subequations}
Determining the contributions from the remaining terms $\mc{F}_{ph}$ and $\mc{F}_{\xph}$, which involves a translation from their respective mixed notation to the $pp$-notation, 
we find that their contributions vanish.
This leads to the final expressions for the asymptotic functions in the $pp$-channel
\begin{subequations}
\begin{align}
   \KPPUD^{ \Omega } & = -\beta \, \delta_{\Omega,0}\frac{U^2}{2} f\left(\frac{U}{2}\right), \\ 
   \KKPPUD^{ \nu \Omega } & = \frac{U^2}{4}\frac{1}{\nu}\frac{1}{\nu-\Omega} \left( \KPPUD^\Omega - U \right),
\end{align}
\end{subequations}
while the $\KKB$ can be acquired by means of the symmetry properties reported in Appendix \ref{app:symmetries}.
Performing the analogous procedure for the remaining two channels yields
\begin{subequations}
\begin{align}
   \KPHUD^{ \Omega } & = -\beta \, \delta_{\Omega,0} \frac{U^2}{4} \left[ f\left(\frac{U}{2}\right) - f\left(-\frac{U}{2}\right) \right], \\ 
   \KKPHUD^{ \nu \Omega } & = \frac{U^2}{4}\frac{1}{\nu}\frac{1}{\nu+\Omega} \left( \KPHUD^\Omega - U \right), 
\end{align}
\end{subequations}
for the $ph$-channel, and
\begin{subequations}
\begin{align}
   \KXPHUD^{ \Omega } & = \beta \, \delta_{\Omega,0} \frac{U^2}{2} f\left(-\frac{U}{2}\right) , \\ 
   \KKXPHUD^{ \nu \Omega } & = \frac{U^2}{4}\frac{1}{\nu}\frac{1}{\nu+\Omega} \left( \KXPHUD^\Omega - U \right).
\end{align}
\end{subequations}
in the $\xph$ case. 

Now that we have determined all asymptotic functions of the atomic limit vertex, let us consider its structures that are localized in the frequency domain. We proceed again in a term-wise fashion, beginning with
$\mc{F}_{\rm odd}$. By subtracting all asymptotic contributions arising from this term, we find that only the fifth order term $-\frac{3U^5}{16}\prod_i \frac{1}{\nu_i}$ survives, while it remains
unclear whether this term can be attributed to the fully irreducible vertex function or the rest functions. 

For the $\mc{F}_r$ terms, let us again consider the $pp$-channel as an example. Here we find 
\begin{equation}
\begin{split}
&\mc{F}_{pp}^{\nu\nu'\Omega} - \lim_{|\nu'|\to\infty}\mc{F}_{pp}^{\nu\nu'\Omega} - \lim_{|\nu|\to\infty}\mc{F}_{pp}^{\nu\nu'\Omega} + \lim_{|\nu|\to\infty} \lim_{|\nu'|\to\infty}\mc{F}_{pp}^{\nu\nu'\Omega}\\
&= \left(\frac{U^2}{4}\frac{1}{\nu}\frac{1}{\nu-\Omega}\right)\times \KPPUD^\Omega \times\left(\frac{U^2}{4}\frac{1}{\nu'}\frac{1}{\nu'-\Omega}\right).
\label{eq:loc_Rpp}
\end{split}
\end{equation}
This term contains three factors, i.e.~a fermion-boson vertex\cite{Ayral2016} that describes the coupling to a pairing field, the bosonic propagator in the $pp$-channel, 
and an additional fermion-boson vertex, as depicted schematically in \cfg{FB_Rpp}.
We can thus argue diagrammatically that this localized term belongs to the rest function $\R_{pp,\ud}$. For the other channels, we find equally that the localized structures belong
to the respective rest function, and hence $\mc{F}_r \in \Phi_{r,\ud}$.
\begin{figure}
\includegraphics[width=0.6\cw]{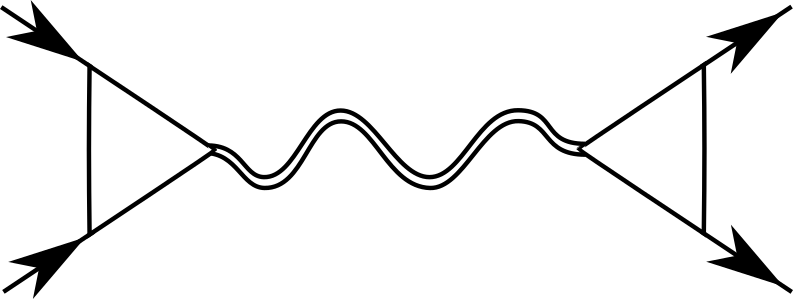}
\caption{Schematic diagrammatic representation of the localized structure presented in \ceq{loc_Rpp}.}
\label{fig:FB_Rpp}
\end{figure}

Note that, to obtain the full rest functions as well as the fully irreducible vertex function, it would require the analytic expressions for all the reducible $\Phi$ functions. The latter have been derived in Ref.~\onlinecite{Thunstroem2018}, where it has been shown that the corresponding expressions for the reducible vertices $\Phi$ and the fully irreducible vertex $\Lambda_{\mathrm{2PI}}$ are very involved. The simplification $\Lambda_{\mathrm{2PI},\ud} + \sum_r \mc{R}_{r,\ud} = -\frac{3U^5}{16}\prod_i \frac{1}{\nu_i}$ is far from obvious, and the complexity of the expressions can be directly linked to the multiple vertex divergencies\cite{Schaefer2013,Kozik2015,Schaefer2016,Ribic2016,Vucicevic2018} appearing in the $\Phi$ functions for $T \leq \frac{\sqrt{3}}{2\pi}U$. These divergencies are however exactly canceled by corresponding terms in $\Lambda_{\mathrm{2PI}}$. These cancellations are in fact a strong argument for vertex-based methods built upon the sum of frequency-localized structures\cite{PhysRevB.100.155149} $\Lambda_{\mathrm{2PI}} + \sum_r \mc{R}_{r}$.

 

  \subsection{Implementation for the fRG solver}	
  \label{subsec:fRG}

The functional renormalization group approach\cite{Metzner2012,Salmhofer1999} implements Wilson's renormalization group idea in a general field-theoretical framework. By introducing a scale-dependence into the quadratic part of the action, i.e.~the non-interacting propagator 
\[ G_0( i\nu ) \to G_0^\Lambda( i\nu ), \]
one can derive an exact functional flow equation\cite{Wetterich1993} for the 1PI generating functional, also named ``effective action". This flow equation describes the gradual evolution of all correlation functions as the scale $\Lambda$ is varied from the initial to the final value. Being an exact reformulation of the initial problem, it serves as a basis for further approximations, and has been used in many different applications ranging from high-energy physics to condensed matter theory. In the fRG, this approximation consists in an expansion in orders of the fields, resulting in an infinite hierarchy of coupled ordinary differential equations for all 1PI $n$-particle vertex functions, e.g.~the self-energy $\Sigma$, the two-particle vertex $F$ and so on. This hierarchy is typically truncated at the two-particle level, rendering the fRG perturbative in the interaction strength\cite{Salmhofer2001}. 

For the flow-parameter dependence, we consider in the following two different schemes: The so-called $\Omega$-flow\cite{Giering2012}:
\begin{equation}
   G_0^\Lambda( i\nu )= \frac{ \nu^2 }{ \nu^2 + \Lambda^2 } G_0( i\nu ),
   \label{eq:omfl}
\end{equation}
and the $U$-flow\cite{Honerkamp2004}:
\begin{equation}
   G_0^\Lambda( i\nu )= \Lambda \cdot G_0( i\nu ).
   \label{eq:intfl}
\end{equation}
The $\Omega$-flow introduces an energy cutoff into the system, that allows to successively integrate out the different energy scales from high to low. This approach is very much in the spirit of other renormalization group approaches.
The $U$-flow on the other hand introduces a frequency-independent regulator into the Green function that treats all energy scales on equal footing. In this sense, the $U$-flow is more similar to common perturbative approaches.

The flow-equations resulting from a second order truncation of the flow-equation hierarchy can be summarized as follows, where, for simplicity, we consider the $SU(2)$ symmetric case.
At the level of the self-energy, the derivative takes the simple form  
\begin{equation} 
   \dot{\Sigma}(k)^\Lambda= \sumint dk' \hspace{0.1cm} S^\Lambda( k' ) \times \left[ F_{ph,\ud}^{\Lambda,k k' (q=0)} + F_{ph,\uu}^{\Lambda,k k'(q=0)} \right],
   \label{eq:flowsig}
\end{equation}  
where we have introduced the so-called single-scale propagator 
\[ S^\Lambda( i\nu ) = \partial_\Lambda G^\Lambda( i\nu )|_{\Sigma^\Lambda { \rm \,fixed} }. \] 
At the level of the 1PI two-particle vertex, the flow-equation is composed of contributions from three scattering channels (particle-particle, particle-hole and transverse particle-hole) 
\begin{equation} 
   \dot{F}^\Lambda=\mathcal{T}^\Lambda_{pp}+\mathcal{T}^\Lambda_{ph}+\mathcal{T}^\Lambda_{\xph},
   \label{eq:flowgam}
\end{equation}  
where
\begin{widetext}
\begin{subequations}
\label{eq:flowChannels}
\begin{align}
   \label{eq:flow_pp}
   \mathcal{T}^{\Lambda,k k' q}_{pp,\ud}&=\phantom{-}\sumint dk'' \hspace{0.1cm} \Big[ S^\Lambda( k'' )G^\Lambda( q-k'' ) + S \leftrightarrow G \Big] \times F^{\Lambda,q k k''}_{pp,\ud}F^{\Lambda,q k'' k'}_{pp,\ud}, \\
   \mathcal{T}^{\Lambda,k k' q}_{ph,\ud}&=-\sumint dk'' \hspace{0.1cm} \Big[S^\Lambda( k''+q )G^\Lambda( k'' ) + S \leftrightarrow G \Big] \times \Big[ F_{ph,\uu}^{\Lambda, k k'' q}F_{ph,\ud}^{\Lambda, k'' k' q} + F_\uu \leftrightarrow F_\ud \Big], \\
   \mathcal{T}^{\Lambda,k k' q}_{\xph,\ud}&=\phantom{-}\sumint dk'' \hspace{0.1cm} \Big[S^\Lambda( k''+q )G^\Lambda( k'' ) + S \leftrightarrow G \Big] \times F_{\xph,\ud}^{\Lambda,k k'' q}F_{\xph,\ud}^{\Lambda,k'' k' q}. 
\end{align} 
\end{subequations}
\end{widetext}
These terms can be depicted diagrammatically as shown in \cfg{PP_Diag} for the $pp$-channel. 
\begin{figure}[b]
   \centering
   \includegraphics[width=0.5\columnwidth]{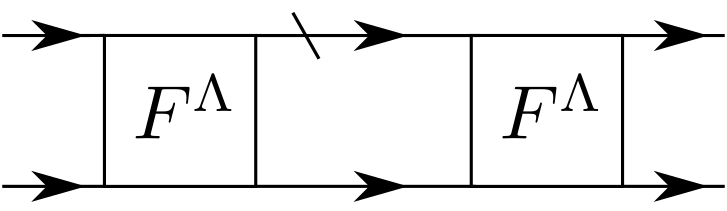}
   \caption{Diagrammatic representation of the particle-particle contribution $\mathcal{T}_{pp}$ (\ref{eq:flow_pp}) in the vertex flow equation. The dashed line denotes the single-scale propagator $S^\Lambda$.}
   \label{fig:PP_Diag}
\end{figure}
To understand the diagrammatic content generated by each channel let us refer to the previously introduced parquet equation, that holds for any scale $\Lambda$ 
\begin{equation} 
   F^\Lambda= \Lambda_{\rm{2PI}}^{\Lambda}+\Phi^{\Lambda}_{pp}+\Phi^{\Lambda}_{ph}+\Phi^{\Lambda}_{\xph}.
\end{equation}  

Considering the vertex flow \ceq{flowgam} (see also \cfg{PP_Diag}), it is obvious that at this level of truncation, the only diagrammatic content than can be generated by the flow is two-particle reducible, meaning $\Lambda_{\rm{2PI}}^{\Lambda}=\overline{\Lambda}_{\rm{2PI}}^{\Lambda_{\rm ini }}$. We can thus separate the different two-particle reducible terms in \ceq{flowgam}, and identify
\begin{align} 
   \dot{\Phi}_{pp}^\Lambda&= \mathcal{T}_{pp}, \quad \dot{\Phi}_{ph}^\Lambda= \mathcal{T}_{ph}, \quad \dot{\Phi}_{\xph}^\Lambda= \mathcal{T}_{\xph}.
\end{align}  
This allows us to make use of the parametrization scheme described in Sec.~\ref{sec:param} during the fRG flow. While keeping track of the reducible vertex functions on a finite frequency grid, we also track the flow of the previously introduced asymptotic functions. In fact, we can directly perform the limits in \ceq{limits_phi} to compute the corresponding derivatives
\begin{subequations}
\label{eq:asympt_fRG}
\begin{align}
\dot{\mathcal{K}}_{1, r, \sigma \sigma'}^{\Lambda,q} &= \lim_{|\nu| \to \infty} \lim_{|\nu'| \to \infty} \dot{\Phi}_{r, \sigma \sigma'}^{\Lambda,k k' q}, \label{eq:k1_dot} \\
\dot{\mathcal{K}}_{2, r, \sigma \sigma'}^{\Lambda,k q}&= \lim_{|\nu'| \to \infty} \dot{\Phi}_{r, \sigma \sigma'}^{\Lambda,k k' q} - \dot{\mathcal{K}}_{1, r, \sigma \sigma'}^{\Lambda,q}, \label{eq:k2_dot} \\
\dot{\overline{\mathcal{K}}}_{2, r, \sigma \sigma'}^{\Lambda,k' q}&= \lim_{|\nu| \to \infty} \dot{\Phi}_{r, \sigma \sigma'}^{\Lambda,k k' q} - \dot{\mathcal{K}}_{1, r, \sigma \sigma'}^{\Lambda,q}. \label{eq:k2b_dot}
\end{align}
\end{subequations}
The limits above are explicitly performed by replacing the associated vertex functions $F^\Lambda_r$ in Eqs.~(\ref{eq:flowChannels}) by the asymptotic forms presented in Eqs.~(\ref{eq:limits_F}b-c). During the fRG flow we then track the flow of the asymptotic functions in addition to the flow of the $\Phi$-functions.

Due to the numerical costs involved in treating the full argument dependence of the vertex function, a simplified parametrization scheme\cite{Karrasch2008}
\begin{equation}
\begin{split}
   &\Phi_{pp}^{ k k' q } \approx \mc{K}_{{\rm eff}, pp}^q = \KPP^q + \KKPP^{ (\lceil\Omega/2\rceil-\nu_0,\textbf{k})q } \\
   & + \KKBPP^{ (\lceil\Omega/2\rceil-\nu_0,\textbf{k}')q }+ \RPP^{ (\lceil\Omega/2\rceil-\nu_0,\textbf{k})(\lceil\Omega/2\rceil-\nu_0,\textbf{k}')q }
   \label{eq:karrasch}
\end{split}
\end{equation} 
has found extensive use in the fRG community. Here, $\nu_0=\frac{\pi}{\beta}$ denotes the first positive Matsubara frequency, and $\lceil \ldots\rceil$ will round up to the next bosonic Matsubara frequency
\footnote{The parametrization scheme presented in \ceq{karrasch} was originally implemented at zero temperature, where the flow of each channel was approximated, in a representation using only bosonic frequencies $(\Omega_{pp}, \Omega_{ph}, \Omega_\xph)$, by setting the transfer frequency of the other two channels to zero. At finite temperature, this choice is only possible for every other transfer frequency, as the condition $\left(\frac{\beta}{2\pi}\sum_r \Omega_r \right) \bmod 2 = 1$ needs to hold.
This leads to ambiguities in the definition.}.
This scheme considers only the dominant transfer frequency dependence of the $\KK$ and $\R$ functions, and will be compared to the full parametrization in Sec.~\ref{subsec:simplified_param}.
Performing the same approximation in the momentum domain limits the scattering to the s-wave type, while higher harmonics can be captured by means of a form-factor expansion\cite{Husemann2009}.

The fRG flow equations in their second order truncated form account for the feedback of $F^\Lambda$ into the flow up to the second order. If we in addition consider partially the neglected contribution of the 1PI three-particle vertex in the flow equations, it is possible to account fully for the feedback up to $\mathcal{O}[( F^\Lambda )^3]$. In practice this is achieved by taking into account both self-energy\footnote{The self-energy correction $S\to \partial_\Lambda G^\Lambda$ is generally referred to as Katanin-substitution\cite{Katanin2004}.}, and vertex corrections from diagrams with overlapping loops\cite{Katanin2004,Eberlein2014}, 
which is possible with a manageable numerical effort\cite{Eberlein2014}. These corrections will in the following be referred to as two-loop ($2\ell$) corrections to distinguish this scheme from the conventional one-loop ($1\ell$) one.

When considering the flow of the asymptotic functions, we find that including the two-loop corrections gives a substantial improvement of the two-particle vertex results. While a quantitative comparison between the one- and two-loop scheme will be presented in Sec.~\ref{subsec:fRG_corrections}, we can already understand from a simple diagrammatic argument that the lowest order contribution to $\KK$ is not fully captured in the one-loop scheme.
Here, the derivative includes four contributions, as depicted in \cfg{K2_corrections}. The one-loop scheme accounts only for the first two diagrams, while the two-loop scheme includes all of them. In particular for the $U$-flow, the contribution from all four diagrams is equal, meaning that in its one-loop implementation the flow reproduces exactly $\frac{1}{2}$ of the exact value for $U \to 0$. This is verified numerically in Sec.~\ref{subsec:fRG_corrections} (see \cfg{fRG_corrections}). A similar argument can be made for the lowest order diagram of $\R$, where the resulting factor is $\frac{1}{3}$. 
\begin{figure}
   \centering
   \begin{minipage}{0.03\columnwidth}
      $\partial_\Lambda$
   \end{minipage}
   \begin{minipage}{0.25\columnwidth}
      \includegraphics[width=0.9\textwidth]{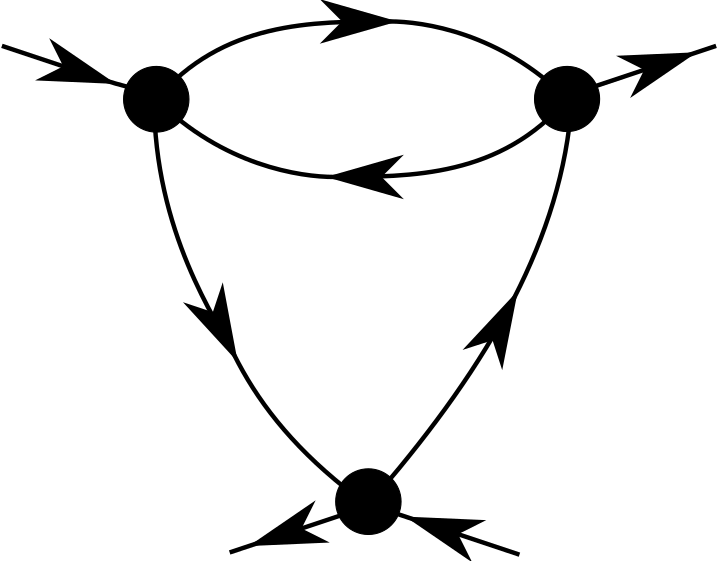}
   \end{minipage}
   \begin{minipage}{0.03\columnwidth}
      $=$
   \end{minipage}
   \begin{minipage}{0.25\columnwidth}
      \includegraphics[width=0.9\textwidth]{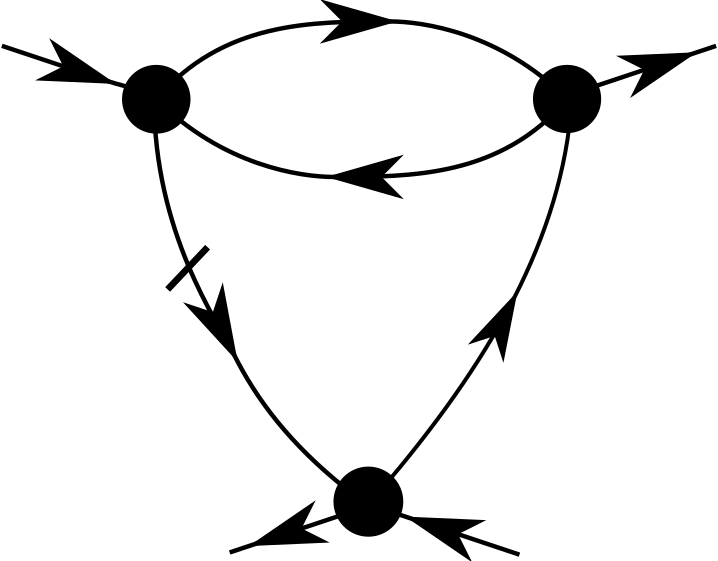}
   \end{minipage}
   \begin{minipage}{0.03\columnwidth}
      $+$
   \end{minipage}
   \begin{minipage}{0.25\columnwidth}
      \includegraphics[width=0.9\textwidth]{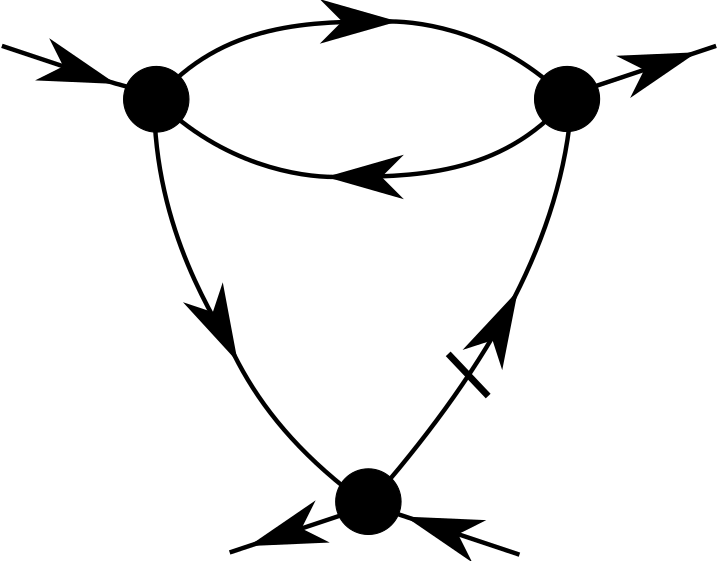}
   \end{minipage}
   \begin{minipage}{0.03\columnwidth}
      $+$
   \end{minipage}
   \\[1.5ex]
   \begin{minipage}{0.29\columnwidth}
      \hspace{1.0\textwidth}
   \end{minipage}
   \begin{minipage}{0.25\columnwidth}
      \includegraphics[width=0.9\textwidth]{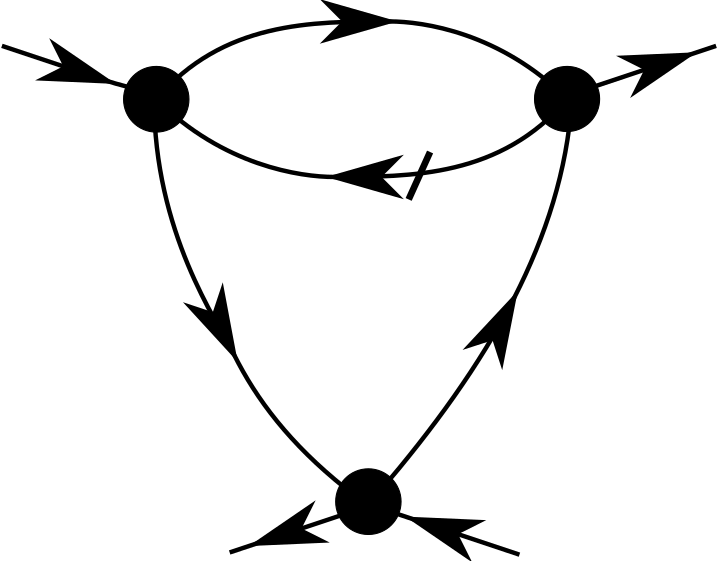}
   \end{minipage}
   \begin{minipage}{0.03\columnwidth}
      $+$
   \end{minipage}
   \begin{minipage}{0.25\columnwidth}
      \includegraphics[width=0.9\textwidth]{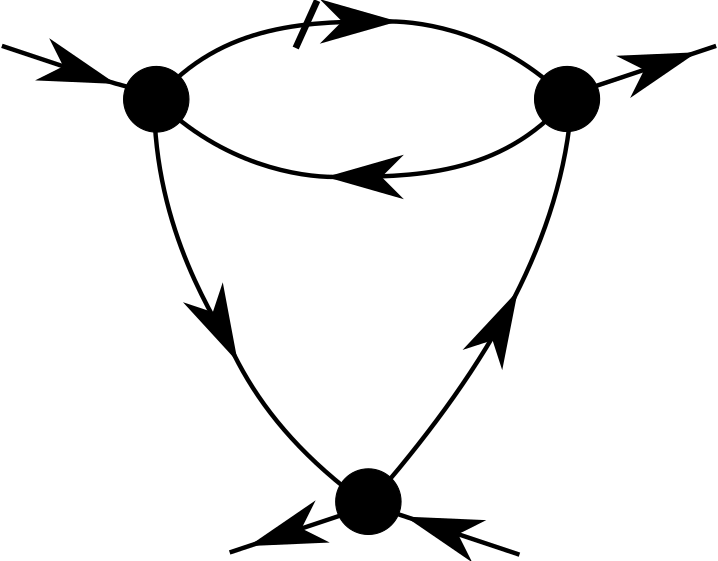}
   \end{minipage}
   \caption{Derivative of the lowest order contribution to $\KKPH^\Lambda$. }
   \label{fig:K2_corrections}
\end{figure}

  \subsection{Implementation for the parquet solver}	
  \label{subsec:parquet}

\begin{figure*}[htbp]
\centering
\includegraphics[width=0.8\linewidth]{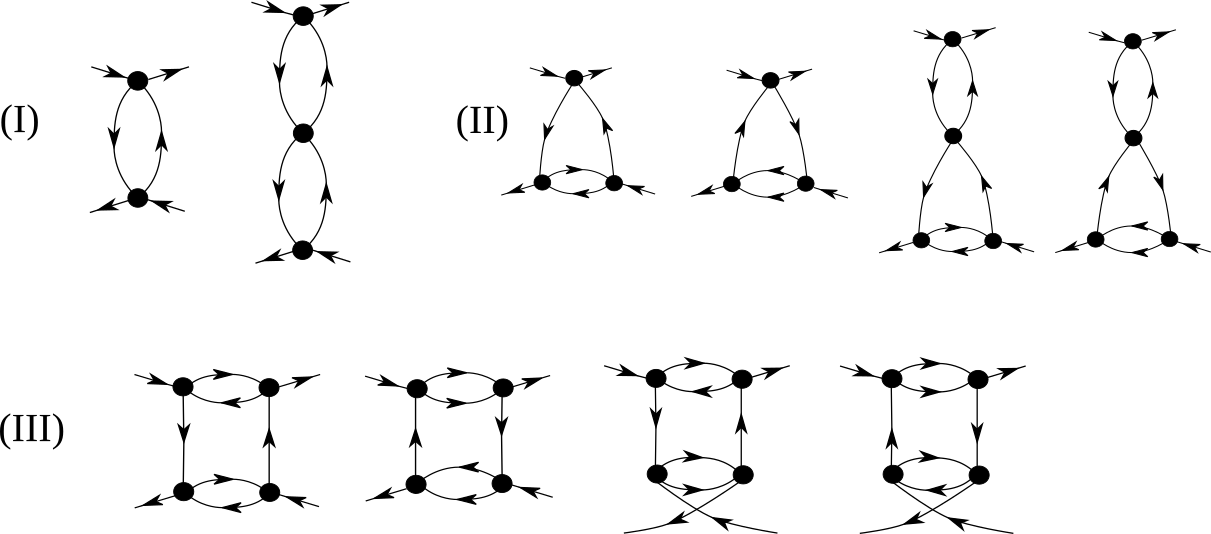}
\caption{The Feynman diagrams for the reducible vertex function $\Phi_{d}^{kk'q}$ generated in the first two iterations of a parquet approximation calculation. They can be attributed to (I) ${\cal K}_{1,d}^{q}$, (II) ${\cal K}_{2,d}^{kq}$ and (III) ${\cal R}_{d}^{kk'q}$ respectively. }
\label{fig:PA2nd}
\end{figure*} 

The self-consistent solution of the parquet equations requires, through the equation $F=\Lambda_{\rm 2PI} + \sum_r \Phi_r$, the repeated recombination of vertex functions with different frequency conventions. The associated transformation to one common convention will lead to the loss of grid points in numerical implementations, making the understanding of the high-frequency behavior an essential component in the iterative solution of the parquet equations. Previous works have neglected the treatment of asymptotics\cite{Yang2009,Tam2013} and were faced with numerical stability problems.
We will discuss in this section how the asymptotic functions $\K^{q}$ and $\KK^{kq}$ can be used during the solution of the parquet equations. While the use of the scanning procedure outlined in Sec.~\ref{sec:implem} was explored\footnote{We note that the 'Kernel functions' $\Phi_r^q$ and $\Phi_r^{qk}$ introduced in Ref.~\onlinecite{Li2016} are related to $\K$ and $\KK$ as $\Phi_r^q = {\cal K}_{1,r}^q$ and $\Phi_r^{qk} = {\cal K}_{1,r}^q + {\cal K}_{2,r}^{kq}$.} in Ref.~\onlinecite{Li2016} we will in the following focus on how the limiting procedure can be used to determine the asymptotic functions during the self-consistent solution of the parquet equations.
Finally we will demonstrate how the diagrams of $\K$, $\KK$ and $\R$, starting from the lowest order, emerge naturally in a self-consistent solution of the parquet approximation.  

Given the fully irreducible vertex function $\Lambda_{\rm 2PI}$, the parquet equations\cite{Bickers2004} form a closed set of equations for the vertex functions $F$, $\Gamma_r$ and $\Phi_r$ as well as the self-energy $\Sigma$. 
They can be summarized schematically as
\begin{subequations}
\label{eq:parquet_equations}
\begin{align}
\label{eq:FGamm}
&F=\Lambda_{\rm 2PI} + \sum_r \Phi_r \qquad \Gamma_r = \Lambda_{\rm 2PI} + \sum_{r'\neq r} \Phi_r \\
\label{eq:PhiSig}
&\Phi_r=\Gamma_r G G F \qquad \hspace{0.8cm}  \Sigma= U \cdot G G G \cdot F,
\end{align}
\end{subequations}
while the detailed explicit forms of these equations are presented in Appendix \ref{app:equations}.

In the following we consider the $SU(2)$ symmetric case, which allows us to decouple the BSE by introducing the density ($d$), magnetic ($m$), singlet ($s$) and triplet ($t$) channel 
\begin{subequations}
\begin{align}
\Phi_d &= \Phi_{ph,\uu} + \Phi_{ph,\ud} \qquad \Phi_m = \Phi_{ph,\uu} - \Phi_{ph,\ud},\\
\Psi_s &= \Phi_{pp,\ud} - \Phi_{pp,\udc} \qquad \Psi_t = \Phi_{pp,\ud} + \Phi_{pp,\udc},
\end{align}
\end{subequations}
and $F_{r}$, $\Gamma_{r}$, $\R_{r}$, ${\cal K}_{1,r}$ and ${\cal K}_{2,r}$ for $r\in\{d,m,s,t\}$ are defined in the same way.

As observed in Ref.~\onlinecite{Li2016} and detailed in the previous sections, the reducible two-particle vertices $\Phi_r$ play a fundamental role in the correct treatment of the two-particle vertex function $F$, 
and thus for the solution of the parquet equations. 
The closed set of equations (\ref{eq:parquet_equations}) is solved self-consistently by using the iterative procedure described in the following. For this solution of the parquet equations, the fact that the asymptotic functions 
$\K$ and $\KK$ allow us to determine the vertex functions on the whole frequency domain is essential, as every step of the solution requires translations between the different notations as e.g.~demonstrated in \ceq{parquet_pp}.
If the vertex functions were known only on a finite frequency grid, these translations would lead to a loss of frequencies with every iteration of the parquet equations.

The only approximation that enters in the solution of the parquet equations is the choice of the fully irreducible vertex $\Lambda_{\rm 2PI}$. In the parquet approximation, it is approximated by its lowest order contribution 
$\Lambda_{\rm 2PI} \sim U$, while the D$\Gamma$A\cite{Toschi2007,Katanin2009,Li2016} and QUADRILEX approximate $\Lambda_{\rm 2PI}$ by the local one of the effective impurity model. A typical procedure for the solution of the parquet approximation can then be outlined as follows: 
\begin{enumerate}
\item Choose a finite but sufficiently large frequency range $[-\lambda, \lambda]$ for the problem studied
\item Initialize $\Sigma$ and the vertex functions $\Phi_r$ to 0, or make some educated guess for their starting values. Initialize $F$ and $\Gamma_r$ according to \ceq{FGamm}.
\item Calculate the reducible vertex functions $\Phi_r^{kk'q}$ with frequency arguments in the range $[-\lambda, \lambda]$ from the BSE [\ceq{PhiSig} left].
\item Update the values of the asymptotic functions using
\begin{equation}
\nonumber
\begin{split}
\mathcal{K}_{1,r}^{q} &= \lim_{|\nu| \to \infty} \lim_{|\nu'| \to \infty} \Phi_{r}^{k k' q} = U G G (U + \mathcal{K}_{1,r} + \mathcal{K}_{2,r}) \\
\mathcal{K}_{2,r}^{k q} &= \lim_{|\nu'| \to \infty} \Phi_{r}^{k k' q} - \mathcal{K}_{1,r}^{q} = (\Gamma_r - U) G G (U + \mathcal{K}_{1,r} + \mathcal{K}_{2,r}),
\end{split}
\end{equation}
The above expressions are found by performing the corresponding limits on the BSE [\ceq{PhiSig} left].\footnote{Note that an alternative implementation based on the scanning procedure outlined in Sec.~\ref{sec:implem} was used in Ref.~\onlinecite{Li2016}.}
\item Compute the vertex functions $F^{kk'q}_{r}$ and $\Gamma^{kk'q}_{r}$ using the updated values of $\Phi_{r}^{kk'q}$ [\ceq{FGamm}]. When any of the three frequency arguments of $k, k'$ or $q$ fall outside of the range $[-\lambda, \lambda]$, we approximate 
\[\Phi^{kk'q}_{pp, {\rm asympt.}} \approx \KPP^q + \KKPP^{ kq } + \KKBPP^{ k'q }.\]
\item Calculate the self-energy from $F$ through the Schwinger-Dyson equation [\ceq{PhiSig} right].
\item Go back to step 3 and iterate until convergence is achieved. 
\end{enumerate}

Let us now consider diagrams (see \cfg{PA2nd} for the $d$-channel) 
generated in the first iterations of the parquet approximation solution (assuming initial conditions $\Phi_r=0$), and attribute them to the diagrammatic classes I ($\K$), II ($\KK,\KKB$) 
and III ($\R$). After the first iteration the reducible vertex functions $\Phi_{r}^{kk'q}$ read
\begin{subequations}
\begin{align}
\Phi_{d}^{kk'q}&=U^{2}\chi^q_{0,ph} \qquad \Phi_{m}^{kk'q}=U^{2}\chi^q_{0,ph},\\
\Psi_s^{kk'q}&=-2U^{2}\chi^{q}_{0,pp} \qquad \Psi_t^{kk'q}=0,
\end{align}
\end{subequations} 
where $\chi_0$ denotes the non-interacting bubble (see first diagram in \cfg{PA2nd} for the $d$ case). 
This corresponds to the lowest order perturbation theory for $\Phi_r$, which is only dependent on the transfer frequency and momentum $q$, and can thus be attributed to ${\cal K}_{1,r}$.
The dependence of $\Phi$ on the fermionic arguments is only generated when the updated vertex functions $F$ and $\Gamma$, e.g.~for the $d$-channel
\begin{subequations}
\begin{align}
F_{d}^{kk'q} &= U - 2U^{2}\chi^{k'-k}_{0,ph} -U^{2}\chi^{k+k'+q}_{0,pp}+U^{2}\chi^{q}_{0,ph}, \\
\Gamma_{d}^{kk'q} &= U - 2U^{2}\chi^{k'-k}_{0,ph}-U^{2}\chi^{k+k'+q}_{0,pp},
\end{align}
\end{subequations} 
are inserted into the BSE in the second iteration. This produces the remaining diagrams in \cfg{PA2nd}, and thus the first contributions to the $\KK$ (II) and $\R$ (III) functions.
Iterating this procedure, we generate successively all reducible diagrams. Results obtained by this approach for a SIAM are presented in the following Sec.~\ref{sec:compare_SIAM}.

\section{Comparison to exact results of the SIAM} 	
\label{sec:compare_SIAM}

In this section we illustrate the high quality of the description of the vertex asymptotics obtained using the algorithmic implementations discussed in the previous sections.

In particular, we present results for the asymptotic functions as obtained from the fRG ($\Omega$-flow including two-loop corrections) and parquet approximation for a single impurity Anderson model and compare them with exact diagonalization data, which were acquired following the procedure outlined in Appendix \ref{app:asympt_ED}. Besides the asymptotic functions, also results for the rest function and the self-energy will be shown. In Sec.~\ref{subsec:fRG_corrections} we will further discuss a detailed comparison between the fRG in its one- and two-loop implementation for both the $\Omega$- as well as the $U$-flow.
We first consider the regime of weak coupling where the fRG and the PA, as approximation schemes, are expected to be quantitatively correct. 
Hence, in this regime, the comparison with the exact results of ED will represent a stringent test for our treatment of the high-frequency asymptotics. After having demonstrated that the error introduced in the high-frequency asymptotics of the vertex function is negligible, we proceed by applying our fRG and PA algorithms, including the high-frequency treatment, to the intermediate to strong coupling regime. In this case, the comparison to the ED will allow us to assess directly the intrinsic performance of the two approximations in the non-perturbative parameter region, since sizable errors, which typically arise from a poor treatment of the high-frequency regime of the vertex functions, are strongly mitigated by the present approach.
\begin{figure*}
   \centering
   \includegraphics[width=0.85\textwidth]{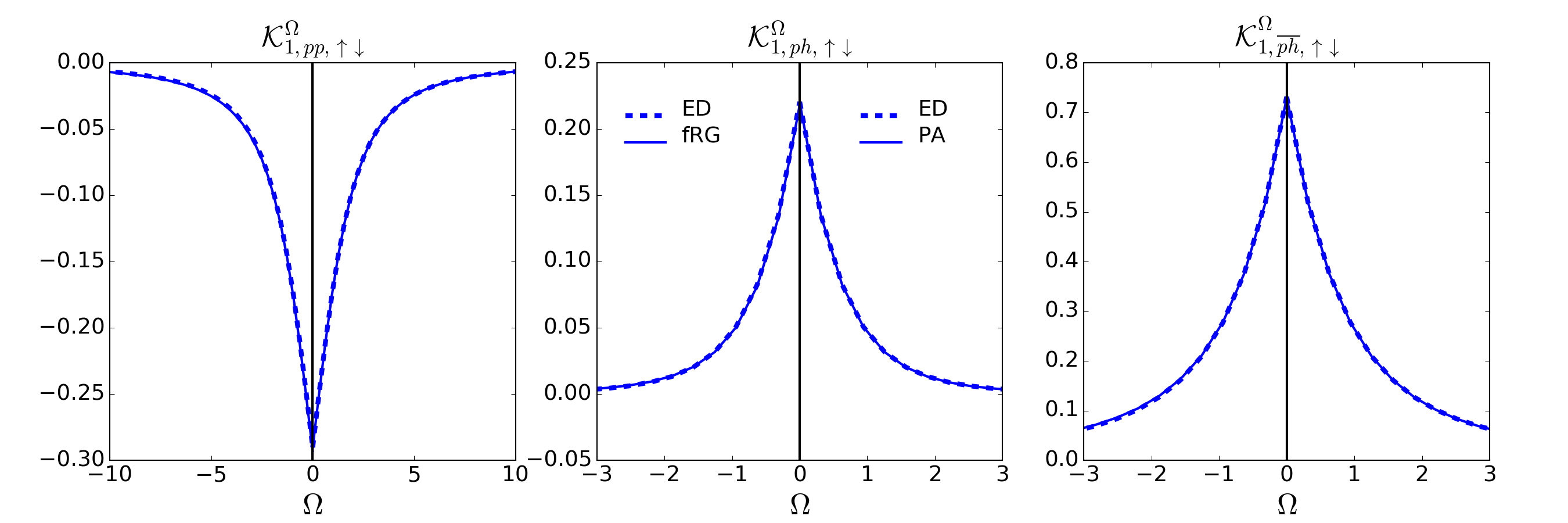}
   \caption{ $\KUD^{\Omega}$ for all three scattering channels. We present results obtained by fRG (left, solid), PA (right, solid) and ED (right, dashed) for the SIAM with $U=1$, $\beta=20$ and $D=1$. }
   \label{fig:u1_k1}
\end{figure*}
\begin{figure*}
   \centering
   \includegraphics[width=0.85\textwidth]{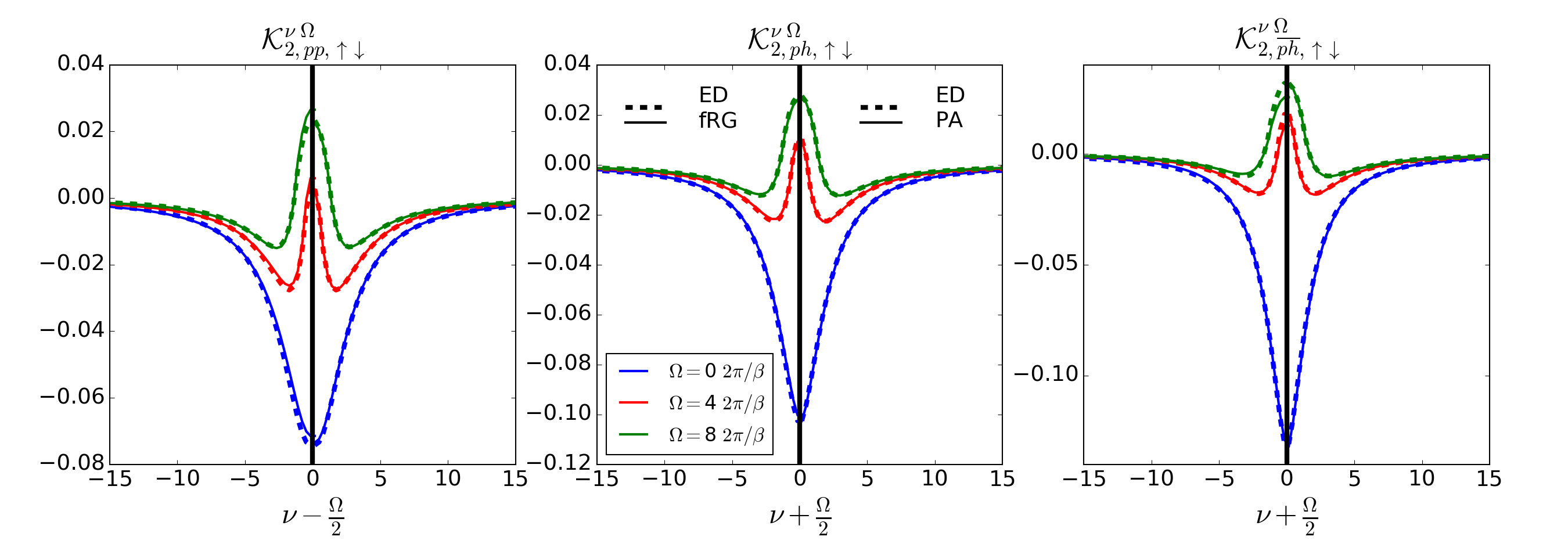}
   \caption{ $\KKUD^{\nu\Omega}$ for all three scattering channels as a function of $\nu$ and for different values of $\Omega$. We present results obtained by fRG (left, solid), PA (right, solid) and ED (right, dashed) for the SIAM with $U=1$, $\beta=20$ and $D=1$. }
   \label{fig:u1_k2}
\end{figure*}
\begin{figure*}
   \centering
   \includegraphics[width=1.0\textwidth]{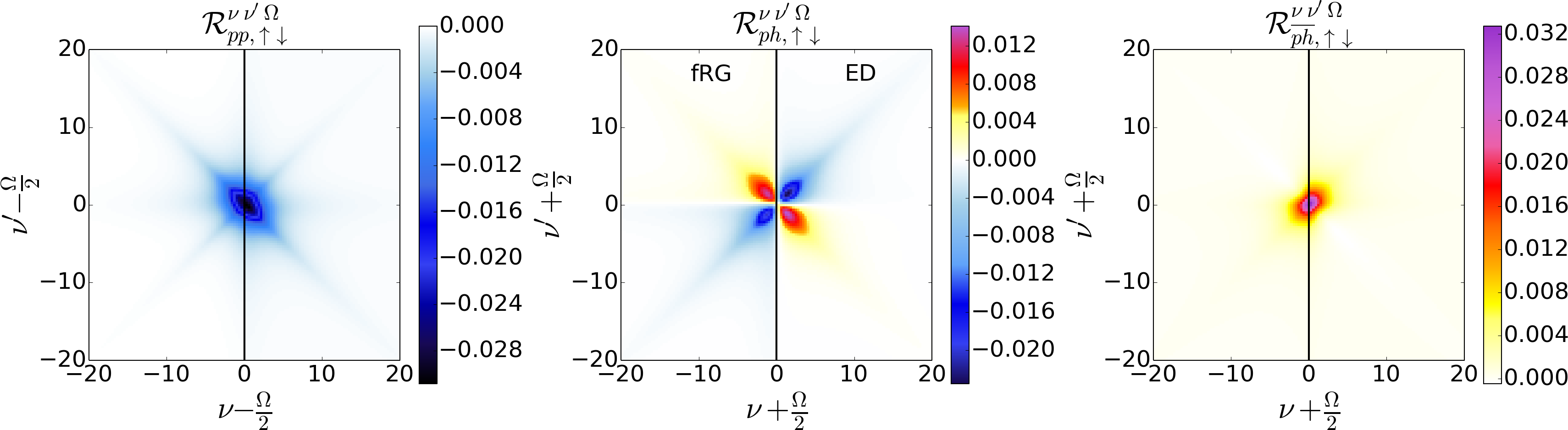}
   \vspace*{0.15cm}\\
   \includegraphics[width=1.0\textwidth]{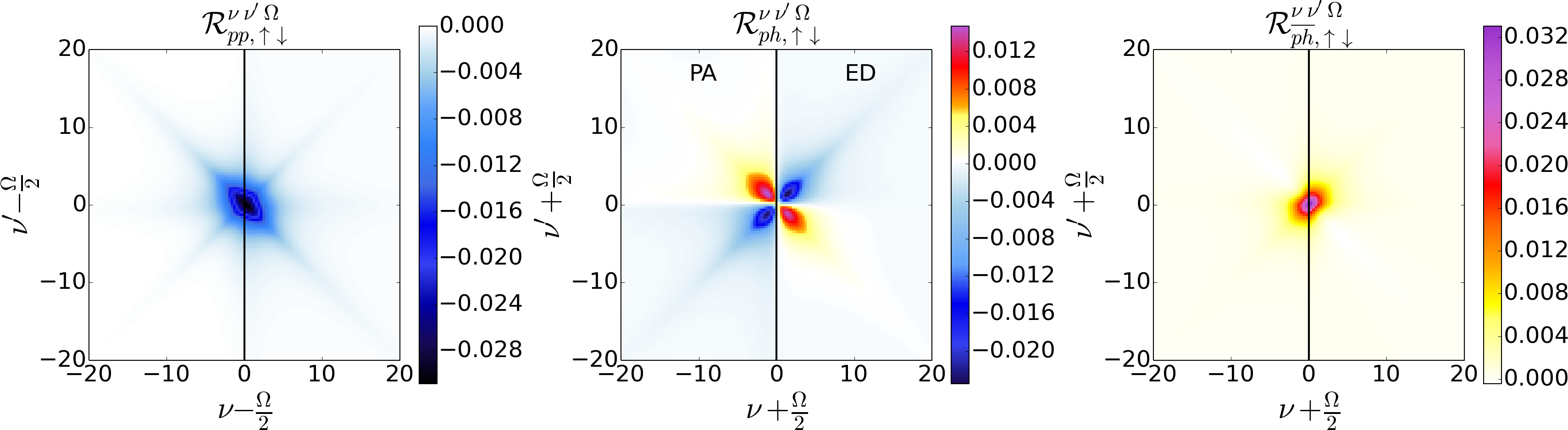}
   \caption{Rest function $\R^{\nu\nu'\Omega}_{\ud}$ for all three scattering channels as a function of $\nu$ and $\nu'$ plotted for $\Omega=0$. We present results obtained by fRG (1st row, left) and PA (2nd row, left) for the SIAM with $U=1$, $\beta=20$ and $D=1$. The right side always shows the corresponding ED result.}
   \label{fig:u1_R_Om0}
\end{figure*}
\begin{figure*}
   \centering
   \includegraphics[width=1.0\textwidth]{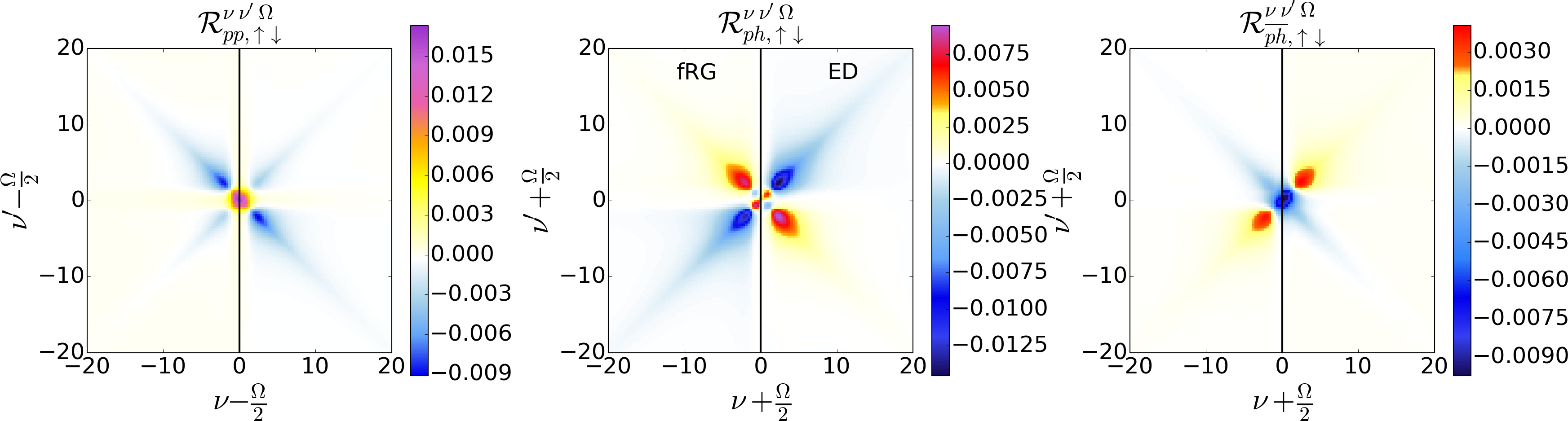}
   \vspace*{0.15cm}\\
   \includegraphics[width=1.0\textwidth]{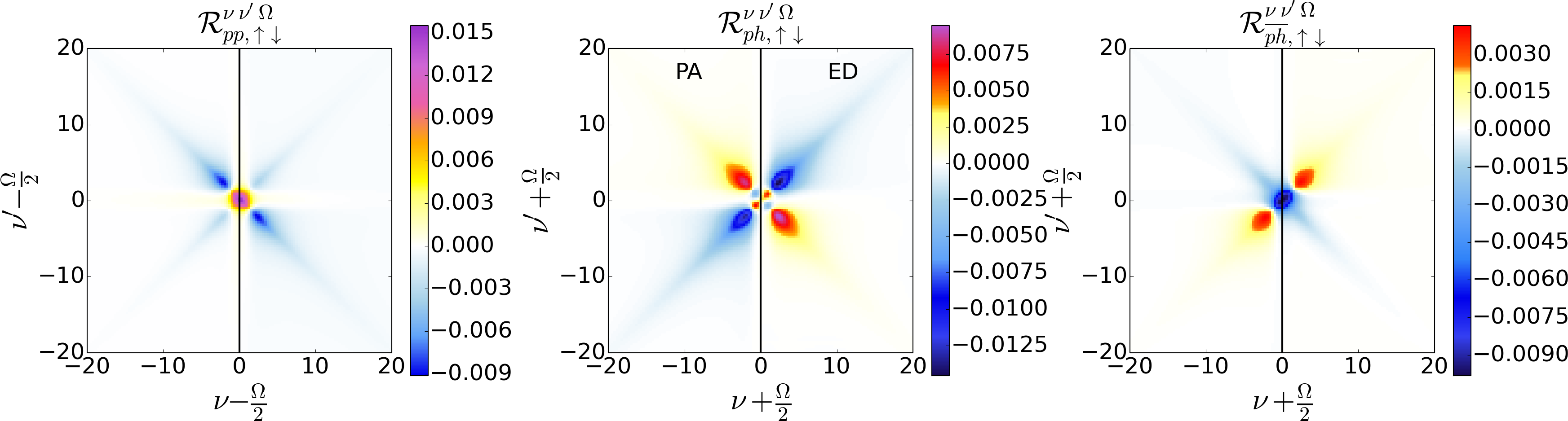}
   \caption{Same as \cfg{u1_R_Om0}, but for a finite transfer frequency $\Omega=8\frac{2\pi}{\beta}$. }
   \label{fig:u1_R_Om8}
\end{figure*}
\begin{figure}
   \centering
   \includegraphics[width=0.8\columnwidth]{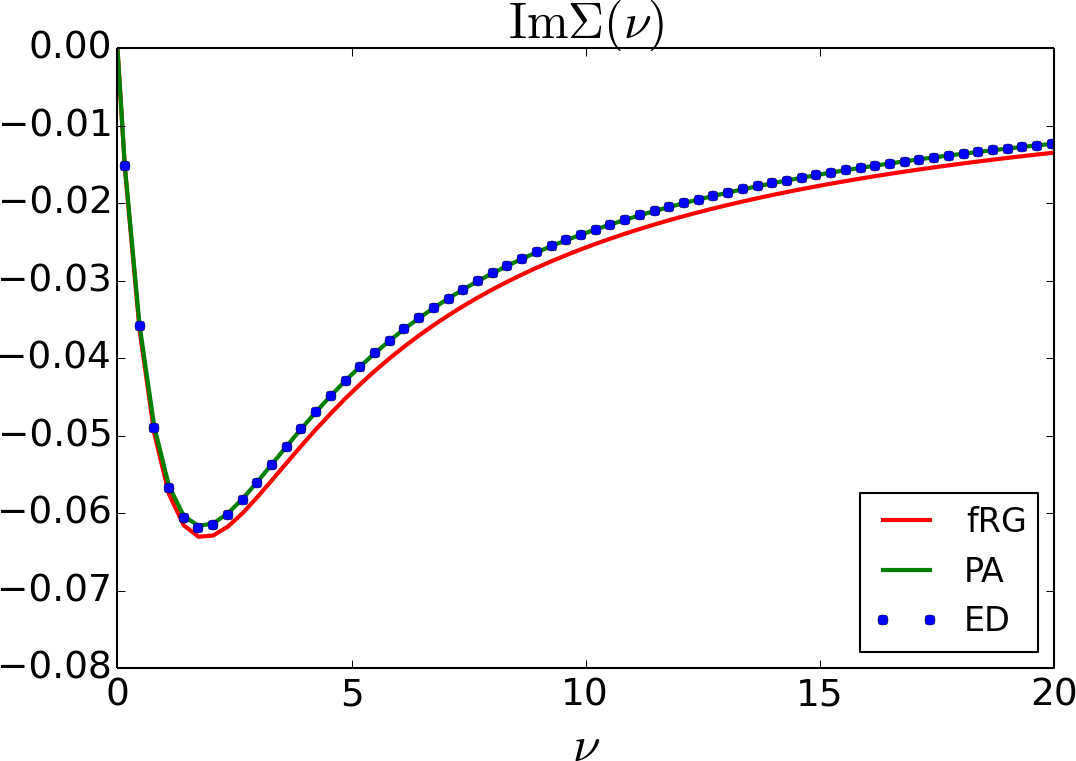}
   \caption{ $\IM \Sigma (i\nu)$ as obtained by fRG~(red, solid), PA~(green, solid) and ED~(blue, dotted) for the SIAM with $U=1$, $\beta=20$ and $D=1$. }
   \label{fig:u1_sig}
\end{figure}

The system of interest in this section is a SIAM, i.e.~a single impurity site with local repulsive Coulomb interaction $U$ coupled to a non-interacting bath [see Sec.~\ref{subsec:formalism}, \ceq{defhamilt}]. In our specific case, we consider a box-like density of states
\begin{equation}
\rho(\omega) = \frac{1}{2D} \Theta(D-|\omega|),
\label{eq:box_like_DOS}
\end{equation}
where $D$ denotes the half-bandwidth, which will be used as our unit of energy, i.e.~$D=1$.
This bath is coupled to our impurity site by means of a hopping $t=\sqrt{2/\pi}$, such that the resulting hybridization function reads 
$\Delta(\omega) = \pi t^{2}  \rho(\omega)=2\rho(\omega)$. This choice results in $\Delta(0)=D=1$, allowing us to directly relate our unit of energy to the one used in wide-band limit calculations\cite{Karrasch2008}, namely the hybridization function evaluated at the chemical potential.

However, the exact diagonalization of the SIAM is not possible for $\rho(\omega)$ of \ceq{box_like_DOS}. Hence, we have determined a set of four optimized bath energy levels ${\epsilon_n}$ and hoppings ${t_n}$ with the resulting hybridization function
\begin{equation}
\Delta^{\text{ED}}(i\nu) = \sum_{n=1}^{4} \frac{t_n^2}{i\nu - \epsilon_n},
\end{equation}
in order to mimic the continuous bath of \ceq{box_like_DOS} in the best way possible within a discretized ED scheme\cite{Caffarel1994,Koch2008}. 
Following a somewhat similar strategy as in the ED algorithms for DMFT, we determine our bath parameters such that the norm
\begin{equation}
   \sum_{i\nu}| \Delta^{\text{ED}}(i\nu) - \Delta(i\nu) |^2
\end{equation}
is minimized. For an inverse temperature $\beta=20$, which was used for all numerical calculations presented in this paper, we have $\epsilon_n ={-0.7, -0.15, 0.15, 0.7}$ and $t_n={0.45,0.34,0.34,0.45}$.
Note also that, since we are considering the particle-hole symmetric case, all two-particle quantities are purely real, while the self-energy is purely imaginary\footnote{Let us stress, however, that our decomposition scheme holds equally in parameter regimes where particle-hole symmetry is broken.}.
Unless mentioned otherwise, calculations are performed with a frequency grid of $128\times128\times256$ Matsubara frequencies for the $\Phi$-functions, while grids of $128\times256$ and $256$ are chosen for $\KK$ and $\K$ respectively.
\begin{figure*}
   \centering
   \includegraphics[width=0.85\textwidth]{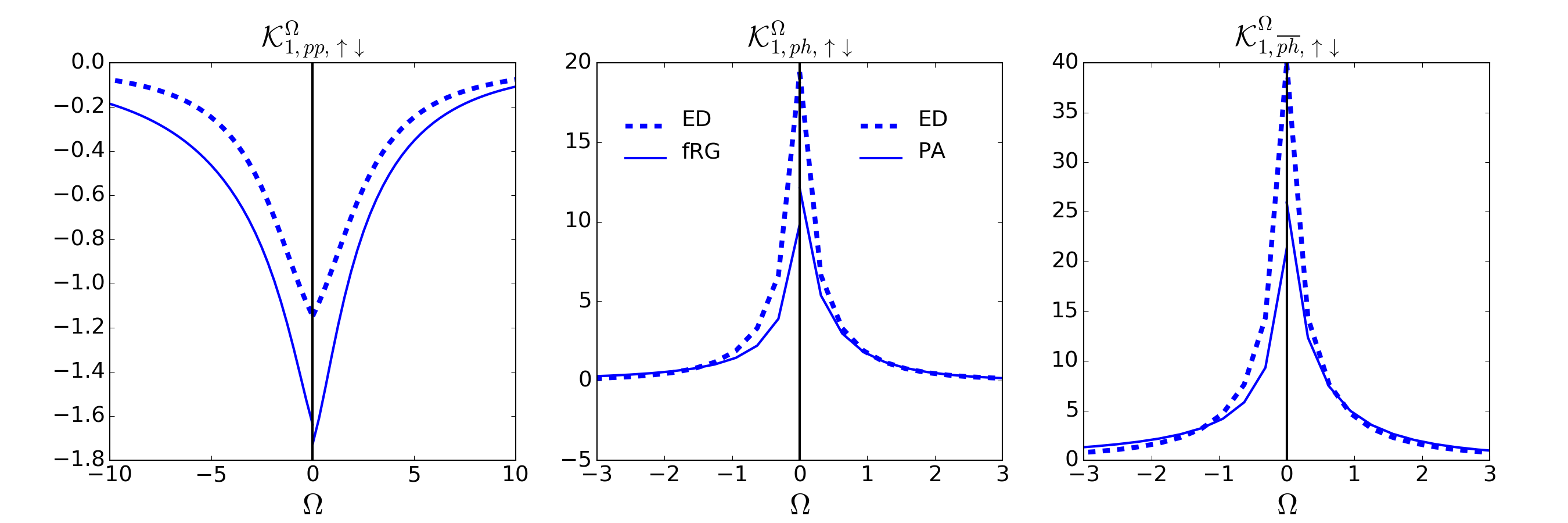}
   \caption{Same as \cfg{u1_k1}, but for $U=4$. }
   \label{fig:u4_k1}
\end{figure*}
\begin{figure*}
   \centering
   \includegraphics[width=0.85\textwidth]{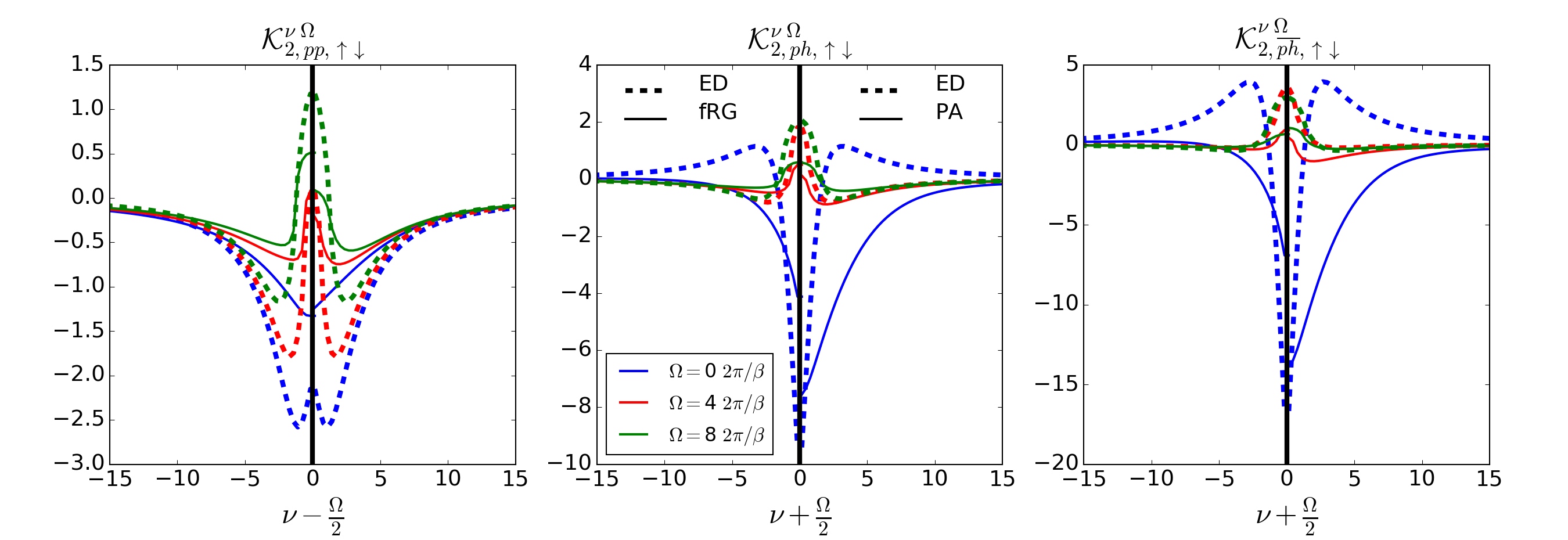}
   \caption{Same as \cfg{u1_k2}, but for $U=4$. }
   \label{fig:u4_k2}
\end{figure*}
\begin{figure}[b]
   \centering
   \includegraphics[width=0.7\columnwidth]{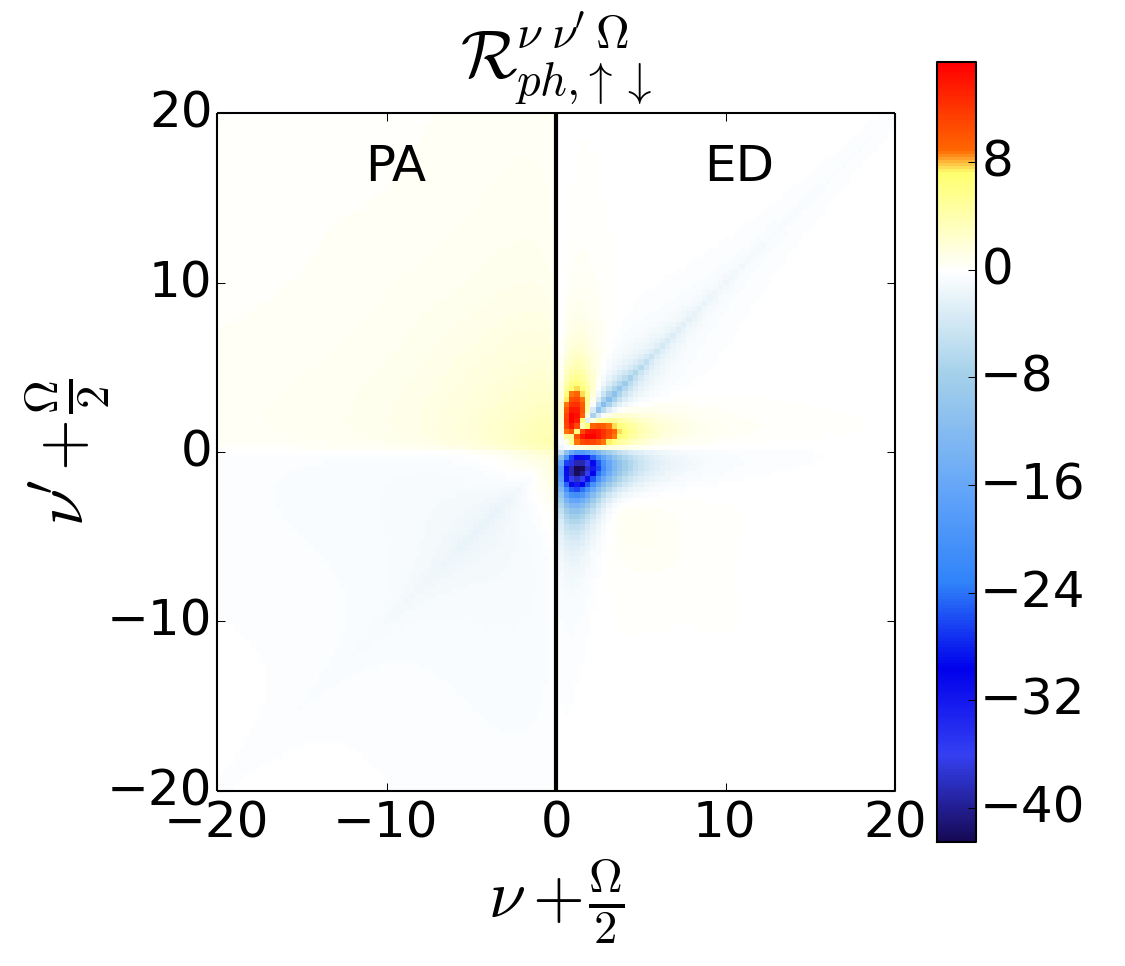}
   \caption{Comparison of $\mathcal{R}_{ph,\ud}$ obtained by means of PA with the exact result. Here, $U=4$.}
   \label{fig:u4_R}
\end{figure}
\begin{figure}[b]
   \centering
   \includegraphics[width=0.8\columnwidth]{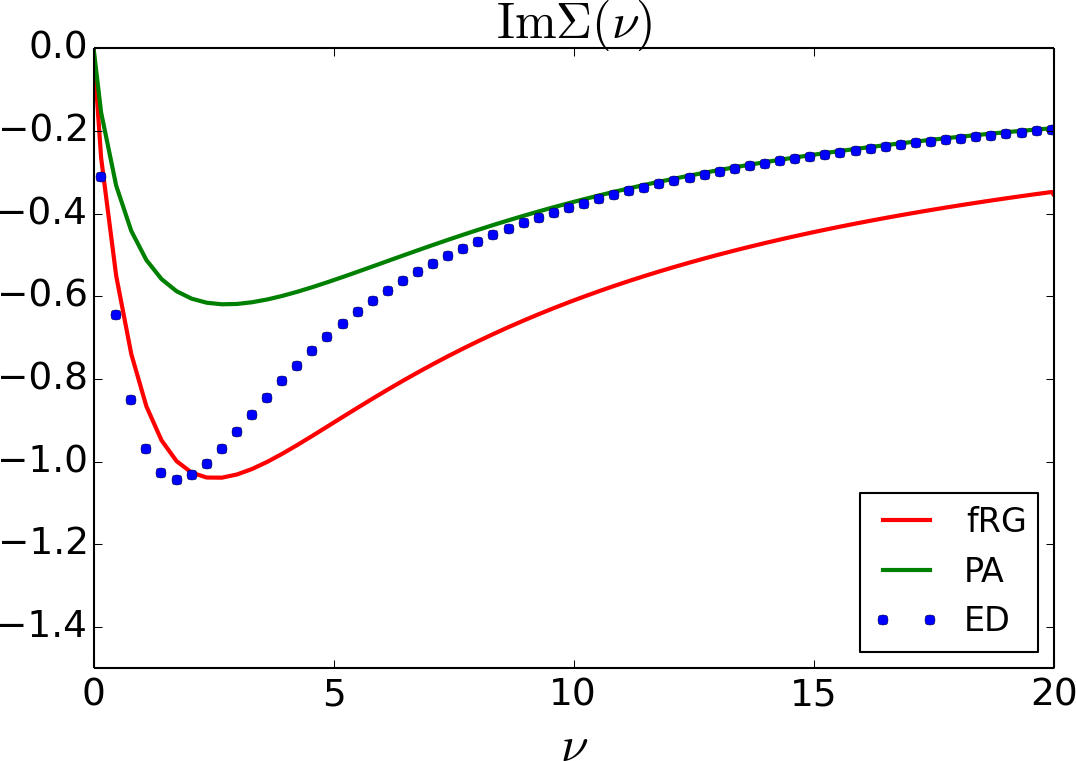}
   \caption{Same as \cfg{u1_sig}, but for $U=4$. }
   \label{fig:u4_sig}
\end{figure}

Let us start considering the weak-coupling case ($U=1$). The data for $\KUD$, $\KKUD$ and $\RUD$ are presented in \cfg{u1_k1}, \cfg{u1_k2} and Figs.~\ref{fig:u1_R_Om0} and \ref{fig:u1_R_Om8} respectively\footnote{The contour plots are created such that every small square of equal color represents the value of the function at the bottom left corner of this square.}. 
For this parameter choice, we find an excellent agreement between the different approaches and the exact solution for all quantities. 
At the level of the asymptotic function $\KUD$, no distinction can be made between the results of the different schemes, while for $\KKUD$ the fRG shows some minor deviations w.r.t.~PA and ED in the $pp$ and $\xph$ channel. 
Even at the level of the rest function $\RUD$, which has as a leading order $U^4$, we find excellent agreement between PA and ED, while only minor deviations are again observed for the fRG.
Note that, contrary to the plotting conventions adopted in previous Refs.~\onlinecite{Rohringer2012, Li2016}, the fermionic frequencies are shifted by $\pm \Omega/2$ for $\KK$ and $\R$, because the main frequency structures move outwards as $\Omega$ is increased.\footnote{We note that both, the fRG and the parquet solver, are making use of this shift by $\pm \Omega/2$ in the implementation to keep the structures of $\KK$, $\R$ and $\Phi$ centered in the frequency grids (in contrast to Ref.~\onlinecite{Li2016}).}
This observation suggests to include a corresponding shift also in the notation used in the numerical implementation, such that the localized frequency structures can be more efficiently captured by means of the finite grid even in the case of finite transfer frequency.
Similar trends are observed for the self-energy shown in \cfg{u1_sig}. While PA and ED agree perfectly, we find that the fRG self-energy deviates from the exact results, especially in its tail.

All this numerical evidence proves the reliability of our treatment of the high frequency asymptotics within the different schemes (see also the results for $U=2$ in the supplements), allowing us to evaluate in an unbiased way their intrinsic performance in the most challenging strong-coupling regime.

Due to the perturbative nature of fRG and PA, the situation changes drastically in the regime of stronger coupling. 
The corresponding results for $U=4$ are presented for $\KUD$ and $\KKUD$ in Figs. \ref{fig:u4_k1} and \ref{fig:u4_k2} respectively. 
Note that, for this value of the interaction, we are clearly in the non-perturbative regime, as divergencies\cite{Yang2011,Schaefer2013,Janis2014,Kozik2015,Schaefer2016} are already present
in the exact vertices obtained by ED.

For both, PA and fRG, $\KUD$ shows already strong deviations from the exact results, while the qualitative structures are still captured. 
These deviations are particularly enhanced in the $ph$ and $\xph$ channel. 
In the case of $\KKUD$ qualitative features are missed by the PA and fRG, in particular for $\Omega=0$, while a qualitative agreement is still achieved for finite transfer frequency.
As the main structures of the rest function $\R$ are neither reproduced by PA nor by fRG, we show only one example for this comparison in \cfg{u4_R} (the full vertices are reported in the supplements).
Since this diagrammatic class is at least fourth order in the interaction, the strongest deviations were to be expected here.
Finally, we note that for the self-energy, shown in \cfg{u4_sig}, strong deviations are observed in both cases.

\subsubsection{Neglecting the asymptotics}

\begin{figure}
   \centering
   \includegraphics[width=0.85\cw]{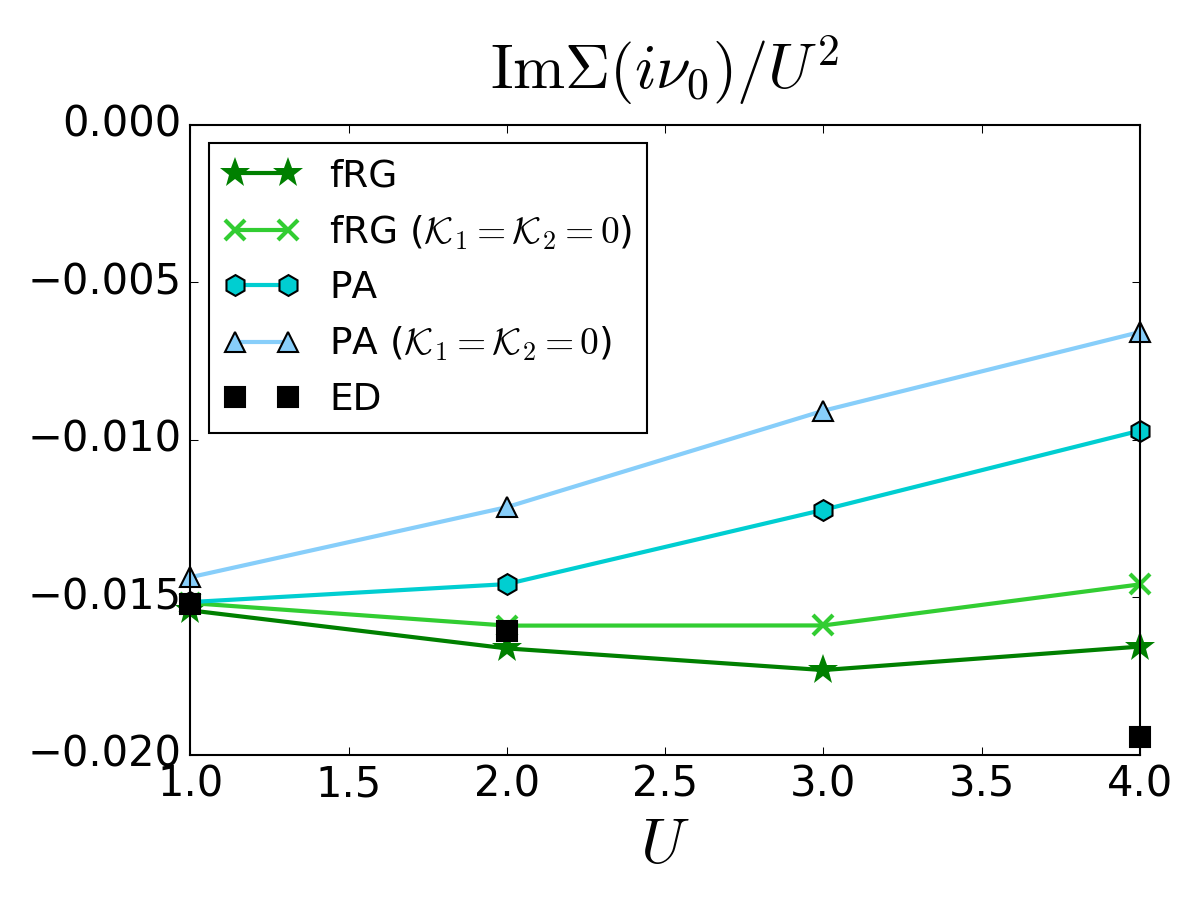}
   \caption{Comparison of $\IM \Sigma(i \nu_0)/U^2$ for the SIAM with $\beta=20$ and $D=1$ as obtained by fRG and PA with and without ($\K=\KK=0$) high-frequency asymptotics, compared with ED for $\nu_0=\frac{\pi}{\beta}$ and different values of $U$.}
   \label{fig:noasympt}
\end{figure}
\begin{figure}
   \centering
   \includegraphics[width=0.90\cw]{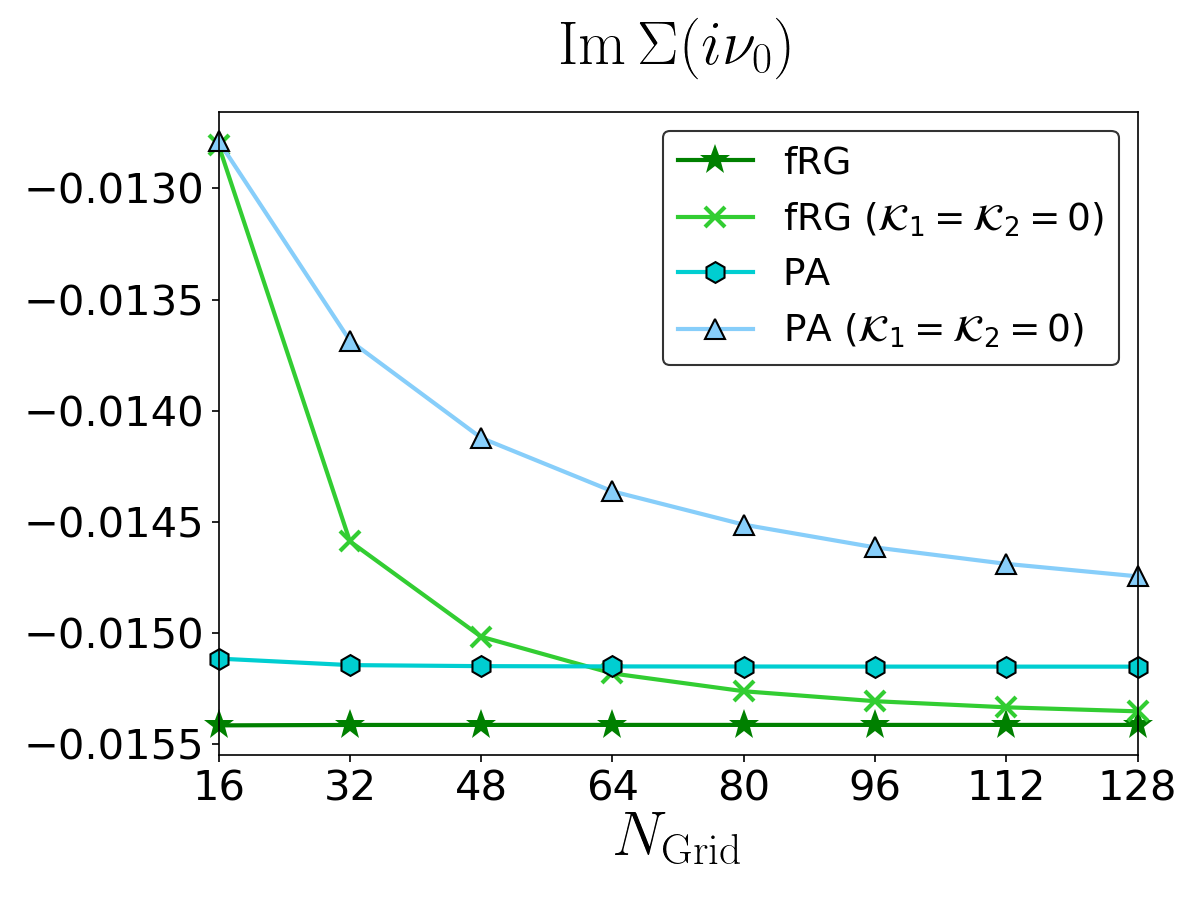}
   \includegraphics[width=0.90\cw]{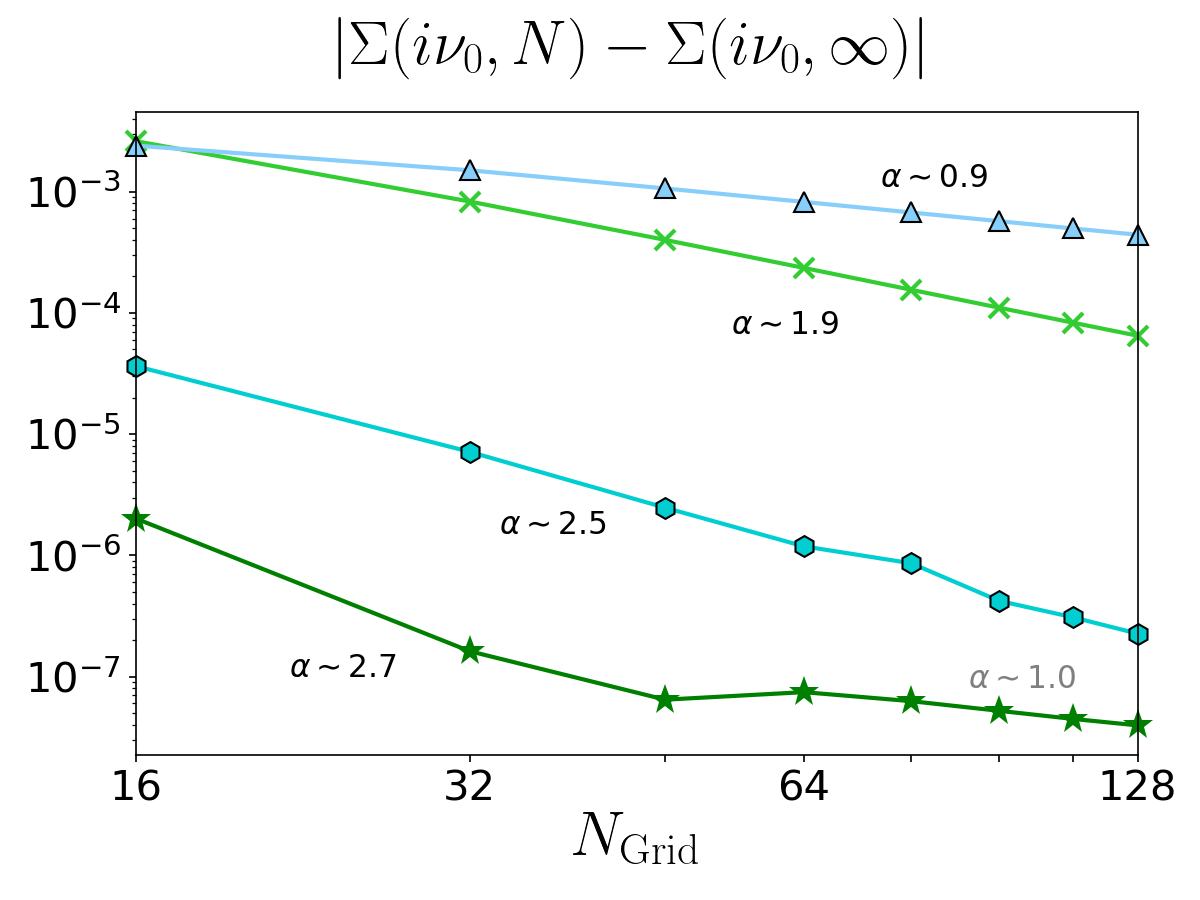}
   \caption {(Top) Comparison of $\IM \Sigma(i \nu_0)$ for the SIAM with $\beta=20$, $U=1$ and $D=1$ as obtained by fRG and PA with and without ($\K=\KK=0$) high-frequency asymptotics for $\nu_0=\frac{\pi}{\beta}$ and different values of the grid-size $N_{\rm Grid}$. (Bottom) Deviation of the data presented in the top-panel from the converged result ($N_{\rm Grid}=\infty$) using a log-log plot. Convergence w.r.t.~the grid size follows a power law $\IM \Sigma(i \nu_0, N_{\rm Grid}) = \IM \Sigma( i\nu_0, \infty) + \mc{O}(N_{\rm Grid}^{-\alpha})$. Estimates for the scaling exponents are shown.}
   \label{fig:convergence}
\end{figure}
Let us now discuss the importance of considering asymptotic functions in numerical implementations. 
In this regard, we present in \cfg{noasympt} results for $\IM \Sigma(i\nu_0)/U^2$ as a function of $U$ 
calculated by fRG and PA, with and without asymptotic functions, and compare them with the exact ED data. 
For these calculations, a frequency grid of $64\times64\times128$ ($N_{\rm Grid} = 64$) Matsubara frequencies was used for the reducible vertex functions.
In the large frequency domain, we used \ceq{asympt} and $\Phi_{r,{\rm asympt.}}\approx0$ respectively.

We observe that the results for both, fRG and PA, are strongly affected if we include the asymptotic functions in the calculations: The comparison with the exact result improves by a substantial amount. This is a strong
indication of the importance of a correct description of the high-frequency part of the vertex function in all vertex-based numerical implementations.

To get a deeper understanding on the convergence properties we show in \cfg{convergence} calculations at fixed $U=1$ for multiple values of $N_{\rm Grid}$. We observe that the results
that treat the asymptotic functions properly are almost converged already for the smallest grid size $N_{\rm Grid} = 16$, while the calculations which approximate $\Phi_{r,{\rm asympt.}}\approx0$ fail to converge even for $N_{\rm Grid}=128$. This convergence is slower for the PA. We expect this to be directly linked to the fully self-consistent nature of PA, which then leads to an enhancement of the rough high-frequency approximation in the iterative solution. 
The bottom panel of \cfg{convergence} analyzes the deviation from the converged result in a log-log plot for the calculations without asymptotic functions. We observe that, through a proper treatment of the high-frequency asymptotics, numerical accuracy can on average be improved by three orders of magnitude. The linear dependence shows that the convergence w.r.t.~the grid size follows a power law. Overall we observe a faster convergence (i.e. larger exponent $\alpha$) when the asymptotics are taken into account. Only for the fRG case (dark green) do we observe a change in the scaling at large grid sizes, possibly attributed to a change in the leading source of error.

\subsubsection{Higher order corrections in fRG}
\label{subsec:fRG_corrections}

\begin{figure*}
   \centering
   \includegraphics[width=0.315\tw]{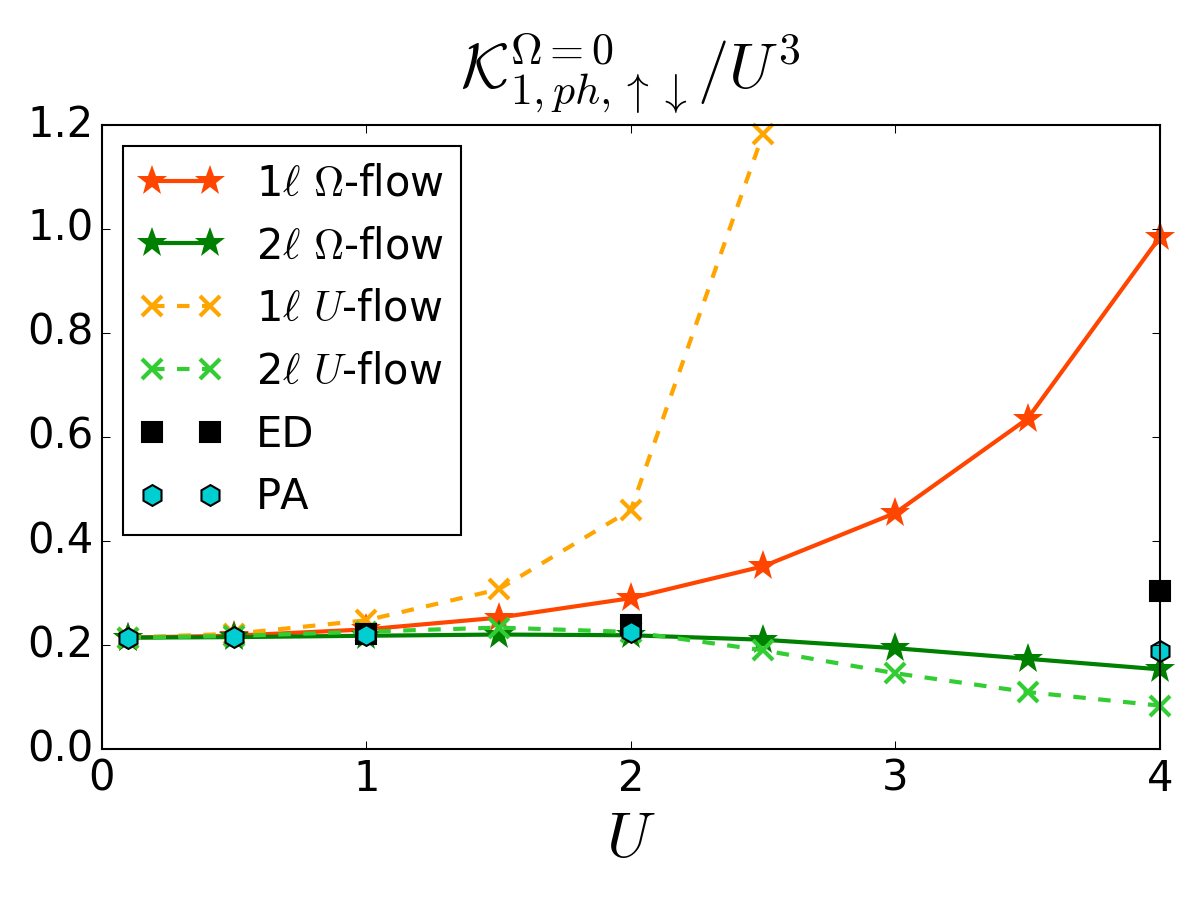}
   \hspace{0.01\textwidth}
   \includegraphics[width=0.32\tw]{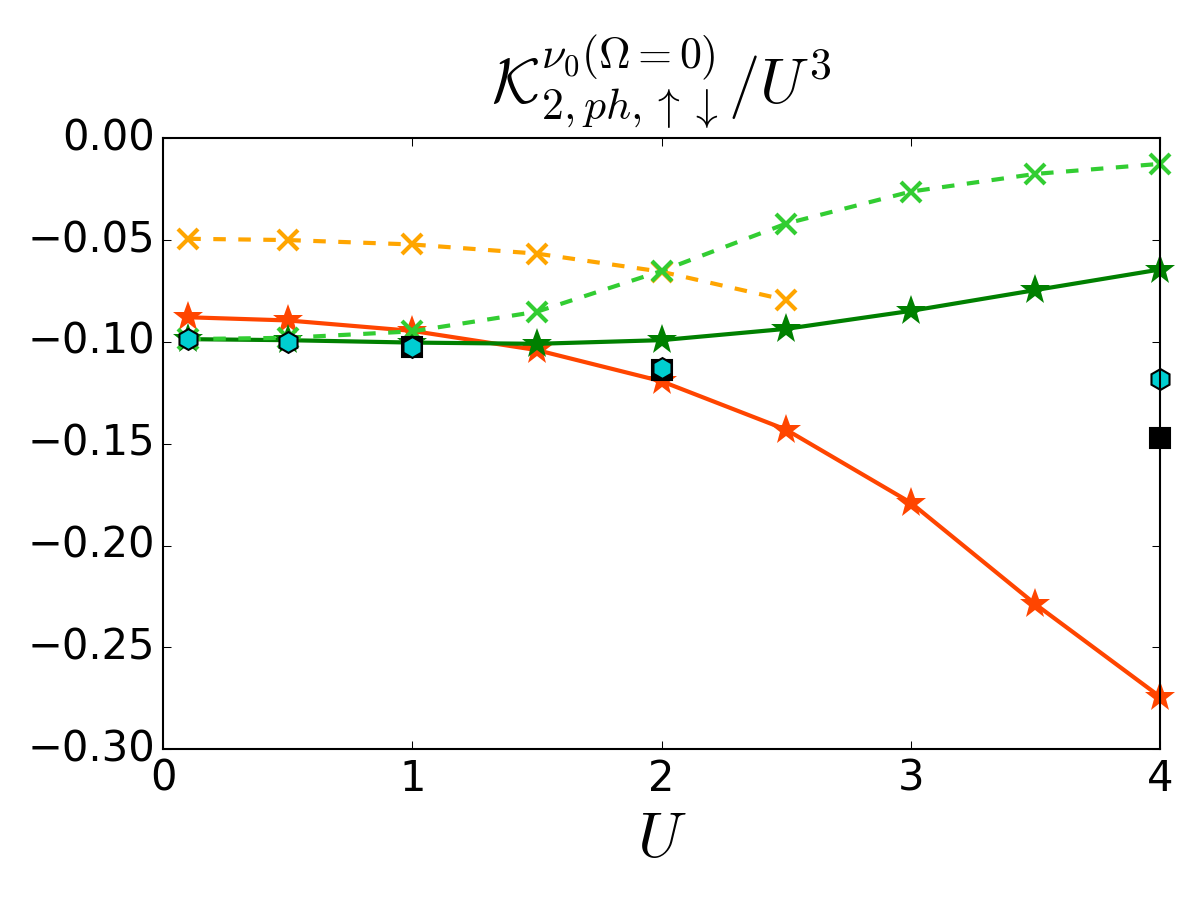}
   \hspace{0.01\textwidth}
   \includegraphics[width=0.315\tw]{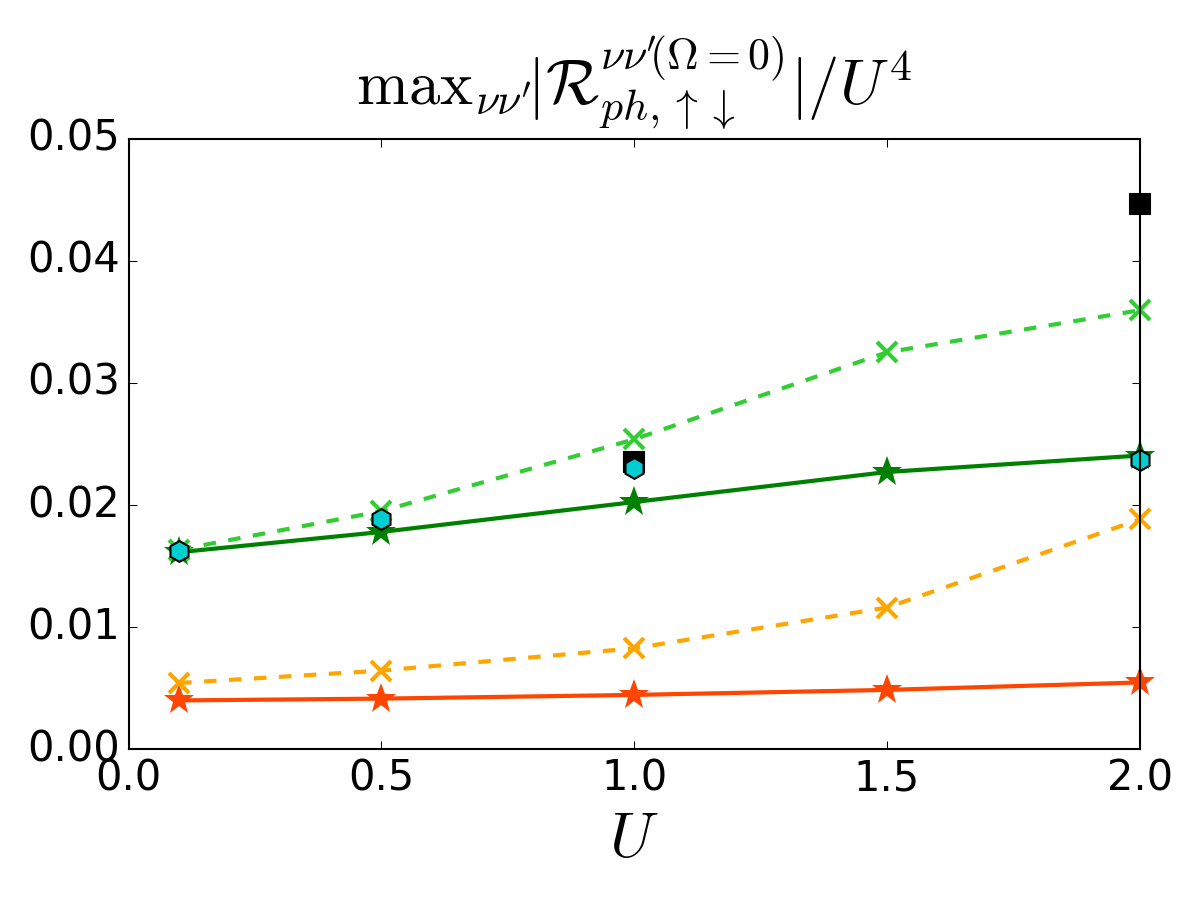}
   \caption{Comparison of $\KPHUD^{\Omega=0}/U^3$, $\KKPHUD^{\nu_0(\Omega=0)}/U^3$, and  $\operatorname{max}_{\nu\nu'}|\R_{ph,\ud}^{\nu\nu'(\Omega=0)}|/U^4$ ($\nu_0=\frac{\pi}{\beta}$) for fRG in the one-loop ($1\ell$) and two-loop ($2\ell$) implementation, for both the $\Omega$- and $U$-flow, with ED and PA. We note that the one-loop $U$-flow diverges for $U=3$ or larger.}
   \label{fig:fRG_corrections}
\end{figure*}

In this subsection, we provide a quantitative comparison between the SIAM results as obtained by means of fRG in its one- and two-loop implementation\cite{Eberlein2014}. To this aim, we compare in \cfg{fRG_corrections} the quantities $\KPHUD^{\Omega=0}$, $\KKPHUD^{\nu_0(\Omega=0)}$, as well as  $\operatorname{max}_{\nu\nu'}|\R_{ph,\ud}^{\nu\nu'(\Omega=0)}|$ normalized by their leading order\footnote{For the particle-hole channel in the $\ud$ spin configuration the bare bubble vanishes, resulting in a leading order $\mathcal{O}(U^3)$.} in $U$, to the exact ED results as well as to the PA.
Consistently to our expectations, we find that the two-loop corrections yield a systematic improvement of the $\K$, $\KK$ and $\R$ functions acquired during the flow, in particular for larger values of the interaction. 

More specifically, for $\KPHUD$ the two-loop corrections have a minor effect in the weak-coupling regime, whereas an excellent agreement with the exact results is achieved already at the one-loop level. 
At larger $U$, the one-loop scheme strongly overestimates $\KPHUD$. Here, the two-loop corrections yield a substantial improvement over the one-loop scheme, while underestimating $\KPHUD$. We also note the strongly improved agreement of the two-loop fRG with the PA, which is a trend to be expected, since the two-loop scheme allows to include higher orders of the reducible diagrams in an exact way.

As for $\KKPHUD$, we observe that already in the limit $U\to0$ the one-loop scheme {\sl fails} to reproduce the exact result. This can be attributed to the fact that the lowest order diagram in $\KKPHUD$ is of order $U^3$, and is thus not captured exactly in the one-loop scheme. 
In particular for the $U$-flow, we numerically verify the factor $\frac{1}{2}$ (w.r.t.~the exact result) already predicted diagrammatically at the end of Sec.~\ref{subsec:fRG}, while for the $\Omega$-flow we find, numerically, a factor of $\sim0.89$ in all channels.
For larger values of $U$, we observe a behavior similar to the one described for $\KPHUD$, that is, a systematic improvement of the results if the two-loop corrections are included in fRG.

For $\R_{ph,\ud}$ the trend is similar, while, being a function of $\mathcal{O}(U^4)$, the relative deviations from the exact results increase substantially. The predicted factor $\frac{1}{3}$ for $U \to 0$ is verified numerically, while for the $\Omega$-flow we find factors $0.78$, $0.25$ and $0.78$ in the $pp$, $ph$ and $\xph$ channel respectively.

As for the comparison between the flow-schemes, consistently with the ratios in the weak-coupling regime, we observe that the simpler $U$-flow performs overall worse than the $\Omega$-flow.

\subsubsection{Efficiency of simplified parametrization schemes}
\label{subsec:simplified_param}

\begin{figure*}
   \centering
   \includegraphics[width=0.315\tw]{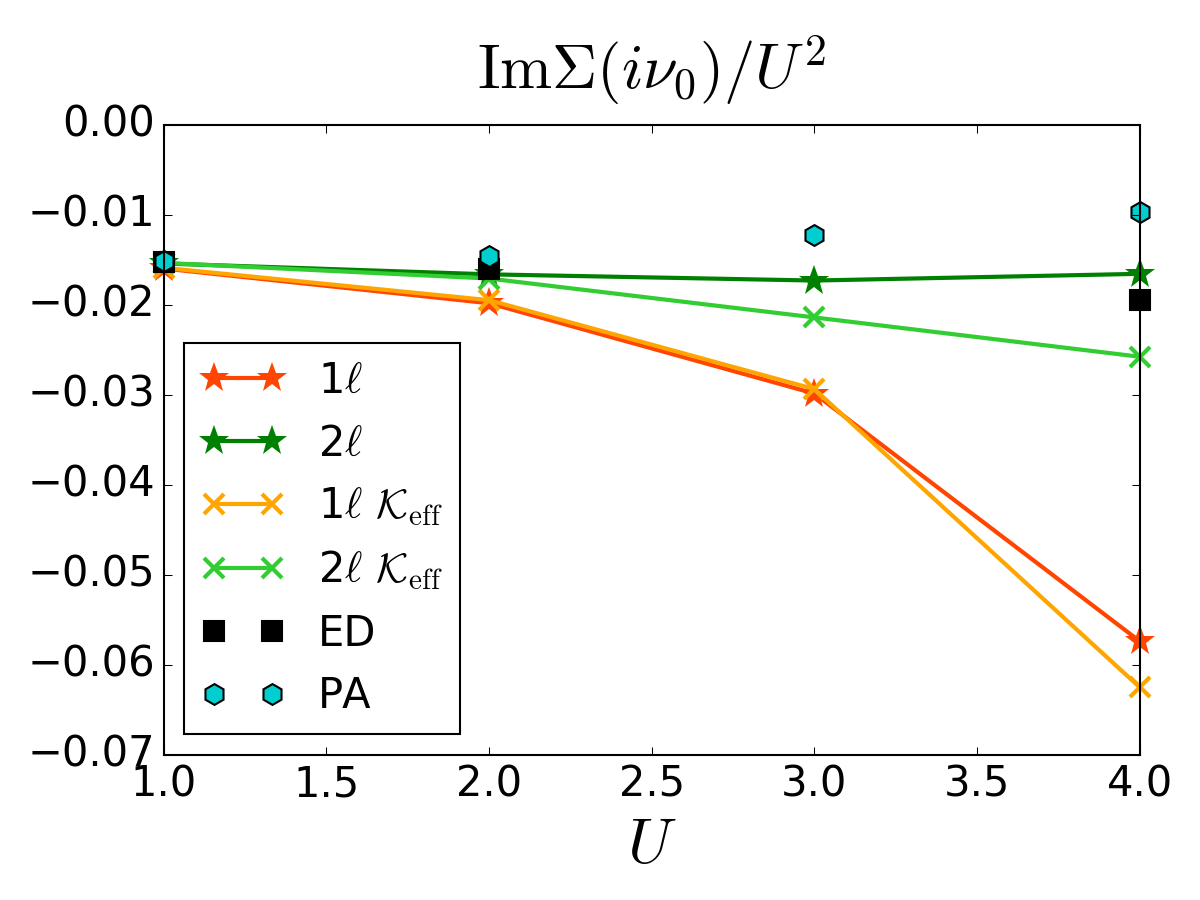}
   \hspace{0.01\textwidth}
   \includegraphics[width=0.315\tw]{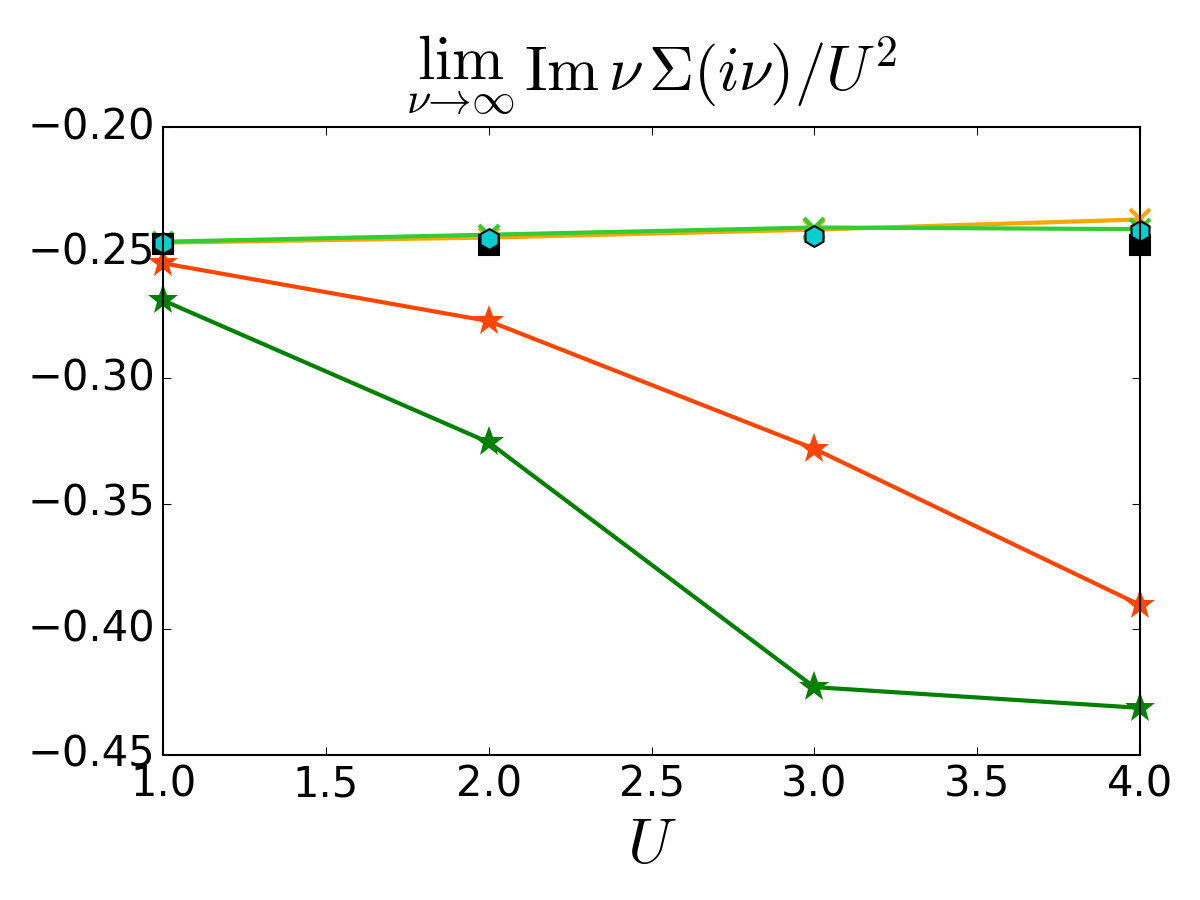}
   \hspace{0.01\textwidth}
   \includegraphics[width=0.315\tw]{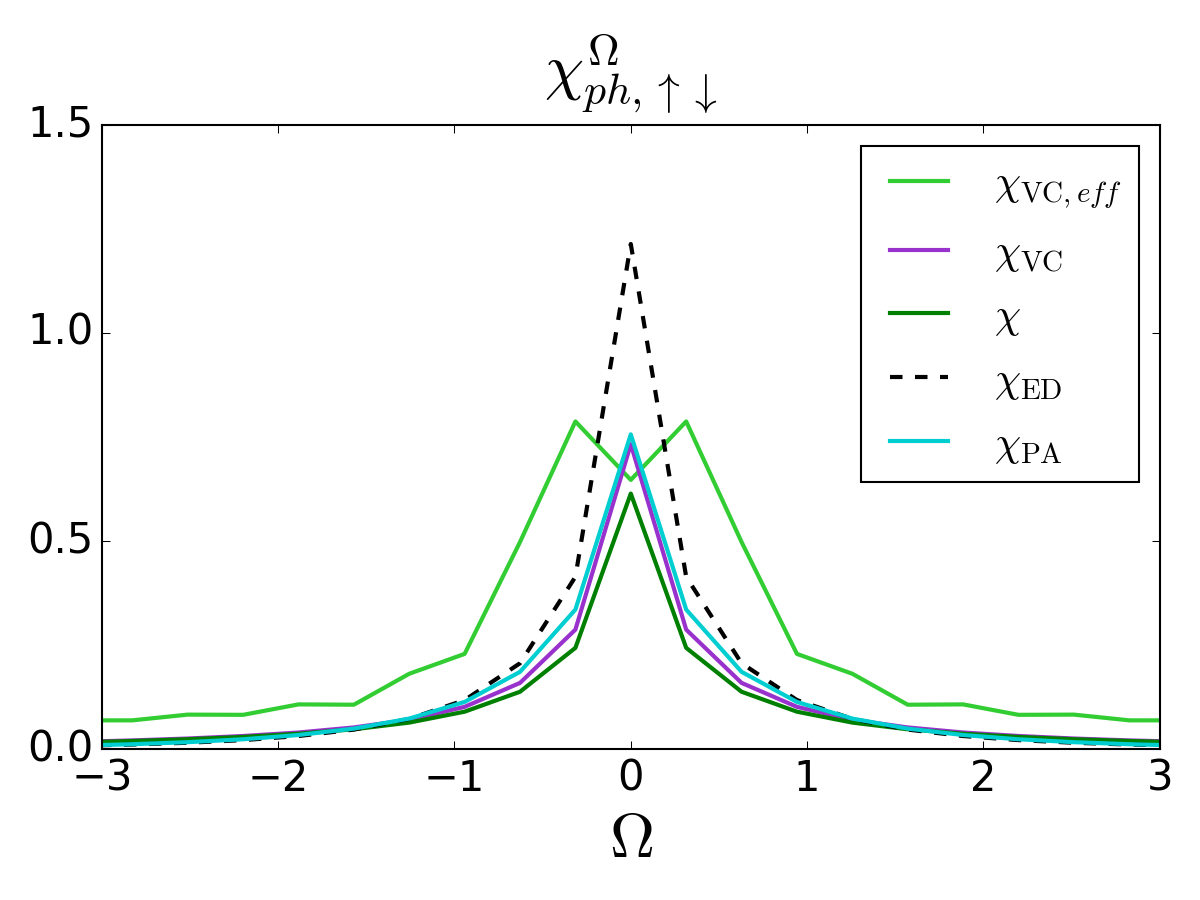}
   \caption{Left two panels: Comparison of $\IM \Sigma(i \nu_0)/U^2$ and $\lim_{\nu\to\infty}\,\IM \nu\Sigma(i\nu)/U^2$ for fRG one-loop and two-loop with the corresponding simplified schemes introduced previously [$\mc{K}_{\rm eff}$, see \ceq{karrasch}], PA and ED. Right panel: Comparison of the susceptibility $\chi$ in the $ph$-channel for $U=4$ as obtained by means of \ceq{k1_ph} after the two-loop fRG flow ($\chi_{\rm VC}$) in the conventional and the simplified scheme (eff). This is compared to the susceptibility $\chi$ obtained directly from \ceq{relation_chi} as well as the PA and ED result. }
   \label{fig:simplified_param}
\end{figure*}
In this subsection we preset results for the simplified parametrization scheme\cite{Karrasch2008} presented in \ceq{karrasch}. 
It has found extensive use in the fRG community\cite{Husemann2009,Eckel2010,Gasenzer2010,Jakobs2010,Reuther2010,Karrasch2011,Floerchinger2012,Giering2012,Goettel2012,Honerkamp2012,Husemann2012,Kennes2012,Kennes2012a,Uebelacker2012,Kinza2013,Kinza2013a,Bauer2014,Eberlein2014,Eberlein2014a,Laakso2014,Taranto2014,Kinza2015,Rentrop2015} and beyond\cite{Nuss2012,Freton2014,Krones2015,Lange2015,Nuss2015,Janis2017}, as it allows for a substantial speedup of numerical calculations.
In the left two panels of \cfg{simplified_param} we compare the self-energy at the first Matsubara frequency $\IM \Sigma(i\nu_0)/U^2$ as well as its tail $\lim_{\nu\to\infty}\,\IM \nu\Sigma(i\nu)/U^2$ 
for fRG one-loop and two-loop in their full and simplified ($\mc{K}_{\rm eff}$) implementation with PA and the exact results from ED.

For the self-energy at the first Matsubara frequency, we find a good agreement between the simplified parametrization scheme and the fully parametrized fRG implementation for both the
one- and two-loop scheme, while the simpler scheme performs slightly worse in reproducing the exact results. 
In the case of the self-energy tail, the situation is reversed. Here, the aforementioned deviations of the fRG from ED are indeed cured by the simplified parametrization scheme. 

To capture the effect of the simplified approximation scheme on the two-particle quantities, we compare in the right panel of \cfg{simplified_param} the corresponding susceptibility in the $ph$-channel for the two-loop case.
It is important to note that $\chi$ cannot be directly extracted from $\mathcal{K}_{\rm eff}$ using \ceq{relation_chi} due to the effective inclusion of $\KK$ and $\R$. 
Instead, we calculate the susceptibility after the flow by calculating explicitly the bare and vertex-corrected bubble according to equation \ceq{k1_ph} (VC). 
This is compared, for an interaction value $U=4$, to $\chi=\K/U^2$ and the corresponding PA and ED data. We find that the simplified parametrization fails to qualitatively reproduce the exact susceptibility, while the other
approximations, although underestimating $\chi$, compare qualitatively well with ED\footnote{We note that $\chi$ calculated after the full two-loop flow by means of \ceq{k1_diag} yields a result different from $\K/U^2$. 
This is connected to the specific approximations introduced in the fRG, and is absent in the fully self-consistent PA.}.

While the parametrization scheme of \ceq{karrasch} performs well for one-particle quantities, we find that the qualitative features of the susceptibility are badly reproduced.
Further, we observe that the ambiguities in the definition of the flow equations for the case of finite temperatures
turn out to have a substantial effect on the results for larger values of the interaction.
These are strong arguments for the fully parametrized schemes, that capture, consistently, all frequency structures of the two-particle vertex function. 

\section{Conclusions and outlook}			
\label{sec:conclusion}

We have presented a detailed analysis of the diagrammatic content of the two-particle vertex functions, focusing on the terms controlling their high frequency structures. This information is extremely valuable, also at a practical level, because the efficient algorithmic treatment of the vertex asymptotics is fundamental for several recently proposed quantum field theoretical approaches based on expansions around a correlated starting point. In particular, by focusing on the two-particle reducible parts of the vertex function, we could identify the different contributions to their high-frequency asymptotics as diagrammatic classes with a reduced frequency (and momentum) dependence, and establish a connection to the (physical) susceptibilities and the fermion-boson vertices. The gained insights are essential in order to devise efficient parametrization schemes for the two-particle vertex functions. We then discussed the algorithmic details necessary for the implementation of these ideas in numerical (and analytical) studies, considering as specific examples the functional renormalization group approach and the parquet approximation. In order to verify the correct treatment of the high frequency asymptotics, we benchmarked our numerical implementations for a SIAM against exact calculations from ED. Finally, we tested the intrinsic performance of the approaches also in the intermediate-to-strong coupling regime.

This algorithmic progress paves the way towards a full numerical treatment of correlations at the two-particle level, which is pivotal for all vertex-based quantum many-body methods. 
In particular, these ideas are directly applicable in the treatment of non-local correlations beyond the dynamical mean-field theory by means of its cutting-edge diagrammatic extensions\cite{Rohringer2018}, such as DF, DMF$^2$RG, D$\Gamma$A, and the recently introduced TRILEX and QUADRILEX approach. They are further crucial for numerically reliable implementations of vertex-based fRG\cite{Tagliavini2019,Hille2020,Chalupa2020a} and its recently proposed multiloop extension\cite{Kugler2018a,Kugler2018b}.
Moreover, our new approach, which has been derived and applied for a purely static and local Hubbard interaction in a half-filled system can be systematically applied to more realistic situations out of half-filling\cite{Kaufmann2017} and longer range interactions. An extension to retarded interactions on the other hand requires a comprehensive analysis of the effect due to the frequency dependence of the coupling which represents an interesting future research direction.


\acknowledgments
We are grateful to A. Antipov, T. Ayral, E. Di Napoli, A. Eberlein, C. Honerkamp, A. Katanin, W. Metzner, O. Parcollet, D. Rohe, T. Sch\"afer, P. Thunstr\"om, A. Valli, and D. Vilardi for valuable discussions. We further thank A. Eberlein and T. Ayral for critical reading of the manuscript.
We acknowledge financial support from the Deutsche Forschungsgemeinschaft (DFG) 
through FOR 723, ZUK 63, Projects No. AN 815/4-1 and 407372336 (G.R.), and SFB/TRR 21, the Austrian Science Fund (FWF) within the Project F41 (SFB ViCoM) and I-2794-N35, and the European Research Council (ERC) under the European
Union’s Seventh Framework Program (FP/2007-2013)/ERC through Grant No. 306447 (G.L., K.H.). Calculations were performed on the Vienna Scientific Cluster (VSC).

\appendix

\section{Explicit equations}				
\label{app:equations}

Here, we want to present some explicit forms of the parquet equations depicted schematically in Sec.~\ref{subsec:parquet}.
Let us begin by showing the BSE in their spin-resolved version.\footnote{We want to point out the fond similarity between Eqs.~(\ref{eq:bethe_salpeter_full}), which are the basis for the iterative parquet approximation solver, and the channel-resolved fRG flow Eqs.~(\ref{eq:flowChannels}), which technically allows for very similar implementations of the two approaches.}
\begin{widetext}
\begin{subequations}
\label{eq:bethe_salpeter_full}
\begin{align}
   \Phi^{k k' q}_{pp,\ud}&=\phantom{-}\sumint dk'' \hspace{0.1cm} G( k'' )G( q-k'' ) \times \Gamma^{k k'' q}_{pp,\ud}F^{k'' k' q}_{pp,\ud}, \\
   \Phi^{k k' q}_{ph,\ud}&=-\sumint dk'' \hspace{0.1cm} G( k''+q )G( k'' )\times \Big[ \Gamma_{ph,\uu}^{k k'' q}F_{ph,\ud}^{k'' k' q} + \Gamma_{ph,\ud}^{k k'' q}F_{ph,\uu}^{k'' k' q} \Big], \\
   \Phi^{k k' q}_{\xph,\ud}&=\phantom{-}\sumint dk'' \hspace{0.1cm} G( k''+q )G( k'' ) \times \Gamma_{\xph,\ud}^{k k'' q}F_{\xph,\ud}^{k'' k' q}.
\end{align} 
\end{subequations}
\end{widetext}
In the $SU(2)$ symmetric case they can be diagonalized by introducing the density (d), magnetic (m), singlet (s) and triplet (t) channel as introduced in Sec.~\ref{subsec:parquet},

\begin{subequations}
\label{eq:bethe_salpeter_diagonal}
\begin{align}
\Phi_{d/m}^{kk'q} &= \phantom{\pm\frac{1}{2}} \sumint dk'' \hspace{0.1cm} G( k''+q )G( k'' ) \times \Gamma^{k k'' q}_{d/m}F^{k'' k' q}_{d/m}, \\
\Psi_{s/t}^{kk'q} &= \pm\frac{1}{2}           \sumint dk'' \hspace{0.1cm} G( k'' )G( q-k'' ) \times \Gamma^{k k'' q}_{s/t}F^{k'' k' q}_{s/t}.
\end{align}
\end{subequations} 
Given the fully irreducible vertex function $\Lambda^{kk'q}$, the other two-particle vertex functions can be calculated through the parquet equations as shown below:
\begin{widetext}
\begin{subequations}
\begin{align}
F_{d}^{kk'q}&=\Lambda_{d}^{kk'q}  + \Phi^{kk'q}_{d} - \frac{1}{2}\Phi_{d}^{k(k+q)(k'-k)} - \frac{3}{2}\Phi_{m}^{k(k+q)(k'-k)} 
+\frac{1}{2}\Psi_{s}^{kk'(k+k'+q)}
+\frac{3}{2}\Psi_{t}^{kk'(k+k'+q)},\label{PA_F_d}\\
F_{m}^{kk'q}&=\Lambda_{m}^{kk'q} + \Phi^{kk'q}_{m} - \frac{1}{2}\Phi_{d}^{k(k+q)(k'-k)} + \frac{1}{2}\Phi_{m}^{k(k+q)(k'-k)}
 - \frac{1}{2}\Psi_{s}^{kk'(k+k'+q)}
+\frac{1}{2}\Psi_{t}^{kk'(k+k'+q)},\label{PA_F_m}\\
F_{s}^{kk'q}&=\Lambda_{s}^{kk'q} + \Psi^{kk'q}_{s}
 +\frac{1}{2}\Phi_{d}^{k(q-k')(k'-k)}
 -\frac{3}{2}\Phi_{m}^{k(q-k')(k'-k)}
 +\frac{1}{2}\Phi_{d}^{kk'(q-k-k')}
 - \frac{3}{2}\Phi_{m}^{kk'(q-k-k')},\label{PA_F_s}\\
F_{t}^{kk'q} &= \Lambda_{t}^{kk'q}  + \Psi^{kk'q}_{t}
 -\frac{1}{2}\Phi_{d}^{k(q-k')(k'-k)}
 -\frac{1}{2}\Phi_{m}^{k(q-k')(k'-k)}
+\frac{1}{2}\Phi_{d}^{kk'(q-k-k')}
+ \frac{1}{2}\Phi_{m}^{kk'(q-k-k')}.\label{PA_F_t}    
\end{align}
\end{subequations} 
\end{widetext}
Similarly, the channel-dependent irreducible two-particle vertex $\Gamma_{r}^{kk'q}$ is obtained as $\Gamma_{r}^{kk'q} = F_{r}^{kk'q}-\Phi_{r}^{kk'q}$.

\section{Extracting asymptotics from $F$}		
\label{app:asympt_ED}

In this part we describe an approach that extracts the asymptotic functions directly from the full vertex function $F$. This procedure was employed to acquire all presented high-frequency results for the ED vertices, and is based on the fact that one can write down explicit Feynman diagrams for all asymptotic functions.
These consist of all possible ways of pinching two external legs of $F$ into one bare vertex $U$\cite{Rohringer2014,Rohringer2016}. 
Since the latter is purely local in space and time,
the dependence on two fermionic arguments is replaced by a single bosonic (transfer) one. The resulting diagrams for $\K$ are shown in \cfg{k1_diag}, and read explicitly 
\begin{widetext}
\begin{subequations}
\label{eq:k1_diag}
\begin{align}
\begin{split}
\hspace{-2cm}\mathcal{K}_{1,pp,\sigma\sigma'}^{q}= \phantom{+} & U^2 (1-\delta_{\sigma\sigma'}) \sumint dk_i \, G(k_1) G(q - k_1) F_{pp, \sigma \sigma'}^{k_1 k_2 q} G(q - k_2) G(k_2) - U^2 (1-\delta_{\sigma\sigma'}) \sumint dk_1 \, G(q-k_1) G(k_1),
\label{eq:k1_pp}
\end{split}
\\[1ex]
\begin{split}
\mathcal{K}_{1,ph,\sigma\sigma'}^{q} = \phantom{-} & U^2 \sumint dk_i \, G(k_1) G(k_1+q) F_{ph, \sigma \sigma'}^{k_1 k_2 q} G(k_2) G(k_2+q) + U^2 \delta_{\sigma,\sigma'}\sumint dk_1 \, G(k_1) G(k_1+q),
\label{eq:k1_ph}
\end{split}
\\[1ex]
\begin{split}
\mathcal{K}_{1,\xph,\sigma\sigma'}^{q} = \phantom{+} & U^2 \sumint dk_i \, G(k_1) G(k_1+q) F_{\xph, \sigma \sigma'}^{k_1 k_2 q} G(k_2) G(k_2+q) - U^2 \sumint dk_1 \, G(k_1) G(k_1+q).
\label{eq:k1_xph}
\end{split}
\end{align}
\end{subequations}
Here $\overline{\sigma}$ denotes the opposite spin of $\sigma$, and $SU(2)$ symmetry is explicitly assumed.
\begin{figure*}
   \centering
   \begin{minipage}{0.18\textwidth}
      $\mathcal{K}_{1, pp, \sigma\sigma'}^{q} = (1-\delta_{\sigma\sigma'})$
   \end{minipage}
   \begin{minipage}{0.33\textwidth}
      \includegraphics[width=1.0\textwidth]{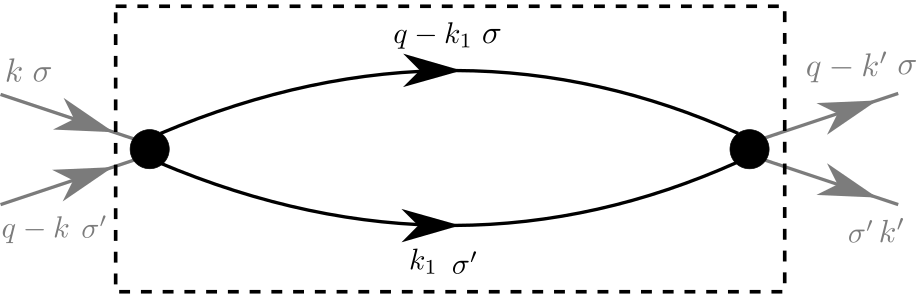}
   \end{minipage}
   \begin{minipage}{0.1\textwidth}
      $+ \,(1-\delta_{\sigma\sigma'})$
   \end{minipage}
   \begin{minipage}{0.33\textwidth}
      \includegraphics[width=1.0\textwidth]{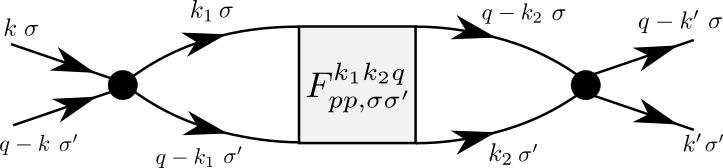}
   \end{minipage}
   \\[4ex]
   \begin{minipage}{0.1\textwidth}
      $\mathcal{K}_{1,ph,\sigma\sigma'}^{q} =$
   \end{minipage}
   \begin{minipage}{0.05\textwidth}
      $\delta_{\sigma\sigma'}$
   \end{minipage}
   \begin{minipage}{0.11\textwidth}
      \includegraphics[width=1.0\textwidth]{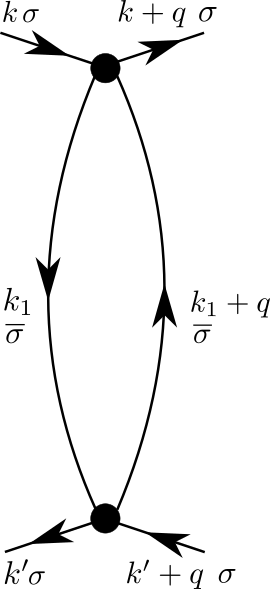}
   \end{minipage}
   \begin{minipage}{0.05\textwidth}
      $+$
   \end{minipage}
   \begin{minipage}{0.105\textwidth}
      \includegraphics[width=1.0\textwidth]{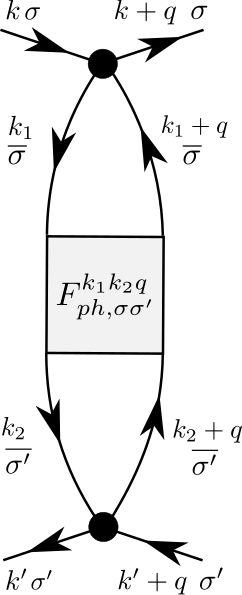}
   \end{minipage}
   \begin{minipage}{0.55\textwidth}
      \begin{minipage}{0.15\textwidth}
	 $\mathcal{K}_{1,\xph, \sigma \sigma'}^{q} =$
      \end{minipage}
      \hspace{0.03\textwidth}
      \begin{minipage}{0.59\textwidth}
	 \includegraphics[width=1.0\textwidth]{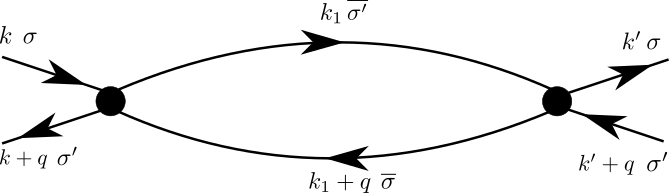}
      \end{minipage}
      \\[2ex]
      \hspace{0.13\textwidth}
      \begin{minipage}{0.05\textwidth}
	 $+$
      \end{minipage}
      \begin{minipage}{0.59\textwidth}
	 \includegraphics[width=1.0\textwidth]{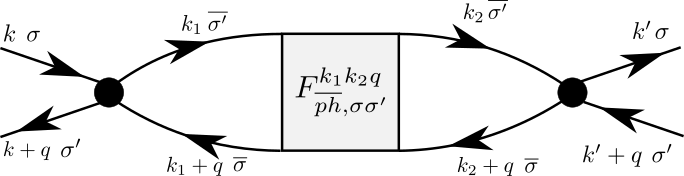}
      \end{minipage}
   \end{minipage}
   \caption{Diagrammatic representation of the $\K$ functions in the three different channels. As denoted in the first diagram, the external lines are to be excluded, making the $k$ and $k'$ arguments redundant. Here, $\overline{\sigma}$ denotes the opposite spin of $\sigma$, and $SU(2)$ symmetry is explicitly assumed. }
   \label{fig:k1_diag}
\end{figure*}
In the case of $\KK$ one introduces just one additional bare vertex, as shown in \cfg{k2_diag}. Note that here, the previously determined $\K$ has to be subtracted. The equations in all scattering channels then read
\begin{subequations}
\label{eq:k2_diag}
\begin{align}
\mathcal{K}_{2,pp, \sigma \sigma'}^{k q} &= - U \sumint dk_1 \, G(q-k_1) F_{pp, \sigma \sigma'}^{k k_1 q} G(k_1)
- {\cal K}_{1,pp, \sigma \sigma'}^{q},
\label{eq:k2_pp}
\\[1ex]
\mathcal{K}_{2,ph, \sigma \sigma'}^{k q} &= \phantom{-} U \sumint dk_1 \, G(k_1) F_{ph, \sigma \overline{\sigma'}}^{k k_1 q} G(k_1+q) 
- {\cal K}_{1, ph, \sigma \sigma'}^{q},
\label{eq:k2_ph}
\\[1ex]
\mathcal{K}_{2,\xph, \sigma \sigma'}^{k q} &= - U \sumint dk_1 \, G(k_1) G(k_1+q) \times 
\left[ \delta_{\sigma\sigma'}F_{\xph,\udc}^{k k_1 q} + (1-\delta_{\sigma\sigma'}) F_{\xph,\ud}^{k k_1 q} \right] - {\cal K}_{1,\xph, \sigma \sigma'}^{q}.
\label{eq:k2_xph}
\end{align}
\end{subequations}
\end{widetext}
Further, by exploiting the symmetry relations shown in Appendix \ref{app:symmetries}, one can easily derive $\KKB$ from $\KK$.
\begin{figure*}
   \centering
   \begin{minipage}{0.1\textwidth}
      $\mathcal{K}_{2,pp, \sigma \sigma'}^{q k} =$
   \end{minipage}
   \begin{minipage}{0.23\textwidth}
      \includegraphics[width=1.0\textwidth]{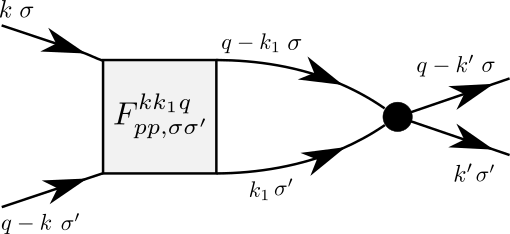}
   \end{minipage}
   \begin{minipage}{0.05\textwidth}
      $-$
   \end{minipage}
   \begin{minipage}{0.05\textwidth}
      $\mathcal{K}_{1, pp, \sigma \sigma'}^{q}$
   \end{minipage}
   \hspace{0.1\textwidth}
   \begin{minipage}{0.1\textwidth}
      $\mathcal{K}_{2,ph,\sigma \sigma'}^{q k} =$
   \end{minipage}
   \begin{minipage}{0.135\textwidth}
      \includegraphics[width=1.0\textwidth]{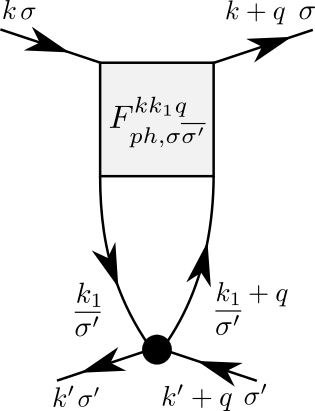}
   \end{minipage}
   \hspace*{-0.02\textwidth}
   \begin{minipage}{0.05\textwidth}
      $-$
   \end{minipage}
   \begin{minipage}{0.05\textwidth}
      $\mathcal{K}_{1,ph,\sigma \sigma'}^{q}$
   \end{minipage}
   \\[1ex]
   \begin{minipage}{0.1\textwidth}
      $\mathcal{K}_{2, \xph, \sigma \sigma'}^{q k} =$
   \end{minipage}
   \begin{minipage}{0.02\textwidth}
      $\delta_{\sigma\sigma'}$
   \end{minipage}
   \begin{minipage}{0.23\textwidth}
      \includegraphics[width=1.0\textwidth]{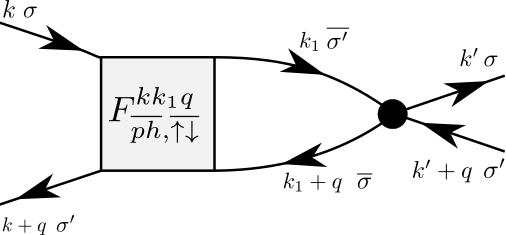}
   \end{minipage}
   \begin{minipage}{0.05\textwidth}
      $+$
   \end{minipage}
   \begin{minipage}{0.1\textwidth}
      $(1-\delta_{\sigma\sigma'})$
   \end{minipage}
   \begin{minipage}{0.23\textwidth}
      \includegraphics[width=1.0\textwidth]{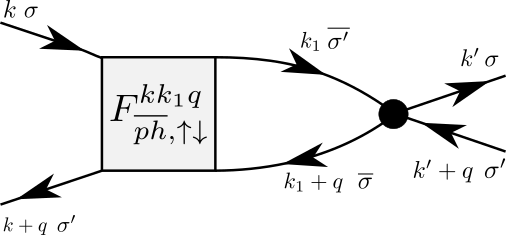}
   \end{minipage}
   \begin{minipage}{0.05\textwidth}
      $-$
   \end{minipage}
   \begin{minipage}{0.05\textwidth}
      $ \mathcal{K}_{1, \xph, \sigma \sigma'}^{q}$
   \end{minipage}
   \hspace{0.1\textwidth}
   \caption{Diagrammatic representation of the $\KK$ functions in the three different channels. Here, $\overline{\sigma}$ denotes the opposite spin of $\sigma$, while $SU(2)$ symmetry is explicitly assumed. }
   \label{fig:k2_diag}
\end{figure*}

As it is typical within an ED algorithm for a SIAM, the values for $F$ are known numerically for a finite grid in the frequency domain. 
Thus, in the first calculation of the aforementioned diagrams according to Eqs.~(\ref{eq:k1_diag}) and (\ref{eq:k2_diag}) we have to make a rough approximation for $F$ (i.e.~$F=U$) 
in the large-frequency domain, which will introduce an error. To improve on this `one-shot' calculation of the diagrams, we exploit a self-consistent scheme:
\begin{itemize}
   \item I: Initialize the $\K$'s and $\KK$'s to 0. Their grids may deviate from the grid for $F$. 
   \item II: Calculate a set of new $\K$'s and $\KK$'s according to Eqs.~(\ref{eq:k1_diag}) and (\ref{eq:k2_diag}).
   \item III: Rebuild the vertex in an arbitrarily large region (as needed) using the updated asymptotic functions. 
   \item IV: Continue from II till convergence
\end{itemize}
Once the asymptotic functions are fully converged, we can directly determine the localized structures using 
\begin{equation}
\label{eq:loc_struct} 
 \Lambda_{\mathrm{2PI}}+\sum_r \mc{R}_r= F - \left(\sum_r \mc{K}_{1,r} + \mc{K}_{2,r} + \mc{\overline{K}}_{2,r}\right).
\end{equation}  
Further, since, at this point, we have $F$ available in the full frequency domain, we can use this additional information to determine also all the $\Phi$ functions on an arbitrarily large frequency grid by means of the BSE~(\ref{eq:bethe_salpeter}). If needed, $\R$ and $\Lambda_{\rm{2PI}}$ can then be determined by \ceq{R} and \ceq{parquet} respectively. 

The approach described in this section was used to compute all the exact asymptotic functions and reducible vertices presented in Sec.~\ref{sec:compare_SIAM}, from ED calculations, originally performed on a fermionic frequency grid of $128\times128\times128$ Matsubara frequencies. While here we are dealing with a purely local vertex, we stress that this approach is equally applicable in the non-local case.

\section{Symmetries}					
\label{app:symmetries}

\subsection{Symmetries of $\K$ and $\KK$}
\begin{table*}
   \renewcommand{\arraystretch}{2.3}
   \centering
  \begin{tabular}{|c|c|c|}
  \hline 
  
   Symmetries & $\mathcal{K}_{1,r}$ & $\mathcal{K}_{2,r}$\\   \hline 
  
   $SU(2)$ & $\mathcal{K}_{1,r, \sigma \sigma'}^{q} = \mathcal{K}_{1,r, \overline{\sigma}\, \overline{\sigma'}}^{q}$ & $\mathcal{K}_{2,r, \sigma \sigma'}^{k q} = \mathcal{K}_{2,r, \overline{\sigma} \,\overline{\sigma'}}^{k q} $ \\ \newline 
   & $\mathcal{K}_{1,r, \sigma \sigma}^{q} = \mathcal{K}_{1,r, \sigma \sigma'}^{q} + \mathcal{K}_{1,r, \overline{\sigma \sigma'}}^{q}$  & $\mathcal{K}_{2,r, \sigma \sigma}^{k q} = \mathcal{K}_{2,r, \sigma \sigma'}^{k q} + \mathcal{K}_{2,r, \overline{\sigma \sigma'}}^{k q}$\\ \hline 
   
    Time reversal & $\mathcal{K}_{1,r, \sigma \sigma'}^{q} = \mathcal{K}_{1,r, \sigma' \sigma}^{q}$ & $\mathcal{K}_{2,r, \sigma \sigma'}^{k q} = {\KKB}_{r, \sigma \sigma'}^{k q}$\\ \hline  
   
    Particle hole & $\left(\mathcal{K}_{1,r, \sigma \sigma'}^{q}\right)^{*} = \mathcal{K}_{1,r, \sigma \sigma'}^{(\Omega,-{\bf q})}$ & $\left(\mathcal{K}_{2,r, \sigma \sigma'}^{k q}\right)^{*} = \mathcal{K}_{2,r, \sigma \sigma'}^{(\nu,{\bf \Pi}-{\bf k}) (\Omega,-{\bf q})}$\\
   
  \hline  
  \end{tabular}
  \caption{Symmetry table for $\K$ and $\KK$. Here, ${\bf \Pi} = (\pi, \pi, ...)$ represents the ($d$-dimensional) 'antiferromagnetic momentum' in the case of a simple (hyper)cubic lattice with lattice constant $a=1$. }
  \label{tab:symm}
\end{table*}

In this section we summarize the symmetries of the previously introduced asymptotic functions. Before addressing the specific 
physical symmetries of the system of our interest, which provide useful relations for $\K$ and $\KK$, we provide some fundamental relations which hold\cite{Bickers2004,Rohringer2012,Rohringer2014} independently of the system under investigation. 
First, we consider the exchange of two (fermionic) annihilation operators in the time-ordered matrix element of \ceq{G2_iw_pf}, which, as a consequence of the Pauli-principle, yields a minus sign (also referred to as `crossing symmetry'\cite{Bickers2004,Rohringer2012,Rohringer2014}). Diagrammatically speaking, this corresponds to an exchange of two outgoing lines.
For $\K$, this operation leads to the following relations:
\begin{subequations}
\begin{align}
&\mathcal{K}_{1,pp,\sigma \sigma'}^{q} =-\mathcal{K}_{1,pp,\overline{\sigma \sigma'}}^{q}\\
&\mathcal{K}_{1,ph,\sigma \sigma'}^{q} = -\mathcal{K}_{1,\xph,\overline{\sigma \sigma'}}^{q}\\ &\mathcal{K}_{1,\xph,\sigma \sigma'}^{q} =-\mathcal{K}_{1,ph,\overline{\sigma \sigma'}}^{q}.
\end{align} 
\end{subequations}
Here, $\overline{\sigma\sigma'}$ denotes a spin configuration of the external indices where spins are crossing.
While for the $pp$ channel one finds relations between different spin configurations within the same channel, the $ph$ and $\xph$-channel are interchanged.
Similarly, for $\KK$ one finds:
\begin{subequations}
\begin{align}
&\mathcal{K}_{2,pp,\sigma \sigma'}^{k q} =-\mathcal{K}_{2,pp,\overline{\sigma \sigma'}}^{k q}\\
&\mathcal{K}_{2,ph,\sigma \sigma'}^{k q} = -\mathcal{K}_{2,\xph,\overline{\sigma \sigma'}}^{k q}\\ 
&\mathcal{K}_{2,\xph,\sigma \sigma'}^{k q} =-\mathcal{K}_{2,ph,\overline{\sigma \sigma'}}^{kq}.
\end{align} 
\end{subequations}

A second generic operation involves the simultaneous exchange of both annihilation and creation operators in 
Eq.~\ref{eq:G2_iw_pf}. Diagrammatically, this corresponds to an exchange of both the incoming and outgoing particles. In this case we end up with the following relations for $\K$:
\begin{subequations}
\begin{align}
&\mathcal{K}_{1,pp,\sigma \sigma'}^{q} =\mathcal{K}_{1,pp,\sigma' \sigma}^{q}\\
&\mathcal{K}_{1,ph,\sigma \sigma'}^{q} =\mathcal{K}_{1,ph,\sigma' \sigma}^{-q}\\
&\mathcal{K}_{1,\xph,\sigma \sigma'}^{q} =\mathcal{K}_{1,\xph,\sigma' \sigma}^{-q}.
\end{align}
\end{subequations}
For $\KK$ one obtains:
\begin{subequations}
\begin{align}
&\mathcal{K}_{2,pp,\sigma \sigma'}^{k q} = \mathcal{K}_{2,pp,\sigma' \sigma}^{(q-k)q}\\
&\mathcal{K}_{2,ph,\sigma \sigma'}^{k q} = \overline{\mathcal{K}}_{2,ph,\sigma' \sigma}^{(k+q)(-q)}\\
&\mathcal{K}_{2,\xph,\sigma \sigma'}^{k q} = \overline{\mathcal{K}}_{2,\xph,\sigma' \sigma}^{(k+q)(-q)}.
\end{align}
\end{subequations}
While for this operation all the corresponding channels are conserved, the diagrammatic class changes from $\KK$ to $\KKB$ in the case of the $ph$ and $\xph$ channel.

To conclude the discussion of the fundamental relations, we consider the complex conjugation operation, that leads to the following relations for $\K$:
\begin{subequations}
\begin{align}
&\left(\mathcal{K}_{1,pp,\sigma \sigma'}^{q}\right)^* =\mathcal{K}_{1,pp,\sigma' \sigma}^{(-\Omega, {\bf q})}\\
&\left(\mathcal{K}_{1,ph,\sigma \sigma'}^{q}\right)^* =\mathcal{K}_{1,ph,\sigma' \sigma}^{(\Omega, {\bf -q})}\\
&\left(\mathcal{K}_{1,\xph,\sigma \sigma'}^{q}\right)^* =\mathcal{K}_{1,\xph,\sigma' \sigma}^{(-\Omega, {\bf q})},
\end{align}
\end{subequations}
and for $\KK$:
\begin{subequations}
\begin{align}
&\left(\mathcal{K}_{2,pp,\sigma \sigma'}^{k q}\right)^* =\mathcal{K}_{2,pp,\sigma' \sigma}^{(-\Omega+\nu', {\bf q}-{\bf k'})(-\Omega, {\bf q})}\\
&\left(\mathcal{K}_{2,ph,\sigma \sigma'}^{k q}\right)^* =\mathcal{K}_{2,ph,\sigma' \sigma}^{(-\Omega - \nu, {\bf q}+{\bf k})(\Omega,-{\bf q})}\\
&\left(\mathcal{K}_{2,\xph,\sigma \sigma'}^{k q}\right)^* =\mathcal{K}_{2,\xph,\sigma' \sigma}^{(-\nu', {\bf k'})(-\Omega,{\bf q})}.
\end{align}
\end{subequations}

Using these fundamental relations, we can formulate the system-related physical symmetries, namely $SU(2)$, time reversal and particle-hole symmetry, in a channel-independent way. The results are summarized in Table \ref{tab:symm}.
Note that for the particle-hole symmetry, the relations differ for the frequency and momentum dependence. While in the purely local case, this symmetry implies a vanishing imaginary part of all two-particle quantities, this holds only for specific lattice-dependent $\bf{k}$ vectors in the non-local case. 
\subsection{Symmetries of $\R$}
\begin{table}
\renewcommand{\arraystretch}{1.8}
\centering
\vspace{0.3cm}
  \begin{tabular}{|c|c|}
   \hline
   Symmetries & ${\R}_r$ \\ \hline 
   $SU(2)$ & ${\R}_{r, \sigma \sigma'}^{k k'q} = {\R}_{r, \overline{\sigma} \,\overline{\sigma'}}^{k k' q} $ \\ \newline
   & ${\R}_{r, \sigma \sigma}^{k k' q} = {\R}_{r, \sigma \sigma'}^{k k' q} + {\R}_{r, \overline{\sigma \sigma'}}^{k k' q}$  \\ \hline
   Time reversal & ${\R}_{r, \sigma \sigma'}^{k k' q} = {\R}_{r, \sigma' \sigma}^{k' k q}$\\ \hline
   Particle hole & $({\R}_{r, \sigma \sigma'}^{k k' q})^{*} = {\R}_{r, \sigma \sigma'}^{(\nu, {\bf \Pi}-{\bf k})(\nu',{\bf \Pi}-{\bf k'})(\Omega,-{\bf q})}$\\
  \hline  
  \end{tabular}
  \caption{Symmetry table for $\R$. Note that the same table holds for the two-particle reducible vertex functions $\Phi_r$. }
  \label{tab:symm_R}
\end{table}
For the sake of completeness we report the symmetries of the remaining diagrammatic class, namely the rest function $\R$, which hold equally for the reducible vertex functions $\Phi$. 
As shown above, the first set of fundamental relations results from exchanging two outgoing particles. We find the following relations for $\R$ in the different channels:
\begin{subequations}
\begin{align}
&\R_{pp, \sigma \sigma'}^{k k' q} = - \R_{pp, \overline{\sigma \sigma'}}^{(k)( q-k')q}\\
&\R_{ph, \sigma \sigma'}^{k k' q} = - \R_{\xph, \overline{\sigma \sigma'}}^{k k' q}\\
&\R_{\xph, \sigma \sigma'}^{k k' q} = - \R_{ph, \overline{\sigma \sigma'}}^{k k'q}.
\end{align} 
\end{subequations}
By means of the simultaneous exchange of both incoming and outgoing particles we obtain:
\begin{subequations}
\begin{align}
& \R_{pp, \sigma \sigma'}^{k k' q} = \R_{pp, \sigma' \sigma}^{(q-k)( q-k')q}\\
& \R_{ph, \sigma \sigma'}^{k k' q} = \R_{ph, \sigma' \sigma}^{(k'+q)(k+q)(-q)}\\
& \R_{\xph, \sigma \sigma'}^{k k' q} = \R_{\xph, \sigma' \sigma}^{(k'+q)( k+q)(-q)}.
\end{align}
\end{subequations}
Finally, the complex conjugation operation leads to the following relations:
\begin{subequations}
\begin{align}
&\left(\R_{pp, \sigma \sigma'}^{k k' q}\right)^* = \R_{pp, \sigma' \sigma}^{(-\Omega+\nu',{\bf q}-{\bf k'} )(-\Omega+\nu,{\bf q}-{\bf k})(-\Omega, {\bf q})}\\
&\left(\R_{ph, \sigma \sigma'}^{k k' q}\right)^* = \R_{ph, \sigma' \sigma}^{(-\Omega-\nu,{\bf q}+{\bf k'} )(-\Omega+\nu',{\bf q}+{\bf k})(\Omega, -{\bf q})}\\
&\left(\R_{\xph, \sigma \sigma'}^{k k' q}\right)^* = \R_{\xph, \sigma' \sigma}^{(-\nu',{\bf k'} )(-\nu,{\bf k})(-\Omega, {\bf q})}.
\end{align}
\end{subequations}
In the same way as for $\K$ and $\KK$, the fundamental relations for $\R$ allow us to express the physical symmetries in a channel-independent way, see Table \ref{tab:symm_R}.

\bibliography{refs}

\end{document}